**Experimental issues in coherent quantum-state manipulation of trapped atomic ions**[*]

D.J. Wineland, C. Monroe, W.M. Itano, D. Leibfried[†], B.E. King, D.M. Meekhof

*NIST, Boulder, CO, 80303*

Methods for, and limitations to, the generation of entangled states of trapped atomic ions are examined. As much as possible, state manipulations are described in terms of quantum logic operations since the conditional dynamics implicit in quantum logic is central to the creation of entanglement. Keeping with current interest, some experimental issues in the proposal for trapped-ion quantum computation by I. Cirac and P. Zoller (University of Innsbruck) are discussed. Several possible decoherence mechanisms are examined and what may be the more important of these are identified. Some potential applications for entangled states of trapped-ions which lie outside the immediate realm of quantum computation are also discussed.

Key words: coherent control; entangled states; laser cooling and trapping; quantum computation; quantum state engineering; trapped ions.

**CONTENTS**

1. Introduction

A number of recent theoretical and experimental papers have investigated the ability to coherently control or "engineer" atomic, molecular, and optical quantum states. This theme is manifested in topics such as atom interferometry, atom optics, the atom laser, Bose-Einstein condensation, cavity QED, electromagnetically induced transparency, lasing without inversion, quantum computation, quantum cryptography, quantum-state engineering, squeezed states, and wavepacket dynamics. In this paper, we investigate a subset of these topics which involve the coherent manipulation of quantum states of trapped atomic ions. The focus will be on a proposal to implement quantum logic and quantum computation using trapped ions [1]. However, we will also consider related work on the generation of nonclassical states of motion and entangled states of trapped ions [2- 39]. Many of these ideas have been summarized in a recent review [40].

Coherent control of spins and internal atomic states has a long history in NMR and rf /laser spectroscopy. For example, the ability to realize coherent "$\pi$ pulses" or "$\pi/2$ pulses" on two-level systems has been routine for decades. In much of what is discussed in this paper, we will consider entangling operations, that is, unitary operations which create entangled states between two or more quantum systems. In particular, we will be interested in situations where the interaction between quantum systems can be selectively turned on and off. For brevity, we will limit discussion to these types of operations in experiments which involve trapped atomic ions; however, many of the discussions, in particular those concerning single trapped ions, will also apply to trapped neutral atom experiments where the atoms can be treated as independent. The aspect of entangling operations is shared by atom optics and atom interferometry [41,42] and, as described below, there are close parallels between the ion trap experiments and those of cavity QED [43].

Earlier experiments on trapped ions, where the zero-point of motion was closely approached through laser cooling, already showed the effects of nonclassical motion in the absorption spectrum [44-46]. These same effects can be used to characterize the average energy of the ion. More recent experiments report the generation of Fock, squeezed, coherent [21], and Schrödinger cat [47] states. These states appear to be of fundamental physical interest and possibly of use for sensitive detection of small forces [26,48]. For comparison, experiments which detect quantized atomic motion in optical lattices are reviewed by Jessen and Deutsch [49]. Also, through the mechanism of Bose-Einstein condensation, which has recently been observed in neutral atomic vapors [50-53], a macroscopic occupation of a single motional state (the ground state of motion) is achieved.

Simple quantum logic experiments have been carried out with single trapped ions [17]; the emphasis of future work will be to implement quantum logic on many ions [1]. The attendant ability to create correlated, or entangled, states of atomic particles appears to be interesting from the standpoint of quantum measurement [54] and, for example, for improved signal-to-noise ratio in spectroscopy of trapped ions (Sec. 3.4).

Therefore, we will be particularly interested in studying the practical limits of applying coherent control methods to trapped ions for (1) the generation and analysis of nonclassical states of motion, (2) the implementation of quantum logic and computation, and (3) the generation of entangled states which can improve signal-to-noise ratio in spectroscopy. We will briefly describe the experimental results in these three areas, but the main purpose of the paper will be to anticipate and characterize decohering mechanisms which limit the ability to produce the desired final quantum states in current and future experiments. This is a particularly important issue for quantum computation where many ions (thousands) and coherent operations (billions) may be required in order for quantum computation to be generally useful. Here, we generalize the meaning of decoherence to include any effect which limits the purity of the desired final states. A fundamental source of decoherence will be the coupling of the ion's motion and internal states to the environment. Also important is induced decoherence caused by, for example, technical fluctuations in the applied fields used to implement the operations. This division between types of decoherence is arbitrary since both effects can be regarded as coupling to the environment; however, the division will provide a useful framework for discussion. As a unifying theme for the paper, we will find it useful to regard, as much as possible, the quantum manipulations we discuss in terms of quantum logic. Of course, the subject of decoherence is much broader than the specific context discussed here; the reader is referred to more general discussions such as the papers by Zurek [55,56,57].

    The paper is organized as follows. In the next section, we briefly discuss ion trapping. In Sec. 3, we consider in somewhat more detail the three areas of application enumerated in the previous paragraph. Since cooling of the ions to their ground state of motion is a prerequisite to the main applications discussed in the paper, we outline methods to accomplish this in the beginning of Sec. 3. Section 4 is the heart of the paper; here, we attempt to identify the most important sources of decoherence. Section 5 briefly discusses some variations on proposed methods for realizing quantum logic in trapped ions. Section 6 suggests some additional applications of the ideas discussed in the paper and Sec. 7 provides a brief summary.

    Such a treatment seems warranted in that several laboratories are investigating the use of trapped ions for quantum logic and related topics; the authors are aware of related experiments being pursued at IBM, Almaden; Innsbruck University; Los Alamos National Laboratory; Max Planck Institute, Garching; NIST, Boulder; and Oxford University. This analysis in this paper necessarily overlaps, but is also intended to complement, other investigations [58-66] and will, by no means, be the end of the story. We hope however, that this paper will stimulate others to do more complete treatments and consider effects that we have neglected.

2. Trapped atomic ions

2.1. Ions confined in Paul traps

    Due to their net charge, atomic ions can be confined by particular arrangements of electromagnetic fields. For studies of ions at low energy, two types of trap are typically used - the Penning trap, which uses a combination of static electric and magnetic fields, and the Paul or rf trap which confines ions primarily through ponderomotive forces generated by inhomogeneous oscillating fields. The operation of these traps is discussed in various reviews (see for example, Refs. [67] - [70]), and in a recent book by Ghosh [71]. For brevity, we discuss one trap configuration, the linear Paul trap, which may be particularly useful in the context of this paper. This choice however, does not rule out the use of other types of ion traps for the

experiments discussed here.

In Fig. 1 we show a schematic diagram of a linear Paul trap. This trap is based on the one described by Raizen, et al. [72] which is derived from the original design of Drees and Paul [73]. It is basically a quadrupole mass filter which is plugged at the ends with static electric potentials. A potential $V_o \cos\Omega_T t + U_r$ is applied between diagonally opposite rods, which are fixed in a quadrupolar configuration, as indicated in Fig. 1. We assume that the rod segments along the z direction are coupled together with capacitors (not shown) so that the rf potential is constant as a function of z. Near the axis of the trap this creates a potential of the form

$$\Phi \simeq \frac{(V_o \cos\Omega_T t + U_r)}{2}\left(1 + \frac{x^2 - y^2}{R^2}\right), \quad (1)$$

where $R$ is equal to the distance from the axis to the surface of the electrode. (Unless the rods conform to equipotentials of Eq. (1), this equation must be multiplied by a constant factor on the order of 1; see for example, Ref. [72].) This gives rise to (harmonic) ponderomotive potentials in the x and y directions. To provide confinement along the z direction, static potentials $U_o$ are applied to the end segments of the rods as indicated. Near the center of the trap, this provides a static harmonic well in the z direction

$$\Phi_s = \kappa U_o\left[z^2 - \frac{1}{2}(x^2 + y^2)\right] = \frac{m}{2q}\omega_z^2\left[z^2 - \frac{1}{2}(x^2 + y^2)\right], \quad (2)$$

where $\kappa$ is a geometric factor, m and q are the ion mass and charge, and $\omega_z = (2\kappa q U_o/m)^{1/2}$ is the oscillation frequency for a single ion or the center-of-mass (COM) oscillation frequency for a collection of identical ions along the z direction. Equations (1) and (2) represents the lowest order terms in the expansion of the potentials for the electrode configuration of Fig. 1. When the size of the ion sample or amplitude of ion motion is comparable to the spacing between electrodes or the spacing between rod segments, higher order terms in $\Phi$ and $\Phi_s$ become important. However for small oscillations of the COM mode, which is relevant here, the harmonic approximation will be valid. In the x and y directions, the action of the potentials of Eqs. (1) and (2) gives the (classical) equations of motion described by the Mathieu equation

$$\frac{d^2x}{d\zeta^2} + [a_x + 2q_x \cos(2\zeta)]x = \frac{d^2y}{d\zeta^2} + [a_y + 2q_y \cos(2\zeta)]y = 0, \quad (3)$$

where $\zeta \equiv \Omega_T t/2$, $a_x = (4q/m\Omega_T^2)(U_r/R^2 - \kappa U_o/z_o^2)$, $a_y = -(4q/m\Omega_T^2)(U_r/R^2 + \kappa U_o/z_o^2)$, $q_x = -q_y = 2qV_o/(\Omega_T^2 mR^2)$. The Mathieu equation can be solved in general using Floquet solutions. Typically, we will have $a_i < q_i^2 \ll 1$, $i \in \{x,y\}$. (Keeping with the usual notation in the ion-trap literature, in this section, the symbols $a_i$ and $q_i$ are defined as above. In all other sections, $a_i$ (or a) will represent the harmonic oscillator lowering operator and $q_i$ will represent the normal mode coordinate for the ith mode). The solution of Eqs. (3) to first order in $a_i$ and second order in $q_i$ is given by

$$u_i(t) = A_i \left( \cos(\omega_i t + \phi_i) \left[ 1 + \frac{q_i}{2} \cos(\Omega_T t) + \frac{q_i^2}{32} \cos(2\Omega_T t) \right] + \beta_i \frac{q_i}{2} \sin(\omega_i t + \phi_i) \sin(\Omega_T t) \right), \quad (4)$$

where $u_i = x$ or $y$, $A_i$ depends on initial conditions, and

$$\omega_i = \beta_i \frac{\Omega_T}{2}, \qquad \beta_i \simeq \left[ \frac{a_i + q_i^2/2}{1 - 3q_i^2/8} \right]^{\frac{1}{2}}. \quad (5)$$

The large amplitude oscillation at frequency $\omega_i$ is typically called the "secular" motion. When $a_i \ll q_i^2 \ll 1$ and $U_r \simeq 0$, if we neglect the micromotion (the terms which oscillate at $\Omega_T$ and $2\Omega_T$), the ion behaves as if it were confined in a harmonic pseudopotential $\Phi_p$ in the radial direction given by

$$q\Phi_p = \frac{1}{2} m \omega_r^2 (x^2 + y^2), \quad (6)$$

where $\omega_r \simeq qV_o/(2^{\frac{1}{2}}\Omega_T mR^2) = q_x \Omega_T/(2\sqrt{2})$ is the radial secular frequency $\omega_r$. For most of the discussions in this paper, we will assume $U_r = 0$; however it may be useful in some cases to make $U_r \neq 0$ to break the degeneracy of the x and y frequencies. Figure 1 also shows an image of a "string" of $^{199}Hg^+$ ions which are confined near the z axis of the trap described in Ref. [74]. This was achieved by making $\omega_r \gg \omega_z$, thereby forcing the ions to the axis of the trap. The spacings between individual ions in this string are governed by a balance of the force along the z direction due to $\Phi_s$ and the mutual Coulomb repulsion of the ions. Example parameters are given in the figure caption.

When this kind of trap is installed in a high-vacuum apparatus, ions can be confined for days with minimal perturbations to their internal structure. Collisions with background gas can be neglected (Sec. 4.1.9). Even though the ions interact strongly through their mutual Coulomb interaction, the fact that the ions are localized necessarily means that the time-averaged value of the electric field they experience is zero; therefore electric field perturbations are small (Sec. 4.2.3). Magnetic field perturbations to internal structure are important; however, the coherence time for superposition states of two internal levels can be very long by operating at fields where the energy separation between levels is at an extremum with respect to field. For example, in a $^9Be^+$ (Penning trap) experiment operating in a field of 0.82 T, a coherence time between hyperfine levels exceeding 10 minutes was observed [75,76]. As described below, we will be interested in coherently exciting the quantized modes of the ions' motion in the trap. Here, not surprisingly, the coupling to the environment is relatively strong because of the ions' charge. One measure of the decoherence rate is obtained from the linewidth of observed motional resonances of the ions; this gives an indication of dephasing times. For example, the linewidths of cyclotron resonance excitation in high resolution mass spectroscopy in Penning traps [77-79] indicate that these coherence times can be at least as long as several tens of seconds. Decoherence can also occur from transitions between the ions' quantized oscillator levels. Transition times out of the zero-point motional energy level have been measured for single $^{198}Hg^+$ ions to be about 0.15 s [44] and for single $^9Be^+$ ions to be about 1 ms [45]. These relatively short times are, so far, unexplained; however, it might be possible to achieve much longer times in the future (Sect. 4.1).

In the linear trap, the radial COM vibration frequency $\omega_r$ must be made sufficiently higher than the axial COM vibrational frequency $\omega_z$ in order for the ions to be collinear along the z axis of the trap. This configuration will aid in addressing individual ions with laser beams and will also suppress rf heating (Sec. 4.1.5). To prevent zig-zag and other complicated shapes of the ion crystal, we require $\omega_r/\omega_z > 1$ for two ions, and $\omega_r/\omega_z > 1.55$ for three ions. For L > 3 ions, the critical ratio $(\omega_r/\omega_z)_c$ for linear confinement has been estimated analytically [80,81] yielding $(\omega_r/\omega_z)_c \simeq 0.73L^{0.86}$ [60]. Other estimates are given in Refs. [82] ($(\omega_r/\omega_z)_c \simeq 0.63 L^{0.865}$) and [64] ($(\omega_r/\omega_z)_c \simeq 0.59L^{0.885}$). An equivalent result is obtained if we consider that as the potential is weakened in the radial direction, ions in a long string which are spaced by distance $s_c$ near the center of the string, will first break into a zig zag configuration. At the point where the ions break into a zig-zag, the net outward force from neighboring ions is equal to the inward trapping force. If we equate these forces, we obtain

$$\omega_r^2 = \frac{7}{8\pi\epsilon_o}\zeta(3)\frac{q^2}{ms_c^3}, \tag{7}$$

where $\zeta$ is the Riemann zeta function. As an example, for a $^9Be^+$ ion, m ≃ 9 u (atomic mass units) and $s_c$ = 3 μm, we must have $\omega_r/2\pi$ > 7.8 MHz to keep the ions along the axis of the trap.

The equilibrium spacing of a linear configuration of trapped ions is not uniform; the middle ions are spaced closer than the outlying ions, as is apparent in Fig. 1. The separation of two ions is $s_2 = 2^{1/3}s$, where $s = (q^2/4\pi\epsilon_o m\omega_z^2)^{1/3}$ is a length scale of ion-ion spacings; the adjacent separation of three ions is $s_3 = (5/4)^{1/3}s$. For L large, estimates of the minimum separation of the center ions are given by $s_c(L) \simeq 2sL^{-0.56}$ [60], $2.018sL^{-0.559}$ [61], $2.29L^{-0.596}$ [83], and $1.92sL^{-2/3}[\ln(0.8L)]^{1/3}$ [59,81]. For typical trapping parameters, the ion-ion separations are on the order of a few μm and the spatial spread of the zero-point vibrational wavepackets are on the order of 10 nm. Thus there is negligible wavefunction overlap between ions and quantum statistics (Bose or Fermi) play no role in the spatial wavefunction of an array.

Of the 3L normal modes of oscillation in a linear trap, we are primarily interested in the L modes associated with axial motion because we will preferentially couple to them with applied laser fields. A remarkable feature of the linear ion trap is that the axial modes frequencies are nearly independent of L [1,60,61,84]. For two ions, the axial normal mode frequencies are at $\omega_z$ and $\sqrt{3}\omega_z$; for three ions they are $\omega_z$, $\sqrt{3}\omega_z$, and $(5.4)^{1/2}\omega_z$. For L > 3 ions, the Lth axial normal mode can be determined numerically [60,61,84].

2.2. Ion motional and internal quantum states

A single ion's motion, or the COM mode of a collection, has a simple description when the ions are trapped in a purely static potential, which is the case for the axial motion in a Penning trap or the axial motion in the trap of Fig. 1. We will assume that the trap potentials are quadratic (Eqs. (1) and (2)). This is a valid approximation when the amplitudes of motion are small, because the local potential, expanded about the equilibrium point of the trap, is quadratic to a good approximation. In this case, motion is harmonic. An ion trapped in a ponderomotive potential (Eq. (6)) can be described effectively as a simple harmonic oscillator, even though the Hamiltonian is actually time-dependent, so no stationary states exist. For practical purposes, the system can be treated as if the Hamiltonian were that of an ordinary, time-independent harmonic oscillator [34,35,85-91] although modifications must be made for laser cooling [92]. The classical micromotion (the terms which vary as $\cos\Omega_T t$ and $\cos 2\Omega_T t$ in Eq. (4)) may be viewed, in the quantum picture, as causing the ion's wavefunction to breathe at the drive frequency $\Omega_T$. This breathing motion is separated spectrally from the secular motion (at frequencies $\omega_x$ and $\omega_y$ in Eq. (4)). Since the operations we will consider rely on a resonant interaction at the secular frequencies, we will average over the components of motion at the drive frequency $\Omega_T$. Therefore, to a good approximation, the pseudopotential secular motion behaves as an oscillator in a static potential. The main consequence of the quantum treatment is that transition rates between quantum levels (Eq. (18), below) are altered [34,35]; however, these changes can be accounted for by experimental calibration. In any case, for most of the applications discussed in this paper, we will be considering the motion of the ions along the axis of a linear Paul trap where this modification is absent.

Therefore, the Hamiltonian describing motion of a single ion (or a normal mode, such as the COM mode, of a collection of ions) in the ith direction is given by

$$H_{osc} = \hbar\omega_i \hat{n}_i, \quad i \in \{x,y,z\}, \qquad (8)$$

where $\hat{n}_i \equiv a_i^\dagger a_i$ and $a_i^\dagger$ and $a_i$ are the usual harmonic oscillator raising and lowering operators and we have suppressed the zero-point energy $\frac{1}{2}\hbar\omega_i$. The operator for the COM motion in the z direction is given by

$$z = z_o(a + a^\dagger), \qquad (9)$$

where $z_o = (\hbar/2m\omega_z)^{1/2}$ is the spread of the zero-point wavefunction and m is the ion mass. That is, $z_o = (\langle 0|z^2|0\rangle)^{1/2}$, where $|n\rangle$ is the nth eigenstate ("number" or Fock state) of the harmonic oscillator. For $^9$Be$^+$ ions in a trap where $\omega_z/2\pi = 10$ MHz, we have $z_o = 7.5$ nm. Therefore, a general pure state of motion for one mode can be written, in the Schrödinger picture, as

$$\Psi_{motion} = \sum_{n=0}^{\infty} C_n e^{-in\omega_i t} |n\rangle , \qquad (10)$$

where $C_n$ are complex and the $|n\rangle$ are time-independent. For applications to quantum logic, we will be interested in motional states of the simple form $\alpha|0\rangle + \beta\exp(-i\omega_i t)|1\rangle$.

We will be interested in the situation where, at any given time, we interact with only two internal levels of an ion. This will be accomplished by insuring that the internal states are nondegenerate and by using resonant excitations to couple only two levels at a time. We will find it convenient to represent a two-level system by its analogy with a spin-½ magnetic moment in a static magnetic field [93,94]. In this equivalent representation, we assume that a (fictitious) magnetic moment $\vec{\mu} = \mu_M \vec{S}$, where $\vec{S}$ is the spin operator (S = ½), is placed in a (fictitious) magnetic field $\vec{B} = B_o \hat{z}$. The Hamiltonian can therefore be written

$$H_{internal} = \hbar\omega_o S_z , \qquad (11)$$

where $S_z$ is the operator for the z component of the spin and $\omega_o \equiv -\mu_M B_o/\hbar$. Typically, the internal resonant frequency will be much larger than any motional mode frequency, $\omega_o \gg \omega_z$. We label the internal eigenstates $|M_z\rangle = |\uparrow\rangle$ and $|\downarrow\rangle$ representing "spin-up" and "spin-down" respectively and, for convenience, will assume $\mu_M < 0$ so that the energy of the $|\uparrow\rangle$ state is higher than the $|\downarrow\rangle$ state. A general pure state of the two-level system is then given by

$$\Psi_{internal} = C_\downarrow e^{i\frac{\omega_o t}{2}} |\downarrow\rangle + C_\uparrow e^{-i\frac{\omega_o t}{2}} |\uparrow\rangle , \qquad (12)$$

where $|C_\downarrow|^2 + |C_\uparrow|^2 = 1$. Of course, the two-level system really could be a S=½ spin such as a trapped electron or the ground state of an atomic ion with a single unpaired outer electron and zero nuclear spin such as $^{24}$Mg$^+$.

2.2.1. Detection of internal states

The applications considered below will benefit from high detection efficiency of the ion's internal states. Unit detection efficiency has been achieved in experiments on "quantum jumps" [95-99] where the internal state of the ion is indicated by light scattering (or lack thereof), correlated with the ion's internal state. (More recently, this type of detection has been used in spectroscopy so that the noise is limited by the fundamental quantum fluctuations in detection of the internal state [100]. In these experiments, detection is accomplished with a laser beam appropriately polarized and tuned to a transition that will scatter many photons if the atom is in one internal state (a "cycling" transition), but will scatter essentially no photons if the atom is in the other internal state. If a modest number of these photons are detected, the efficiency of our ability to discriminate between these two states approaches 100%. We note that for a string of ions in a linear trap, the scattered light from one ion will impinge on the other ions; however, since this light has nearly the same frequency of as the light used for detection, this should not affect the overall detection efficiency.

The overall efficiency can be explained as follows. Suppose the atom scatters N total photons if it is measured to be in state $|\downarrow\rangle$ and no photons if it is measured to be in state $|\uparrow\rangle$. In practice N will be limited by optical pumping but can be $10^6$ or higher [101]. Here, we assume that N is large enough that we can neglect its fluctuations from experiment to experiment. We typically detect only a small fraction of these photons due to small solid angle collection and small detector efficiency. Therefore, on average, we detect $n_d = \eta_d N$ photons where $\eta_d \ll 1$ is the net photon detection efficiency. If we can neglect background, then for each separate experiment, if we detect at least one scattered photon, we can assume the ion is in state $|\downarrow\rangle$. If we detect no photons, the probability of a false reading, that is, the probability the ion is in state $|\downarrow\rangle$ but we simply did not detect any photons, is given by $P_N(0) = (1 - \eta_d)^N \approx \exp(-n_d)$. For $n_d = 10$, $P_N(0) \approx 4.5 \times 10^{-5}$, for $n_d = 100$, $P_N(0) \approx 4 \times 10^{-44}$. Therefore for $n_d > 10$, detection can be highly efficient.

Detection of ion motion can be accomplished directly by observing the currents induced in the trap electrodes [102-104]. However, the sensitivity of this method is limited by electronic detection noise. Because the detection of internal states can be so efficient, motional states can be detected by mapping their properties onto internal states which are then detected (Sec. 3.2).

2.3. Interaction with additional applied electromagnetic fields

2.3.1. Single ion, single applied field, single mode of motion

We first consider the situation where a single, periodically varying, (classical) electromagnetic field propagating along the z direction is applied to a single trapped ion which is constrained to move in the z direction in a harmonic well with frequency $\omega_z$. We consider situations where fields resonantly drive transitions between internal or motional states <u>and</u> when they drive transitions between these states simultaneously (entanglement). If we assume that the internal levels are coupled by electric fields, then the interaction Hamiltonian is

$$H_I = -\vec{\mu}_d \cdot \vec{E}(\vec{z}, t),  \quad\quad (13)$$

where $\vec{\mu}_d$ is the electric dipole operator for the internal transition and $\vec{E}$ is from a uniform wave propagating along the z direction and polarized in the x direction, $\vec{E} = E_1 \hat{x} \cos(kz - \omega t + \phi)$, where $\omega$ is the frequency, k is the wavevector $2\pi/\lambda$, and $\lambda$ is the wavelength. In the equivalent spin-½ analog, we assume that a traveling wave magnetic field propagates along the z direction, is polarized in the x direction ($\vec{B} = B_1 \hat{x} \cos(kz - \omega t + \phi)$), and interacts with the fictitious spin ($\vec{\mu} = \mu_M \vec{S}$). Therefore, for the spin analog, Eq. (13) is replaced by

$$H_I = -\vec{\mu} \cdot \vec{B}(z,t) = \hbar \Omega (S_+ + S_-)(e^{i(kz - \omega t + \phi)} + e^{-i(kz - \omega t + \phi)}), \qquad (14)$$

where $\hbar\Omega \equiv -\mu_M B_1/4$ (or $-\mu_d E_1/4$ for an electric dipole), $S_+ \equiv S_x + iS_y$, $S_- \equiv S_x - iS_y$ and z is given by Eq. (9). We will assume that the lifetimes of the levels are long; in this case, the spectrum of the transitions excited by the traveling wave is well resolved if $\Omega$ is sufficiently small.

It will be useful to transform to an interaction picture where we assume $H_o = H_{internal} + H_{osc}$ and $V_{interaction} = H_I$. In this interaction picture, if we make the rotating-wave approximation (neglecting $\exp(\pm i(\omega + \omega_o)t)$ terms), the wavefunction can be written

$$\Psi = \sum_{M_z = \downarrow, \uparrow} \sum_{n=0}^{\infty} C_{M_z, n}(t) |M_z\rangle |n\rangle, \qquad (15)$$

where $|M_z\rangle$ and $|n\rangle$ are the time-independent internal and motional eigenstates. In general, this wavefunction will be entangled between the two degrees of freedom; that is, we will not be able to write the wavefunction as a product of internal and motional wavefunctions. We have $H \rightarrow H_I' = U_o^\dagger(t) H_I U_o(t)$ where $U_o(t) = \exp(-i(H_o/\hbar)t)$, resulting in

$$H_I' = \hbar \left[ \Omega S_+ \exp(i[\eta(ae^{-i\omega_z t} + a^\dagger e^{i\omega_z t}) - \delta t + \phi] + h.c. \right] \quad (\delta \equiv \omega - \omega_o), \qquad (16)$$

where $\eta \equiv kz_o$ is the Lamb-Dicke parameter and $S_+$, $S_-$, a, and $a^\dagger$ are time independent. We will be primarily interested in resonant transitions, that is, where $\delta = \omega_z(n'-n)$ where n' and n are integers. However, since we want to consider nonideal realizations, we will assume $\delta = (n'-n)\omega_z + \Delta$ where $\Delta \ll \omega_z, \Omega$. If we can neglect couplings to other levels (see Sec. 4.4.6), transitions are coherently driven between levels $|\downarrow,n\rangle$ and $|\uparrow,n'\rangle$ and the coefficients in Eq. (15) are given by Schrödinger's equation $i\hbar \partial \Psi / \partial t = H_I' \Psi$ to be

$$\dot{C}_{\uparrow,n'} = -i^{(1+|n'-n|)} e^{-i(\Delta t - \phi)} \Omega_{n',n} C_{\downarrow,n}, \quad \dot{C}_{\downarrow,n} = -i^{(1-|n'-n|)} e^{i(\Delta t - \phi)} \Omega_{n',n} C_{\uparrow,n'}, \qquad (17)$$

where $\Omega_{n',n}$ is given by [105,106]

$$\Omega_{n',n} \equiv \Omega | \langle n' | e^{i\eta(a + a^\dagger)} | n \rangle |$$

$$= \Omega \exp[-\eta^2/2](n_<!/n_>!)^{1/2} \eta^{|n'-n|} L_{n_<}^{|n'-n|}(\eta^2), \quad (18)$$

where $n_<$ ($n_>$) is the lesser (greater) of n' and n, and $L_n^\alpha$ is the generalized Laguerre polynomial

$$L_n^\alpha(X) = \sum_{m=0}^{n} (-1)^m \binom{n+\alpha}{n-m} \frac{X^m}{m!}. \quad (19)$$

Since we will be particularly interested in small values of n and $\alpha$, for convenience, we list a few values of $L_n^\alpha(X)$.

$$L_0^0(X) = 1, \quad L_1^0(X) = 1 - X, \quad L_2^0(X) = 1 - 2X + \frac{X^2}{2}, \quad L_3^0(X) = 1 - 3X + \frac{3}{2}X^2 - \frac{1}{6}X$$

$$L_0^1(X) = 1, \quad L_1^1(X) = 2 - X, \quad L_2^1(X) = 3 - 3X + \frac{1}{2}X^2, \quad L_3^1(X) = 4 - 6X + 2X^2 - \frac{1}{6}X^3 \quad (20)$$

$$L_0^2(X) = 1, \quad L_1^2(X) = 3 - X, \quad L_2^2(X) = 6 - 4X + \frac{1}{2}X^2, \quad L_3^2(X) = 10 - 10X + \frac{5}{2}X^2 - \frac{1}{6}X$$

Equations (17) can be solved using Laplace transforms. The solution shows sinusoidal "Rabi oscillations" between the states $|\uparrow,n'\rangle$ and $|\downarrow,n\rangle$, so over the subspace of these two states we have

$$\psi(t) = \begin{bmatrix} e^{-i\frac{\Delta}{2}t}\left[\cos(\frac{X_{n',n}}{2}t) + i\frac{\Delta}{X_{n',n}}\sin(\frac{X_{n',n}}{2}t)\right] & -2i\frac{\Omega_{n',n}}{X_{n',n}}e^{-i\left(\frac{\Delta}{2}t - \phi - \frac{\pi}{2}|n'-n|\right)}\sin(\frac{X_{n',n}}{2}t) \\ -2i\frac{\Omega_{n',n}}{X_{n',n}}e^{i\left(\frac{\Delta}{2}t - \phi - \frac{\pi}{2}|n'-n|\right)}\sin(\frac{X_{n',n}}{2}t) & e^{i\frac{\Delta}{2}t}\left[-i\frac{\Delta}{X_{n',n}}\sin(\frac{X_{n',n}}{2}t) + \cos(\frac{X_{n',n}}{2}t)\right] \end{bmatrix} \Psi(0), \quad (21)$$

where $X_{n',n} \equiv (\Delta^2 + 4\Omega_{n',n}^2)^{1/2}$, $\Delta = \omega - \omega_0 - (n' - n)\omega_z$, and $\Psi$ is given by

$$\Psi = C_{\downarrow,n}|\downarrow,n\rangle + C_{\uparrow,n'}|\uparrow,n'\rangle = \begin{bmatrix} C_{\uparrow,n'} \\ C_{\downarrow,n} \end{bmatrix}. \tag{22}$$

For the resonance condition $\Delta = 0$, Eq. (21) simplifies to

$$\psi(t) = \begin{bmatrix} \cos\Omega_{n',n}t & -ie^{i[\phi + \frac{\pi}{2}|n'-n|]}\sin\Omega_{n',n}t \\ -ie^{-i[\phi + \frac{\pi}{2}|n'-n|]}\sin\Omega_{n',n}t & \cos\Omega_{n',n}t \end{bmatrix} \Psi(0). \tag{23}$$

For each value of n' - n, the phase factor $\phi + \pi|n'\text{-}n|/2$ can be chosen arbitrarily for the first application of $H_I$; however once chosen, it must be kept track of if subsequent applications of $H_I$ are performed on the same ion. For convenience, we can choose it to be zero, although in most of what follows we will include a phase factor as a reminder that we must keep track of it. In these expressions, we assume $\Omega_{n',n}$ to be constant during a given application time t; this condition can be relaxed as discussed in Sec. 4.3.2. A special case of interest is when the Lamb-Dicke criterion, or Lamb-Dicke limit, is satisfied. Here, the amplitude of the ion's motion in the direction of the radiation is much less than $\lambda/2\pi$ which corresponds to the condition $\langle\psi_{motion}|k^2z^2|\psi_{motion}\rangle^{\frac{1}{2}} \ll 1$. This should not be confused with the less restrictive condition where the Lamb-Dicke parameter is less than 1 ($\eta \ll 1$); if the Lamb-Dicke criterion is satisfied, then $\eta \ll 1$, but the inverse is not necessarily true. If the Lamb-Dicke criterion is satisfied, we can evaluate $\Omega_{n',n}$ to lowest order in $\eta$ to obtain

$$\Omega_{n',n} = \Omega_{n,n'} = \Omega\eta^{|n'-n|}(n_>!/n_<!)^{\frac{1}{2}}(|n'-n|!)^{-1}. \tag{24}$$

We will be primarily interested in three types of transitions - the carrier (n' = n), the first red sideband (n' = n-1), and the first blue sideband (n' = n+1) whose Rabi frequencies, in the Lamb-Dicke limit, are given from Eq. (24) by $\Omega$, $\eta n^{\frac{1}{2}}\Omega$, and $\eta(n+1)^{\frac{1}{2}}\Omega$ respectively.
    In general, the Lamb-Dicke limit is not rigorously satisfied and higher order terms must be accounted for in the interaction [16,21,106]. As a simple example, suppose $\Psi(0) = |\downarrow\rangle|n\rangle$ and we apply radiation at the carrier frequency ($\delta = 0$). From Eq. (23), the wavefunction evolves as

$$\Psi(t) = \cos\Omega_{n,n}t|\downarrow,n\rangle - ie^{i\phi}\sin\Omega_{n,n}t|\uparrow,n\rangle. \tag{25}$$

For n = 0, we have $\Omega_{0,0} = \Omega\exp(-\eta^2/2)$. The exponential factor in this expression is the Debye-

Waller factor familiar from studies of X-ray scattering in solids; for a discussion in the context of trapped atoms see Ref. [106] and Sec. 4.4.5. This factor indicates that the matrix element for absorption of a photon is reduced due to the averaging of the electromagnetic wave (averaging of the $e^{ikz}$ factor in Eq. (14)) over the spread of the atom's zero-point wavefunction.

As a second simple example, we consider $\Psi(0) = |\downarrow,n\rangle$ and $\delta = +\omega_z$ (first blue sideband). Equation (23) implies

$$\Psi(t) = \cos\Omega_{n+1,n}t |\downarrow,n\rangle + e^{i\phi}\sin\Omega_{n+1,n}t |\uparrow,n+1\rangle . \tag{26}$$

At any time $t \neq m\pi/(2\Omega_{n+1,n})$ (m an integer), $\Psi$ is an entangled state between the spin and motion. If the excitation is left on continuously, the atom sinusoidally oscillates between the state $|\downarrow,n\rangle$ and $|\uparrow,n+1\rangle$. This oscillation has been observed by [21] and is reproduced in Fig. 2.

When the Lamb-Dicke confinement criterion is met and when the radiation is tuned to the red sideband ($\delta = -\omega_z$), we find (choosing $\phi = -\pi/2$)

$$H_I = \hbar\eta\Omega(S_+ a + S_- a^\dagger) . \tag{27}$$

This Hamiltonian is the same as the "Jaynes-Cummings Hamiltonian" [107] of cavity QED [43], which describes the coupling of a two-level atom to a single mode of the (quantized) radiation field. The problem we have described here, the coupling of a single two-level atom to the atom's (harmonic) motion is entirely analogous; the difference is that the harmonic oscillator associated with a single mode of the radiation field in cavity QED is replaced by that of the atom's motion. The suggestion to realize this type of Hamiltonian (in the context of cavity-QED) with a trapped ion was outlined in Refs. [2], [3], and [7]; however its use was already employed in the g-2 single electron experiments of Dehmelt [108].

Driving transitions between the $|\downarrow\rangle$ and $|\uparrow\rangle$ states will create entangled states between the internal and motional states since, in general, the Rabi frequency will depend on the motional states (Eq. (18)). This "conditional dynamics," where the dynamics of one system is conditioned on the state of another system, provides the basis for quantum logic (Sec. 3.3).

In this section, we have assumed that the atom interacts with an electromagnetic wave (Eq. (14)), which will usually be a laser beam. However, the essential physics which gives rise to entanglement is that the atom's internal levels are coupled to its motion through an inhomogeneous applied field. In the spin-½ analog, the magnetic moment $\vec{\mu}$ couples to a magnetic field $\vec{B} = B(z,t)\hat{x}$, yielding the Hamiltonian

$$H_I = -\mu_x B(z,t) = -\mu_x \left[ B(z=0,t) + \frac{\partial B}{\partial z}z + \frac{1}{2}\frac{\partial^2 B}{\partial z^2}z^2 + \cdots \right], \tag{28}$$

where, as above, $\mu_x \propto S_+ + S_-$ and z is the position operator. The key term is the gradient $\partial B/\partial z$. From the atom's oscillatory motion in the z direction, it experiences, in its rest frame, a modulation of $\vec{B}$ at frequency $\omega_z$. This oscillating component of $\vec{B}$ can then drive the spin-flip transition. As a simple example, suppose $\vec{B}$ is static (but inhomogeneous along the z direction so that $\partial B/\partial z \neq 0$) and $\omega_z$ is equal to the resonance frequency $\omega_o$ of the internal state transition. In its reference frame, the atom experiences an oscillating field due to the motion through the inhomogeneous field. Since $\omega = \omega_o$, this field resonantly drives transitions between the internal states. Because this term is resonant, it is the dominant term in Eq. (28), so $H_I \simeq -\mu_x(\partial B/\partial z)z \propto (S_+ + S_-)(a + a^\dagger) \simeq S_+a + S_-a^\dagger$, where the last equality neglects nonresonant terms. If the extent of the atom's motion is small enough that we need only consider the first two terms on the right hand side of Eq. (28), $H_I$ is given by the Jaynes-Cummings Hamiltonian (Eq. (27)). This Hamiltonian is also obtained if $\vec{B}$ is sinusoidally time varying (frequency $\omega$), we satisfy the resonance condition $\delta = \omega - \omega_o = -\omega_z$, and we make the rotating-wave approximation. This situation was realized in the classic electron g - 2 experiments of Dehmelt, Van Dyck, and coworkers to couple the spin and cyclotron motion [108]. Higher-order sidebands are obtained by considering higher order terms in the expansion of Eq. (28).

One reason to use optical fields is that the field gradients (for example, $\partial/(\partial z)[e^{ikz}] = ike^{ikz}$) can be large because of the smallness of $\lambda$. Stated another way, single-photon transitions between levels separated by rf or microwave transitions, which are driven by plane waves, may not be of interest because k is small ($\lambda$ large) and $\exp(ikz) \simeq 1$ which implies $\partial/(\partial z)[e^{ikz}] \simeq 0$. This makes interactions which couple the internal and external states as in Eqs. (27) and (28) negligibly small. This is not a fundamental restriction because electrode structures whose dimensions are small compared to the wave length can be used to achieve much stronger gradients than are achieved with plane waves. Microwave or rf transitions can also be driven by using stimulated-Raman transitions as discussed in Sec. 2.3.3 below. A second reason to use laser fields is they can be focused so that, to a good approximation, they interact only with a selected ion in a collection.

The unitary transformations of Eqs. (21) and (23) form the basic operations upon which most of the manipulations discussed in this paper are based. In this section, they were used to describe transitions between two states labeled $|\downarrow\rangle|n\rangle$ and $|\uparrow\rangle|n'\rangle$. In what follows, we will include other internal states of the atom which will take on different labels; however, the transitions between selected individual levels can still be described by Eqs. (21) and (23). Sequences of these basic operations can be used to construct more complicated operations such as logic gates (Sec. 3.3).

2.3.2. State dynamics including multiple modes of motion

In what follows, we will generalize the interaction with electromagnetic fields to consider motion in all 3L modes of motion for L trapped ions. Here, as was assumed by Cirac and Zoller [1], we consider that, on any given operation, the laser beam(s) interacts with only the jth ion; however, that ion will, in general, have components of motion from all modes. In this case Eq. (14) for the jth ion becomes

$$H_{Ij} = \hbar\Omega(S_{+j} + S_{-j})\left[e^{i(\vec{k}\cdot\vec{x}_j - \omega t + \phi_j)} + h.c.\right], \quad (29)$$

where we now assume $\vec{k}$ has some arbitrary direction. We will write the position operator of the jth ion (which represents the deviation from its equilibrium position) as

$$\vec{x}_j = u_j\hat{x} + u_{L+j}\hat{y} + u_{2L+j}\hat{z}, \quad j \in \{1, 2, ...L\}. \quad (30)$$

We can express the $u_j$ in terms of normal mode coordinates $q_k$ ($k \in \{1, 2, ... 3L\}$) through the matrix $D_k^p$, by the following relations [109]

$$u_p = \sum_{k=1}^{3L} D_k^p q_k, \quad q_k = \sum_{p=1}^{3L} D_k^p u_p, \quad q_k \equiv q_{ko}(a_k + a_k^\dagger), \quad (31)$$

where $q_k$ is the operator for the kth normal mode and $a_k$ and $a_k^\dagger$ are the lowering and raising operators for the kth mode. We have assumed that all normal modes are harmonic, which is a reasonably good assumption as long as the amplitude of normal mode motion is small compared to the ion spacing. (For two ions, the axial stretch mode's frequency is approximately equal to $\omega_z(\sqrt{3} - 9(a_z/a_s)^2)$ where $a_z$ is the (classical) amplitude of one ion's motion for this mode and $a_s$ is the ion spacing). Following the procedure of the last section, we take $H_o$ to be the Hamiltonian of the jth ion's internal states and all of the motional (normal) modes

$$H_{oj} = \hbar\omega_o S_{zj} + \sum_{k=1}^{3L} \hbar\omega_k \hat{n}_k, \quad (32)$$

where $\hat{n}_k \equiv a_k^\dagger a_k$. In the interaction picture (and making the rotating wave approximation), we have $H_{Ij}' = U_{oj}^\dagger H_{Ij} U_{oj}$ where $U_{oj} = \exp(-i(H_{oj}/\hbar)t)$, yielding

$$H_{Ij}' = \hbar\Omega\left(S_{+j}\exp\left[i\sum_{k=1}^{3L}\eta_k^j(a_k e^{-i\omega_k t} + a_k^\dagger e^{i\omega_k t}) - i(\delta t - \phi_j)\right] + h.c.\right), \quad (33)$$

where $\eta_k^j \equiv (\vec{k}\cdot\hat{x}D_k^j + \vec{k}\cdot\hat{y}D_k^{L+j} + \vec{k}\cdot\hat{z}D_k^{2L+j})q_{ko}$. (For the linear trap case, motion will be separable in the x, y, and z directions and $\eta_k^j$ will consist of one term.) In this interaction picture, the wavefunction is given by

$$\Psi_j = \sum_{M_z = \downarrow,\uparrow} \sum_{\{n_k\}=0}^{\infty} C^j_{M_z,\{n_k\}}(t) |M_z\rangle_j |\{n_k\}\rangle, \tag{34}$$

where the coefficients are slowly varying and $|\{n_k\}\rangle$ are the normal mode eigenstates (we have used the shorthand notation $\{n_k\} = n_1, n_2, ..., n_{3L}$). In analogy with the previous section, we will be primarily interested in a particular resonance condition, that is, where $\delta \approx \omega_k(n_k' - n_k)$ and $\Omega$ is sufficiently small that coupling to other internal levels and motional modes can be neglected. In this case, Eqs. (21) and (23) apply to the subspace of states $|\downarrow\rangle_j|n_k\rangle$ and $|\uparrow\rangle_j|n_k'\rangle$ if we make the definitions

$$\Psi = \Psi_j = C^j_{\downarrow,n_k}|\downarrow\rangle_j|n_k\rangle + C^j_{\uparrow,n_k'}|\uparrow\rangle_j|n_k'\rangle = \begin{bmatrix} C^j_{\uparrow,n_k'} \\ C^j_{\downarrow,n_k} \end{bmatrix}, \quad \Delta \equiv \delta - (n_k' - n_k)\omega_k, \tag{35}$$

and

$$X^j_{n_k',n_k} \equiv (\Delta^2 + 4(\Omega^j_{n_k',n_k})^2)^{1/2}, \quad \Omega^j_{n_k',n_k} \equiv \Omega \left| \langle \{n_{p \neq k}\}, n_k' | \prod_{l=1}^{3L} e^{i\eta^j_l(a_l + a_l^\dagger)} | \{n_{p \neq k}\}, n_k \rangle \right|. \tag{36}$$

The last expression is the Rabi frequency for particular values of the $\{n_{p \neq k}\}$. More likely, the other mode states ($p \neq k$) will correspond to a statistical distribution; this is discussed in Sec. 4.4.5. For the application to quantum logic (Sec. 3.3) the COM mode appears to be a natural choice since $\eta^j_k$ will be independent of j. The dependence of $\eta^j_k$ on j for the other modes is not a fundamental problem, but requires accurate bookkeeping when addressing different ions. The values of $\eta^j_k$ can be obtained from the normal mode coefficients as described by James [61].

2.3.3. Stimulated-Raman transitions

As indicated in the discussion following Eq. (28), we want strong field gradients to couple the internal states to the motion. If the internal state transition frequency $\omega_o$ is small, one way we can achieve strong field gradients is by using two-photon stimulated-Raman transitions [48,45] through a third, optical level as indicated in Fig. 3. In this case, as we outline below, the effective Hamiltonian corresponding to that in Eq. (14) is replaced by

$$H_I = \hbar\Omega(S_+ + S_-)\left[e^{i[(\vec{k}_1 - \vec{k}_2)\cdot\vec{x} - (\omega_{L1} - \omega_{L2})t + \phi]} + h.c.\right], \tag{37}$$

where $\vec{k}_1$, $\vec{k}_2$ and $\omega_{L1}$, $\omega_{L2}$ are the wavevectors and frequencies of the two laser beams and the resonance condition between internal states corresponds to $|\omega_{L1} - \omega_{L2}| = \omega_o$. Even if $\omega_o$ is small compared to optical frequencies, $|\vec{k}_1 - \vec{k}_2|$ can correspond to the wavevector of an optical frequency by choosing different directions for $\vec{k}_1$ and $\vec{k}_2$; this choice can thereby provide the desired strong field gradients.

In Fig. 3 we consider that a transition is driven between states $|\downarrow\rangle$ and $|\uparrow\rangle$ through state $|3\rangle$ by stimulated-Raman transitions using plane waves. Typically, we consider coupling with electric dipole transitions in which case

$$\vec{E}_i = \hat{\epsilon}_i E_i \cos(\vec{k}_i \cdot \vec{x} - \omega_{Li} t + \phi_i), \quad i \in \{1,2\}. \tag{38}$$

For simplicity, we assume laser beam 1 has a coupling only between intermediate state $|3\rangle$ and state $|\downarrow\rangle$. Similarly, laser beam 2 has a coupling only between state $|3\rangle$ and state $|\uparrow\rangle$. Not shown in Fig. 3 are the energy levels corresponding to the 3L motional modes. Laser detunings are indicated in the figure, so $\omega_{L1} - (\omega_{L2} + \delta) = \omega_o$, and we assume $\Delta_R \gg \delta, \{\omega_k\}$ where $\{\omega_k\}$ are the 3L mode frequencies. We will assume the Raman beams are focussed so that they interact only with the jth ion. In the Schrödinger picture, the wavefunction is written

$$\Psi_j = \sum_{p=\downarrow,\uparrow,3} \sum_{\{n_k\}=0}^{\infty} C_{p,\{n_k\}}^j \exp[-i(\omega_p + n_1\omega_1 + n_2\omega_2 + \ldots + n_{3L}\omega_{3L})t] \times |p\rangle|\{n_{3L}\}\rangle, \tag{39}$$

Since $\Delta_R$ is large, state $|3\rangle$ can be adiabatically eliminated in a theoretical treatment (see, for example, Refs. [32], [48], [110], and Sec. 4.4.6.2). If we assume the difference frequency is tuned to a particular resonance $\delta = \omega_k(n_k' - n_k)$, we can neglect rapidly varying terms and obtain

$$\dot{C}_{\uparrow,n_1,\ldots n_k',\ldots n_{3L}}^j = i\frac{|g_2|^2}{\Delta_R} C_{\uparrow,n_1,\ldots n_k',\ldots n_{3L}}^j - i\Omega_{n_k',n_k}^j C_{\downarrow,n_1,\ldots n_k,\ldots n_{3L}}^j,$$

$$\dot{C}_{\downarrow,n_1,\ldots n_k,\ldots n_{3L}}^j = i\frac{|g_1|^2}{\Delta_R} C_{\downarrow,n_1,\ldots n_k,\ldots n_{3L}}^j - i\left(\Omega_{n_k',n_k}^j\right)^* C_{\uparrow,n_1,\ldots n_k',\ldots n_{3L}}^j, \tag{40}$$

where

$$\Omega_{n_k',n_k}^j \equiv -\frac{g_1^* g_2}{\Delta_R}\langle n_k'|e^{i\eta_k^j(a_k + a_k^\dagger)}|n_k\rangle, \tag{41}$$

where $g_1 \equiv E_1 e \langle \downarrow | \hat{\epsilon} \cdot \vec{r} | 3 \rangle \exp(-i\phi_1)/(2\hbar)$, $g_2 \equiv E_2 e \langle \uparrow | \hat{\epsilon} \cdot \vec{r} | 3 \rangle \exp(-i\phi_2)/(2\hbar)$, $\eta_k^j \equiv (\Delta\vec{k}\cdot\hat{x}D_k^j + \Delta\vec{k}\cdot\hat{y}D_k^{L+j} + \Delta\vec{k}\cdot\hat{z}D_k^{2L+j})q_{ko}$ and $\Delta\vec{k} \equiv \vec{k}_1 - \vec{k}_2$. The terms $|g_1|^2/\Delta_R$ and $|g_2|^2/\Delta_R$ are the optical Stark shifts of levels $|1\rangle$ and $|2\rangle$, respectively. They can be eliminated from Eqs. (40) by including them in the definitions of the energies for the $|\downarrow\rangle$ and $|\uparrow\rangle$ states or, equivalently, tuning the Raman beam difference frequency $\delta$ to compensate for these shifts. If the Stark shifts are equal, both the $|\downarrow\rangle$ and $|\uparrow\rangle$ states are shifted by the same amount, and there is no additional phase shift to be accounted (Sec. 4.4.3). Equations (40) for stimulated-Raman transition amplitudes are the same as for the two-level system (Sec. 2.3.1) if we make the identifications $\phi_1 - \phi_2 \Leftrightarrow \phi$ and $\Delta\vec{k} \Leftrightarrow \vec{k}$. Although the experiments can benefit from use of stimulated-Raman transitions, for simplicity, we will assume single photon transitions below except where noted.

Another advantage of using stimulated-Raman transitions on low frequency transitions, as opposed to single-photon optical transitions, is the difference frequency between Raman beams can be precisely controlled using an acousto-optic modulator (AOM) to generate the two beams from a single laser beam. If the laser frequency fluctuations are much less than $\Delta_R$, phase errors on the overall Raman transitions can be negligible [111]. Other advantages (and some disadvantages) are noted below.

3. Quantum-state manipulation

3.1 Laser cooling to the ground state of motion

As a starting point for all of the quantum-state manipulations described below, we will need to initialize the ion(s) in known pure states. Using standard optical pumping techniques [112], we can prepare the ions in the "$|\downarrow\rangle$" internal state. Laser cooling in the resolved sideband limit [106, 113] can, for single ions, generate the $|n=0\rangle$ motional state with reasonable efficiency [44,45]. This type of laser cooling is usually preceded by a stage of "Doppler" laser cooling [106,114,115] which cools the ion to an equivalent temperature of about 1 mK. For Doppler cooling, we have $\langle \hat{n} \rangle \geq 1$, so an additional stage of cooling is required.

Resolved sideband laser cooling for a single, harmonically-bound atom can be explained as follows: For simplicity, we assume the atom is confined by a 1-D harmonic well of vibration frequency $\omega_z$. We use an optical transition whose radiative linewidth $\gamma_{rad}$ is relatively narrow, $\gamma_{rad} \ll \omega_z$ (Doppler laser cooling applies when $\gamma_{rad} \geq \omega_z$). If a laser beam (frequency $\omega$) is incident along the direction of the atomic motion, the bound atom's absorption spectrum is composed of a "carrier" at frequency $\omega_o$ and resolved frequency-modulation sidebands that are spaced by $\omega_z$, that is, at frequencies $\omega_o + (n' - n)\omega_z$ (Sec. 2.3). These sidebands in the spectrum are generated from the Doppler effect (like vibrational substructure in a molecular optical spectrum). Laser cooling can occur if the laser is tuned to a lower (red) sideband, for example, at $\omega = \omega_o - \omega_z$. In this case, photons of energy $\hbar(\omega_o - \omega_z)$ are absorbed, and spontaneously emitted photons of average energy $\hbar\omega_o$ - R return the atom to its initial internal state, where R $\equiv$ $(\hbar k)^2/2m = \hbar\omega_R$ is the photon recoil energy of the atom. Overall, for each scattering event, this reduces the atom's kinetic energy by $\hbar\omega_z$ if $\omega_z \gg \omega_R$, a condition which is satisfied for ions in strong traps. Since $\omega_R/\omega_z = \eta^2$ where $\eta$ is the Lamb-Dicke parameter, this simple form of sideband cooling requires that the Lamb-Dicke parameter be small. For example, in $^9Be^+$, if the recoil corresponds to spontaneous emission from the 313 nm 2p $^2P_{\frac{1}{2}} \rightarrow$ 2s $^2S_{\frac{1}{2}}$ transition (typically used for laser cooling), $\omega_R/2\pi \simeq 230$ kHz. This is to be compared to trap oscillation frequencies in some laser-cooling experiments of around 10 MHz [45]. Cooling proceeds until the atom's mean vibrational quantum number in the harmonic well is given by $\langle \hat{n} \rangle_{min} \simeq (\gamma/2\omega_z)^2 \ll 1$ [106,115,116].

In experiments, we find it convenient to use two-photon stimulated Raman transitions for sideband cooling [45,117], but the basic idea for, and limits to, cooling are essentially the same as for single-photon transitions. The steps required for sideband laser cooling using stimulated-Raman transitions are illustrated in Fig. 4. This figure is similar to Fig. 3, except we include the quantum states of the harmonic oscillator for one mode of motion. Part (a) of this figure shows how, when the ion starts in the $|\downarrow\rangle$ internal state, a stimulated-Raman transition tuned to the first red sideband $|\downarrow\rangle|n\rangle \rightarrow |\uparrow\rangle|n-1\rangle$ reduces the motional energy by $\hbar\omega_z$. In part (b), the atom is reset to the $|\downarrow\rangle$ internal state by a spontaneous-Raman transition from a third laser beam tuned to the $|\uparrow\rangle \rightarrow |3\rangle$ transition. We assume that there is a reasonable branching ratio from state $|3\rangle$ to state $|\downarrow\rangle$, so that even if the atom decays back to level $|\uparrow\rangle$ after being excited to level $|3\rangle$, after a few scattering events, the atom decays to state $|\downarrow\rangle$. If $\omega_R \ll \omega_z$, step (b) accomplishes the transition $|\uparrow\rangle|n-1\rangle \rightarrow |\downarrow\rangle|n-1\rangle$ with high efficiency. Steps (a) and (b) are repeated until the atom is optically pumped into the $|\downarrow\rangle|0\rangle$ state. When this condition is reached, neither step (a) or (b) is active and the process stops. In this simple discussion, we have assumed the transition $|\downarrow\rangle|n\rangle \rightarrow |\uparrow\rangle|n-1\rangle$ is accomplished with 100% efficiency. However since, in general, the atom doesn't start in a given motional state $|n\rangle$, and since the Rabi frequencies (Eq. (18)) depend on n, this process is not 100% efficient; nevertheless, the atom will still be pumped to the $|\downarrow\rangle|0\rangle$ state. The only danger is having the stimulated-Raman intensities and pulse time t adjusted so that for a particular n, $\Omega_{n-1,n}t = m\pi$ (m an integer), in which case the atom is "trapped" in the $|\downarrow\rangle|n\rangle$ level. This is avoided by varying the laser beam intensities from pulse to pulse; one particular strategy is described in Ref. [17].

So far, laser cooling to the $|n=0\rangle$ state has been achieved only with single ions [44,45]; therefore an immediate goal of future work is to laser cool a collection of ions (or, at least one mode of the collection) to the zero-point state. Cooling of any of the 3L modes of motion of a collection of ions should, in principle, work the same as cooling of a single ion. To cool a particular mode, we tune the cooling radiation to its first lower sideband. If we want to cool all modes, sideband cooling must be cycled through all 3L modes more than once, or applied to all 3L modes at once, since recoil will heat all modes. For the COM mode, the cooling is essentially the same as cooling a single particle of mass Lm; however, the recoil energy upon re-emission is distributed over the 3L-1 other modes. Other methods to prepare atoms in the $|n=0\rangle$ state are discussed in Refs. [5], [10], [12]. Morigi, *et al.* [118] show that it is not necessary to satisfy the condition $\omega_R \ll \omega_z$ ($\eta \ll 1$) to achieve cooling to n = 0.

3.2. Generation of nonclassical states of motion of a single ion

We begin with a discussion of the generation of nonclassical motional states of a single trapped ion. This seems appropriate because the other applications discussed in this paper incorporate similar techniques.  Much of the original interest in nonclassical states of mechanical motion grew out of the desire to make sensitive detectors of gravitational waves using (macroscopic) mechanical resonators [119,120].  For example, parametric amplification of mechanical harmonic oscillations can lead to quantum mechanical squeezing of the oscillation. In the meantime, nonclassical states of the radiation field were observed [43].  The close relationship of these two problems was pointed out above: in quantum optics, $H_{osc}$ of Eq. (8) represents a single mode of the radiation field, and $H_I$ of Eq. (14) represents the coupling between the (quantized) field and atom.  The nonclassical states of motion considered here, such as squeezed states, are the direct analogs of the nonclassical photon states in quantum optics. They appear to be of intrinsic interest because, as in cavity QED, they allow the rather complicated dynamics of the simple quantum system (described by the Hamiltonian in Eq. (27)) to be studied.  Before discussing some methods to create nonclassical states, we consider one method for analyzing them.

3.2.1. Population analysis of motional states

As described below in this section, from the $|\downarrow\rangle|0\rangle$ state, it is possible to coherently create states of the form $|\downarrow\rangle\Psi_{motion}$ where $\Psi_{motion}$ is given by Eq. (10).  One way we can analyze the motional state created is as follows [21]:   To the state $|\downarrow\rangle\Psi_{motion}$, we apply radiation on the first blue sideband (n' = n+1 in Eq. (23)) for a time $\tau$.  We then measure the probability $P_\downarrow(\tau)$ that the ion is in the $|\downarrow\rangle$ internal state.  In the experiments of Meekhof, et al. [21], the internal state $|\downarrow\rangle$ is the 2s $^2S_{1/2}$ (F = 2, $M_F$ = 2) state of $^9Be^+$, and $|\uparrow\rangle$ corresponds to the 2s $^2S_{1/2}$ (1,1) state as shown in Fig. 5.  The $|\downarrow\rangle$ state is detected by applying nearly resonant $\sigma^+$-polarized laser radiation between the $|\downarrow\rangle$ and $^2P_{3/2}$(F=3, $M_F$=3) energy levels.  Because the only decay channel of the $^2P_{3/2}$(F=3, $M_F$=3) state is back to the $|\downarrow\rangle$ state, this is a cycling transition, and detection efficiency is near 1 (Sec. 2.2.1).  The experiment is repeated many times for each value of $\tau$, and for a range of $\tau$ values.  We find

$$P_\downarrow(\tau) = \frac{1}{2}\left(1 + \sum_{n=0}^{\infty} P_n e^{-\gamma_n \tau} \cos(2\Omega_{n+1,n}\tau)\right), \tag{42}$$

where $P_n \equiv |C_{\downarrow,n}|^2$ is the probability of finding the ion in state $|\downarrow\rangle|n\rangle$.  The phenomenological decay constants $\gamma_n$ are introduced to model decoherence that occurs during the application of the blue sideband. The  measured signal $P_\downarrow(\tau)$ can be inverted (Fourier cosine transform), allowing the extraction of the probability distribution of vibrational state occupation $P_n$.

3.2.2. Fock states

In Fig. 2, we show an experimental plot [21] of the probability $P_\downarrow(\tau)$ of finding the ion in the $|\downarrow\rangle$ internal state after first preparing it in the $|\downarrow\rangle|0\rangle$ state, and applying the first blue sideband for a time $\tau$.  From Eq. (23), we would expect  $P_\downarrow(\tau) = \cos^2\Omega_{1,0}\tau$; however, we clearly see the effects of some decoherence process which we can represent adequately by the first term in Eq. (42)

$$P_\downarrow(\tau) = \frac{1}{2}\left(1 + e^{-\gamma_0 \tau} \cos 2\Omega_{1,0}\tau\right). \tag{43}$$

In this experiment, we think the decoherence is not simply caused by fundamental (radiative) decoherence but has contributions from fluctuations in laser power (which cause fluctuations in $\Omega_{1,0}$), fluctuations in trap drive voltage $V_o$ (which cause fluctuations in $\omega_z$), and fluctuations in the ambient magnetic field (which cause fluctuations in $\omega_o$).

Neglecting for the moment the effects of decoherence, we see that for times $\tau = m\pi/(2\Omega_{1,0})$ (m an integer), the ion is in a nonentangled state ($|\downarrow\rangle|0\rangle$ or $|\uparrow\rangle|1\rangle$). Therefore, if the ion starts in the $|\downarrow\rangle|0\rangle$ state, we can prepare the atom in the $|\uparrow\rangle|1\rangle$ state (the $|n=1\rangle$ Fock state) by applying the blue sideband for a time $\tau = \pi/2\Omega_{1,0}$, a so called Rabi $\pi$ pulse. For other times, the ion is in an entangled state given by Eq. (26). This operation and the analogous operation on the first red sideband will form key elements of quantum logic using trapped ions.

We can generate higher-n Fock states of motion by a sequence of similar operations. For example, to generate the $|\downarrow\rangle|2\rangle$ state, we start in the $|\downarrow\rangle|0\rangle$ state, apply a $\pi$ pulse on the first blue sideband, followed by a $\pi$ pulse on the first red sideband. This leads to the sequence $|\downarrow\rangle|0\rangle \rightarrow |\uparrow\rangle|1\rangle \rightarrow |\downarrow\rangle|2\rangle$ (neglecting overall phase factors). In a similar fashion, Fock states up to $|n=16\rangle$ have been created [21]. Other methods for creating Fock states have been suggested in Refs. [5], [8], [12], and [13].

3.2.3. Coherent states

We can also create coherent states of motion; these states are closest in character to classical states of motion. This can be accomplished if the atom is subjected to a spatially uniform classical force, or any force derived from a potential $-\vec{f}(t)\cdot\vec{z}$, where $\vec{f}$ is a real c-number vector. For an ion which starts in the $|n=0\rangle$ state, this force creates a displacement leading to a coherent state $|\alpha\rangle$ defined by $a|\alpha\rangle = \alpha|\alpha\rangle$ where $\alpha$ is a complex number [121]. This classical force can be realized by applying an electric field which oscillates at frequency $\omega_z$. For example, if we apply a (classical) electric field $\vec{E}(t) = \hat{z}E_z\sin(\omega t - \phi)$, the corresponding interaction Hamiltonian (in the interaction frame for the motion) is given by

$$H_I = -qE_z z_o(ae^{-i\omega_z t} + a^\dagger e^{i\omega_z t})\sin(\omega t - \phi). \tag{44}$$

If we express the motional wavefunction as in Eq. (10), Schrödinger's equation yields for the coefficients of the wavefunction

$$\dot{C}_n = \Omega_1 \sqrt{n}\, C_{n-1} - \Omega_1^* \sqrt{n+1}\, C_{n+1}, \tag{45}$$

where $\Omega_1 \equiv -qE_z z_o e^{i\phi}/(2\hbar)$. Equivalently, in the interaction picture for the motion, the Hamiltonian of Eq. (44) leads to the evolution operator

$$U(t) = e^{[(\Omega_1 t)a^\dagger - (\Omega_1 t)^* a]} = D(\Omega_1 t), \tag{46}$$

where D is the displacement operator with argument $\Omega_1 t$ [121].

We can achieve the same evolution if we superimpose two traveling wave fields which drive stimulated-Raman transitions between different $|n\rangle$ levels of the same internal state, and we make the difference frequency between the Raman beams is equal to the trap oscillation frequency. For example, assume the ion is subjected to two lasers fields given by Eq. (38), where $\omega_{L1} - \omega_{L2} \simeq \omega_z \ll \omega_o$. The dynamics can be obtained following the analysis in Sec. 2.3.3, except we replace level $|\downarrow\rangle|n\rangle$ ($|\uparrow\rangle|n'\rangle$) with level $|g\rangle|n\rangle$ ($|g\rangle|n'\rangle$) where $|g\rangle$ can be any ground state which has a matrix element with level $|3\rangle$. For the coefficients of Eq. (10), we find

$$\dot{C}_n = i\left[\frac{|g_1|^2 + |g_2|^2}{\Delta_R}C_n - \sum_{n'=0}^{\infty} \langle n|\left(\Omega \exp[i\eta(ae^{-i\omega t} + a^\dagger e^{i\omega t}) - i(\omega_{L1} - \omega_{L2})t] + h.c.\right)|n'\rangle C_{n'}\right], \tag{47}$$

where $\Omega \equiv -g_1^* g_2/\Delta_R$, $g_i \equiv qE_i\langle g|\hat{\epsilon}_i \cdot \vec{r}|3\rangle\exp(-i\phi_i)/(2\hbar)$, $\eta \equiv (\vec{k}_1 - \vec{k}_2)\cdot\hat{z}\, z_o$, and $\Delta_R \equiv \omega_{3g} - \omega_{L1}$. The first term on the right side of Eq. (47) corresponds to a Stark shift of level g; this Stark shift can be absorbed into the definition of the ground state energy (see, for example, Sec. 4.4.6.2). If this is done, the same equations for the $C_n$ are obtained from the Hamiltonian (in the oscillator interaction picture)

$$H_I' = \hbar\Omega\exp\left(i[\eta(ae^{-i\omega_z t} + a^\dagger e^{i\omega_z t}) - (\omega_{L1} - \omega_{L2})t]\right) + h.c., \tag{48}$$

In the Lamb-Dicke limit, if we choose the resonance condition where $\omega_{L1} - \omega_{L2} = \omega_z$ and assume $\Omega \ll \omega_z$, Eqs. (47) are the same as Eqs. (45) where $\Omega_1 = \eta\Omega$. Therefore, in the Lamb-Dicke limit, the standing wave laser fields act like a uniform oscillating electric field which oscillates at frequency $\omega_{L1} - \omega_{L2}$. This can be understood if we consider that the two laser fields give rise to an optical dipole force which is modulated in such a way to resonantly excite the ion motion. To see this, assume for simplicity that, in Eq. (38), $E_1 = E_2 = E_o$, $\hat{\epsilon}_1 = \hat{\epsilon}_2 = \hat{\epsilon}$, and $\vec{k}_1 - \vec{k}_2 = \hat{z}\Delta k$. It is useful to write the total electric field as

$$\vec{E} = \vec{E}_1 + \vec{E}_2 = \hat{\epsilon}\, 2E(t)\cos((\vec{k}_1 + \vec{k}_2)\cdot\vec{z}/2 - \bar{\omega}t + \bar{\phi}), \tag{49}$$

where $\bar{\omega} \equiv (\omega_{L1} + \omega_{L2})/2$, $\bar{\phi} \equiv (\phi_1 + \phi_2)/2$, and E(t) is a slowly varying function

$$E(t) = E_o \cos\left(\frac{\Delta k}{2} z - \frac{\omega_{L1} - \omega_{L2}}{2} t + \frac{\phi_1 - \phi_2}{2}\right). \tag{50}$$

On a time scale long compared to $1/\Delta_R$ but short compared to $1/(\omega_{L1} - \omega_{L2})$, the atom experiences a nonresonant electric field of amplitude E(t) which is nearly constant in time. If we consider coupling of this electric field between the ground state and state $|3\rangle$ (for example, see Sec. 4.4.6.2), this electric field leads to a spatially-dependent Stark shift of the ground state equal to $\Delta E_{Stark} = -4\hbar |g(z,t)|^2/\Delta_R$ where

$$|g(z,t)|^2 = |g|^2 \cos^2\left(\frac{\Delta k}{2} z - \frac{\omega_{L1} - \omega_{L2}}{2} t + \frac{\phi_1 - \phi_2}{2}\right), \quad g \equiv \frac{qE_o \langle g|\hat{\mathbf{e}}\cdot\vec{r}|3\rangle e^{-i\bar{\phi}}}{2\hbar}. \tag{51}$$

This Stark shift leads to an optical dipole force [122-124] $F_z = -\partial(\Delta E_{Stark})/\partial z$. On a longer time scale, this dipole force is modulated at frequency $\omega_{L1} - \omega_{L2}$ which can resonantly excite the ion's motion when $(\omega_{L1} - \omega_{L2}) = \omega_z$. This leads to Eqs. (45). When $\vec{k}_1$ and $\vec{k}_2$ are both directed along the $\hat{z}$ axis (but in opposite directions), the dipole force potential can be viewed as a "moving standing wave" in the $\hat{z}$ direction which slips over the ion and whose accompanying dipole force resonantly excites the ion's motion [125]. Both methods have been used to excite coherent states in Refs. [21] and [47]. Other methods for generating coherent states are suggested in Refs. [4] and [48].

In the experiment of Ref. [47], a dipole force oscillating at the ion oscillation frequency was created with particular polarizations of the laser fields. This led to a force which was dependent on the ion's internal state, enabling the generation of entangled "Schrödinger-cat" states of the form $\Psi = \frac{1}{2}(|\downarrow\rangle|\alpha e^{i\phi}\rangle + |\uparrow\rangle|\alpha e^{-i\phi}\rangle)$.

3.2.4. Other nonclassical states

When $(\omega_{L1} - \omega_{L2}) = 2\omega$, a similar analysis shows that the moving standing wave potential discussed in the last section has a component which acts like a parametric excitation of the ion's harmonic well at frequency $2\omega$ [21]. This can produce quantum mechanical squeezing of the ion's motion. Squeezing could also be achieved by amplitude modulating $V_o$, $U_o$, or $U_r$ (Eqs. (1) and (2)) at frequency $2\omega_z$, by a nonadiabatic change in the trap spring constant [48], or through a combination of standing and traveling-wave laser fields [4]. A quantum mechanical treatment of the motion in an rf trap shows the effects of squeezing from the applied rf trapping fields [34,35,91]. More general nonlinear effects in the interaction can lead to higher "nonlinear coherent states" as discussed by de Matos Filho and Vogel [126].

Other methods for generation of Schrödinger-cat like states in ions are suggested in Refs.

[8], [19], [29], [31], [33], [36], and [38]. Additional nonclassical states are investigated theoretically by Gou and Knight [23], Gou, *et al.* [30], and Gerry, *et al.* [37]. Schemes which can generate arbitrary states of the single-mode photon field [127,128] can also be applied directly to generate arbitrary motional states of a trapped atom and perform quantum measurements of an arbitrary motional observable [129]. A scheme which can generate arbitrary entanglement between the internal and motional levels of a trapped ion is discussed by Kneer and Law [130].

The procedure for analyzing motional states outlined in Sec. 3.2.1 yields only the populations of the various motional states and not the coherences. Coherences must be verified separately [21,47]. The most complete characterization is achieved with a complete state reconstruction or tomographic technique; a description of how this has been implemented to measure the density matrix or Wigner function for trapped atoms is Leibfried, *et al.* [131,132]. These experiments represent the first measurement of negative values of the Wigner function in position-momentum space. Wigner functions for free atoms have also been recently determined experimentally by Kurtsiefer, *et al.* [133]. Other methods for trapped atoms have been suggested in Refs. [15], [22], [24], [27], [34], and [35]. These techniques can also be extended to characterize entangled motional states [39] and states which are entangled between the motional and internal states.

3.3. Quantum logic

Significant attention has been given recently to the possibility of quantum computation. Although this field is about 15 years old [134-138], interest has intensified because of the discovery of algorithms, notably for prime factorization [139-142], which could provide dramatic speedup over conventional computers. Quantum computation may also find other applications [142, 143,144,145,146,147,148,149,150,151,152]. Schemes for implementing quantum computation have been proposed by Teich [153], Lloyd [143,144], Berman, *et al.*.[154], DiVincenzo [155,156], Cirac and Zoller [1], Barenco, *et al.* [157], Sleator and Weinfurter [158], Pellizzari, *et al.* [159], Domokos, *et al.* [160,161] Turchette, *et al.* [162], Lange, *et al.* [163], Törmä and Stenholm [164], Gershenfeld and Chuang [165], Cory, *et al.* [166], Privman, *et al.* [167], Loss and DiVincenzo [168], and Bocko, *et al.* [169]. In this paper, we focus on a scheme suggested by Cirac and Zoller [1], which uses trapped ions. Since, in general, any quantum computation can be composed of a series of single-bit rotations and two-bit controlled-not operations [140,155,156,170,171], we will focus our attention on these operations.

In the parlance of quantum computation, we say that two internal states of an ion can form a quantum bit or "qubit" whose levels are labeled $|0\rangle$ and $|1\rangle$ or, equivalently, $|\downarrow\rangle$ and $|\uparrow\rangle$. Single-bit rotations on ion j can be characterized by the transformation (Eq. (23) for n' = n)

$$R(\theta,\phi)(C_{\downarrow,j}|\downarrow\rangle_j + C_{\uparrow,j}|\uparrow\rangle_j) = \begin{bmatrix} \cos(\theta/2) & -ie^{i\phi}\sin(\theta/2) \\ -ie^{-i\phi}\sin(\theta/2) & \cos(\theta/2) \end{bmatrix} \cdot \begin{bmatrix} C_{\uparrow,j} \\ C_{\downarrow,j} \end{bmatrix}. \quad \textbf{(52)}$$

In the spin-½ model, this transformation is realized by application of a magnetic field $B_1/2$ which rotates at frequency $\omega_o$ and in the same sense as $\bar{\mu}$, which is applied along the direction $\hat{x}\cos\phi - \hat{y}\sin\phi$ in the rotating frame. This is equivalent to application of the field $B_1\hat{x}\cos(kz-\omega t+\phi)$ in Eq. (14). In this expression, $\theta = 2\Omega_{n,n}t$ is the angle of rotation about the axis of this field. For $\theta = \pi$ and $\phi = 0$, $R(\theta,\phi)$ is a logical "not" operation (within an overall phase factor). $R(\pi/2,-\pi/2)$ (plus a rotation about z) is essentially a Hadamard transform.

A fundamental two-bit gate is a controlled-not (CN) gate [142,155,156,157]. This provides the transformation

$$|\epsilon_1\rangle|\epsilon_2\rangle \rightarrow |\epsilon_1\rangle|\epsilon_1 \oplus \epsilon_2\rangle , \qquad (53)$$

where $\epsilon_1,\epsilon_2 \in \{0,1\}$ and $\oplus$ is addition modulo 2. Although Eq. (53) is written in terms of eigenstates, the transformation is assumed to apply to arbitrary superpositions of states $|\epsilon_1\rangle|\epsilon_2\rangle$. In this expression, $\epsilon_1$ is the called the control bit and $\epsilon_2$ is the target bit. If $\epsilon_1 = 0$, the target bit remains unchanged; if $\epsilon_1 = 1$, the target bit flips.

A spectroscopy experiment on any four-level quantum system, where the level spacings are unequal, shows this type of logic structure if we make the appropriate labeling of the levels. For example, we could label these four levels as in Eq. (53). If we tune radiation to the $|1,0\rangle \rightarrow |1,1\rangle$ resonance frequency and adjust its duration to make a $\pi$ pulse, we realize the logic of Eq. (53). Similarly, an eight-level quantum system with unequal level spacings realizes a Toffoli gate [142], where the flip of a third bit is conditioned upon the first two bits being 1's - and so on (see Sec. 5.2). This basic idea can be applied to molecules composed of many interacting spins such as in the proposals of Gershenfeld and Chuang [165] and Cory, *et al.* [166]. For quantum computation to be most useful, however, we need to perform a series of logic operations between an arbitrary number of qubits in a system which can be scaled to large numbers, such as the scheme of Cirac and Zoller [1].

Another type of fundamental two-bit gate is a phase gate, which could take the form

$$|\epsilon_1\rangle|\epsilon_2\rangle \rightarrow e^{i\phi\epsilon_1\epsilon_2}|\epsilon_1\rangle|\epsilon_2\rangle . \qquad (54)$$

This type of conditional dynamics has been demonstrated in the context of cavity QED [162,172] and for a trapped ion [17] (step (1b) below).

The Cirac/Zoller scheme assumes that an array of ions are confined in a common ion trap. The ions are held apart from one another by mutual Coulomb repulsion as shown, for example, in Fig. 1. They can be individually addressed by focusing laser beams on the selected ion. Ion motion can be described in terms of normal modes of oscillation which are *shared* by all of the ions; a particularly useful mode might be the COM axial mode. When quantized, this mode can form the "bus qubit" through which all gate operations are performed. We first describe how logic is accomplished between this COM mode qubit and the internal-state qubit of a single trapped ion. In particular, the transformation in Eq. (53) has been realized for a single trapped ion [17]. In that experiment, performed on a trapped $^9$Be$^+$ ion, the control bit was the quantized state of one mode of the ion's motion (labeled the x mode). If the motional state was $|n=0\rangle$, this was taken to be a $|\epsilon_1=0\rangle$ state; if the motional state was $|n=1\rangle$, this was taken to be a $|\epsilon_1=1\rangle$ state. The target states were two ground-hyperfine states of the ion, the $|F=2, M_F=2\rangle$ and $|F=1, M_F=1\rangle$ states, labeled $|\downarrow\rangle$ and $|\uparrow\rangle$ (Fig. 5), with the identification here $|\downarrow\rangle \Leftrightarrow |\epsilon_2=0\rangle$ and $|\uparrow\rangle \Leftrightarrow |\epsilon_2=1\rangle$. Transitions between levels were produced using two laser beams to realize stimulated-Raman transitions. The wavevector difference $\vec{k}_1 - \vec{k}_2$ was chosen to be aligned along the x direction. The CN operation between these states was realized by applying three pulses in succession:

(1a)  A $\pi/2$ pulse ($\Omega t = \pi/4$ in Eq. (25), where we assume $\Omega_{0,0} = \Omega_{1,1} = \Omega$) is applied on the carrier transition. For a certain choice of initial phase, this corresponds to the operator $V^{1/2}(\pi/2)$ of Cirac and Zoller [1].

(1b)  A $2\pi$ pulse is applied on the first blue sideband transition between levels $|\uparrow\rangle$ and an auxiliary level $|aux\rangle$ in the ion (the $|F=2, M_F=0\rangle$ level in $^9$Be$^+$; see Fig. 5). This operator is analogous to the operator $U_n^{2,1}$ of Cirac and Zoller [1]. This operation provides the "conditional dynamics" for the CN operation. It changes the sign of the $|\uparrow\rangle|n=1\rangle$ component of the wavefunction but leaves the sign of the $|\uparrow\rangle|n=0\rangle$ component of the wavefunction unchanged; that is, the sign change is conditioned on whether or not the ion is in the $|n=0\rangle$ or $|n=1\rangle$ motional state. Therefore, this step is the phase gate of Eq. (54) with $\phi = \pi$, where we make the identifications $(\epsilon_1 = 0,1) \Leftrightarrow (n = 0,1)$ and $(\epsilon_2 = 0,1) \Leftrightarrow$ (internal state $= \downarrow, \uparrow$).

(1c)  A $\pi/2$ pulse is applied to the spin carrier transition with a 180° phase shift relative to step (a). This corresponds to the operator $V^{1/2}(-\pi/2)$ of Cirac and Zoller [1].

Steps (1a) and (1c) can be regarded as two resonant pulses (of opposite phase) in the Ramsey separated-field method of spectroscopy [173]. If step (1b) is active (thereby changing the sign of the $|\uparrow\rangle|n=1\rangle$ component of the wavefunction), then a state change (spin flip) is induced by the Ramsey fields. If step (1b) is inactive, step (1c) reverses the effect of step (1a).

Instead of the three pulses (1a-1c above), a simpler CN gate scheme between an ion's internal and motional states can be achieved with a single laser pulse, while eliminating the requirement of the auxiliary internal electronic level [174], as described below. These simplifications can be important for several reasons:

(1) Several popular ion candidates, including $^{24}$Mg$^+$, $^{40}$Ca$^+$, $^{88}$Sr$^+$, $^{138}$Ba$^+$, $^{174}$Yb$^+$, $^{172}$Yb$^+$, and $^{198}$Hg$^+$, do not have a third electronic ground state available for the auxiliary level. These ions have zero nuclear spin with only two Zeeman ground states ($M_z = \downarrow, \uparrow$). Although excited optical metastable states may be suitable for auxiliary levels in some of these ion species, use of such states places stringent requirements on the frequency stability of the exciting optical field to preserve coherence (see Sec. 4.4.3).

(2) The elimination of an auxiliary ground state level removes "spectator" internal atomic levels, which can act as potential "leaks" from the two levels spanned by the quantum bits (assuming negligible population in excited electronic metastable states). This feature may be important to the success of quantum error-correction schemes [142, 175,176,177,178,179,180,181,182,183,184,185,186,187] which can be degraded when leaks to

spectator states are present [188]. (Specific error-correction schemes for ions are suggested in Refs. [182] and [187].)

(3) The elimination of the need for an auxiliary level can dramatically reduce the sensitivity of a CN quantum logic gate to external magnetic fields fluctuations. It is generally impossible to find three atomic ground states whose splittings are all magnetic field insensitive to first order. However, for ions possessing hyperfine structure, the transition frequency between two levels can be made magnetic field independent to first order at particular values of an applied magnetic field (see Sec. 4.2.2).

(4) Finally, a reduction of laser pulses simplifies the tuning procedure and may increase the speed of the gate. For example, the gate realized in Ref. [17] required the accurate setting of the phase and frequency of three laser pulses, and the duration of the gate was dominated by the transit time through the auxiliary level.

The CN quantum logic gate can be realized with a single pulse tuned to the carrier transition which couples the states $|n\rangle|\downarrow\rangle$ and $|n\rangle|\uparrow\rangle$ with Rabi frequency $\Omega_{n,n}$ (see Eqs. (18), (36), and (41)). Considering only a single mode of motion,

$$\Omega_{n,n} = \Omega |\langle n| e^{i\eta(a+a^\dagger)} |n\rangle| = \Omega e^{-\eta^2/2} \mathcal{L}_n(\eta^2) , \qquad (55)$$

where $\mathcal{L}_n(\eta^2) \equiv L_n^0(\eta^2)$ Eq. (18). Specializing to the $|n\rangle = |0\rangle$ and $|n\rangle = |1\rangle$ vibrational levels relevant to quantum logic, we have

$$\begin{aligned}\Omega_{0,0} &= \Omega e^{-\eta^2/2} , \\ \Omega_{1,1} &= \Omega e^{-\eta^2/2}(1-\eta^2) .\end{aligned} \qquad (56)$$

The CN gate can be achieved in a single pulse by setting $\eta$ so that $\Omega_{1,1}/\Omega_{0,0} = (2k+1)/2m$, with k and m positive integers satisfying $m > k \geq 0$. Setting $\Omega_{1,1}/\Omega_{0,0} = 2m/(2k+1)$ will also work, with the roles of the $|0\rangle$ and $|1\rangle$ motional states switched in Eq. (55). By driving the carrier transition for a duration $\tau$ such that $\Omega_{1,1}\tau = (k+\frac{1}{2})\pi$, or a "$\pi$-pulse" (mod $2\pi$) on the $|n\rangle=|1\rangle$ component, this forces $\Omega_{0,0}\tau = m\pi$. Thus the states $|\downarrow\rangle|1\rangle \leftrightarrow |\uparrow\rangle|1\rangle$ are swapped, while the states $|\downarrow\rangle|0\rangle$ and $|\uparrow\rangle|0\rangle$ remain unaffected. The net unitary transformation, in the $\{\downarrow 0, \uparrow 0, \downarrow 1, \uparrow 1\}$ basis is

$$\begin{bmatrix} 1 & 0 & 0 & 0 \\ 0 & 1 & 0 & 0 \\ 0 & 0 & 0 & ie^{i\phi}(-1)^{k-m} \\ 0 & 0 & ie^{-i\phi}(-1)^{k-m} & 0 \end{bmatrix}. \qquad (57)$$

This transformation is equivalent to the reduced CN of Eq. (53), apart from phase factors which can be eliminated by the appropriate settings of the phase of subsequent logic operations [157].

The "magic" values of the Lamb-Dicke parameter which allow the above transformation satisfy $\mathcal{L}_1(\eta^2) = 1 - \eta^2 = (2k+1)/2m$, and are tabulated in Table I of Ref. [174] for the first few values. (For rf (Paul) trap confinement along the COM motional mode, the Rabi frequencies of Eqs. (55) and (56) must be altered to include effects from the micromotion at the rf drive frequency $\omega_{rf}$. In the pseudopotential approximation ($\omega \ll \omega_{rf}$), this correction amounts to replacing the Lamb-Dicke parameter $\eta$ in this paper by $\eta[1-\omega/(2\sqrt{2}\omega_{rf})]$, as pointed out by Bardroff, *et al.* [34,35]. However, there is no correction if the COM motional mode is confined by static fields (such as the axial COM mode of a linear trap).) It may be desirable for the reduced CN gate to employ the $|n\rangle=|2\rangle$ or $|n\rangle=|3\rangle$ state instead of the $|n\rangle=|1\rangle$ state for error-correction of motional state decoherence [182]. In these cases, the "magic" Lamb-Dicke parameters satisfy $\mathcal{L}_2(\eta^2) = 1-2\eta^2+\eta^4/2 = (2k+1)/2m$ for quantum logic with $|n\rangle=|0\rangle$ and $|2\rangle$, or $\mathcal{L}_3(\eta^2) = 1-3\eta^2+3\eta^4/2-\eta^6/6 = (2k+1)/2m$ for quantum logic with $|n\rangle=|0\rangle$ and $|3\rangle$.

This scheme places a more stringent requirement on the accuracy of $\Omega$ and $\eta$, roughly by a factor of m. In the two-photon Raman configuration (Sec. 2.3.3), the Lamb-Dicke parameter $\eta = |\Delta\vec{k}|z_o$ can be controlled by both the frequency of the trap (appearing in $z_o$) and by the geometrical wavevector difference $\Delta\vec{k}$ of the two Raman beams. Accurate setting of the Lamb-Dicke parameter should therefore not be difficult. Both CN-gate schemes are sensitive to excitation in other modes as discussed in Sec. 4.4.5.

The CN operations between a motional and internal state qubit described above can be incorporated to provide an overall CN operation between two ions in a collection of L ions. Here, we choose the particular ion oscillator mode to be a COM mode of the collection. Specifically, to realize a controlled-not $\hat{C}_{c,t}$ between two ions (c = control bit, t = target bit), we first assume the COM mode is prepared in the zero-point state. The initial state of the system is therefore given by

$$\Psi = \left( \sum_{M_1=\downarrow,\uparrow} \sum_{M_2=\downarrow,\uparrow} \cdots \sum_{M_L=\downarrow,\uparrow} C_{M_1,M_2,\ldots M_L} |M_1\rangle_1 |M_2\rangle_2 \cdots |M_L\rangle_L \right) |0\rangle . \qquad (58)$$

$\hat{C}_{c,t}$ can be accomplished with the following steps:

(2a) Apply a $\pi$-pulse on the red sideband of ion c. This accomplishes the mapping $(\alpha|\downarrow\rangle_c + \beta|\uparrow\rangle_c)|0\rangle \rightarrow |\downarrow\rangle_c(\alpha|0\rangle + \beta|1\rangle)$, and corresponds to the operator $U_m^{1,0}$ of Cirac and Zoller [1]. We note that in the NIST experiments [17], we prepare the state $(\alpha|\downarrow\rangle + \beta|\uparrow\rangle)|0\rangle$ from the $|\downarrow\rangle|0\rangle$ state using the carrier transition. We can then implement the mapping $(\alpha|\downarrow\rangle + \beta|\uparrow\rangle)|0\rangle \rightarrow |\downarrow\rangle(\alpha|0\rangle + \beta|1\rangle)$ by applying a $\pi$-pulse on the red sideband. This is the "keyboard" operation for preparation of arbitrary motional input states for the CN gate of steps 1a - 1c above. Analogous mapping of internal state superpositions to motional state superpositions are reported in Refs. [47], [131], and [132].
(2b) Apply the CN operation (steps 1a - 1c or, the single carrier pulse described above) between the COM motion and ion t.
(2c) Apply the inverse of step (2a).

Overall, $\hat{C}_{c,t}$ provides the unitary transformation (in the $\{|\downarrow\rangle_c|\downarrow\rangle_t, |\downarrow\rangle_c|\uparrow\rangle_t, |\uparrow\rangle_c|\downarrow\rangle_t, |\uparrow\rangle_c|\uparrow\rangle_t\}$ basis)

$$\hat{C}_{c,t} = \begin{bmatrix} 1 & 0 & 0 & 0 \\ 0 & 1 & 0 & 0 \\ 0 & 0 & 0 & 1 \\ 0 & 0 & 1 & 0 \end{bmatrix}, \tag{59}$$

which is the desired logic of Eq. (53). Effectively, $\hat{C}_{c,t}$ works by mapping the internal state of ion c onto the COM motion, performing a CN between the motion and ion t, and then mapping the COM state back onto ion c. The resulting CN between ions c and t is the same as the CN described by Cirac and Zoller [1], because the operations $V^{\frac{1}{2}}(\theta)$ and $U_m^{1,0}$ commute.

A third possibility, which also uses only one internal state transition on each ion, is the following. We employ two nondegenerate motional modes, which we label here as 1 and 2. These might be the COM modes in two different directions. We first map the internal state information from two qubits j and k onto the separate motional modes (which are both initially in the $|n=0\rangle_1|n=0\rangle_2$ zero-point state). This can be accomplished as described in step (2a) above. We then apply a conditional phase gate (Eq. (54) with $\phi = \pi$) to the two motional modes. This could be accomplished by driving a $2\pi$ transition on a second order red sideband, at frequency $\omega_o - \omega_1 - \omega_2$, on a particular (extra) ion "g" which is initially in the $|\downarrow\rangle_g$ state. This ion is not used to store information; it is only used for this one particular purpose. This would be followed by operations which map the motional states back onto the internal states of ions j and k (like step (2c) above). Overall, this provides a phase gate (Eq. (54) with $\phi = \pi$) between ions j and k. To make a CN gate between ions j and k, we need to precede the above operations with a $\pi/2$ pulse on the internal state of ion j (or k) and follow the above operations with a $-\pi/2$ pulse on the internal state of ion j (or k).

In this section, we have assumed that each ion can be addressed independently. Also, since very many such operations will be desired for a quantum computer, the accuracy or fidelity of these operations is of crucial importance. These issues are confronted in Sec. 4. As noted in Sec. 2.3, in each separate operation involved in a quantum computation, such as application of the red sideband in step 2(a) to ion j, a definite phase of the applied fields is assumed. This phase for each ion can be chosen arbitrarily for the first operation, but upon successive applications of the same operation to the same ion, it must be held fixed, or at least be known, relative to the initial phase. An exception to this is application of $2\pi$-pulses as in step 1(b) where the phase of the fields does not enter into the final result of the operation.

3.4. Entangled states for spectroscopy

A collection of atoms, whose internal states are entangled through the use of quantum logic, can improve the quantum-limited signal-to-noise ratio in spectroscopy. Compared to the factorization problem, this application has the advantage of being useful with a relatively small number of ions and logic operations. For example, for high-accuracy, ion-based frequency standards [74,189,190,191], use of a relatively small number of trapped ions (L ≤ 100) appears optimum. As outlined here, the states involving L ions which are useful for spectroscopy and frequency standards can be generated with L logic gates. For L ≃ 100, a significant improvement in performance in atomic clocks could be expected.

In spectroscopy experiments on L atoms, in which the signal relies on detecting changes in atomic populations, we can view the problem in the following way using the spin-½ analog for two-level atoms. The total angular momentum of the system is given by $\vec{J} = \sum_{i=1}^{L} \vec{S}_i$ where $\vec{S}_i$ is the spin of the ith atom. The basic task is to measure $\omega_o$, the frequency of transitions between the $|\downarrow\rangle$ and $|\uparrow\rangle$ states Eq. (11). We first prepare an initial state for the spins. We assume spectroscopy is performed by applying (classical) fields of frequency $\omega_R$ for a time $T_R$ according to the method of separated fields by Ramsey [173]. The same field is applied to all atoms. After applying these fields, we measure the final state populations; that is, we detect, for example, the number of atoms $N_\downarrow$ in the $|\downarrow\rangle$ state. In the spin-½ analog, this is equivalent to measuring the operator $J_z$, since $N_\downarrow = J\mathbb{I} - J_z$ where $\mathbb{I}$ is the identity operator. We assume the internal states can be detected with 100% efficiency (Sec. 2.2.1). If all sources of technical noise are eliminated, the signal-to-noise ratio (for repeated measurements) is fundamentally limited by the quantum fluctuations in the number of atoms which are observed to be in the $|\downarrow\rangle$ state. These fluctuations can be called quantum projection noise [100]. Spectroscopy is typically performed on L initially nonentangled atoms (for example, $\Psi(t=0) = \Pi_{i=1}^{L}|\downarrow\rangle_i$). With the application of the Ramsey fields, the atoms remain nonentangled. For this case, the imprecision in a determination of the frequency of the transition is limited by projection noise to $(\Delta\omega)_{meas.} = 1/(LT_R\tau)^{½}$ where $\tau \gg T_R$ is the total averaging time [100]. If the atoms can be initially prepared in particular entangled states, it is possible to achieve $(\Delta\omega)_{meas.} < 1/(LT_R\tau)^{½}$. Initial theoretical investigations for ions [3,9] examined the use of correlated states which could achieve $(\Delta\omega)_{meas.} < 1/(LT_R\tau)^{½}$ when the population ($\hat{J}_z$) was measured. These states are analogous to those previously considered for interferometers [192,193]. More recent theoretical investigations [194] consider the initial state to be one where, after the first Ramsey pulse, the internal state is the maximally entangled state

$$\Psi = \frac{1}{\sqrt{2}}(|\downarrow\rangle_1|\downarrow\rangle_2...|\downarrow\rangle_L + e^{i\phi(t)}|\uparrow\rangle_1|\uparrow\rangle_2...|\uparrow\rangle_L), \tag{60}$$

where $\phi(t) = \phi_o + L\omega_o t$. After applying the Ramsey fields, we measure the operator $\hat{O} = \Pi_{i=1}^{L} S_{zi}$ instead of $J_z$. This gives $(\Delta\omega)_{meas.} = 1/(L^2 T_R \tau)^{½}$, which is the maximum signal-to-noise ratio possible and corresponds to the Heisenberg limit [194]. (In the language of quantum error correction, if we express $\Psi$ in terms of the basis states $|0\rangle' = (|0\rangle + |1\rangle)/2^{½}$ and $|1\rangle' = (|0\rangle - |1\rangle)/2^{½}$, we find that $\phi(t)$ is determined from a parity check of the total state in this second basis [142]. For an atomic clock where the interrogation time $T_R$ is fixed by other constraints, this means that the time $\tau$ required to reach a certain measurement precision is reduced by a factor of L relative to the nonentangled atom case. This improvement is of significant practical importance since, to achieve high measurement precision, atomic clocks are run for averaging times τ of weeks, months, and even longer.

Cirac and Zoller [1] have outlined a scheme for producing the state in Eq. (60) using quantum logic gates. Starting with the state $\Psi(t=0) = \Pi_{i=1}^{L}|\downarrow\rangle_i|0\rangle$, we first apply a π/2 rotation ($\Omega_{o,o}t = \pi/4$, $\phi_1 = +\pi/2$ in Eq. (23)) to ion 1 to create the state $\Psi = 2^{-\frac{1}{2}}(|\downarrow\rangle_1 + |\uparrow\rangle_1)|\downarrow\rangle_2|\downarrow\rangle_3...|\downarrow\rangle_L|0\rangle$. We then apply the CN gate $\hat{C}_{1,i}$ sequentially between ion 1 and ions i = 2 through L to achieve the state of Eq. (60). An alternative method for generating this state is described in Ref. [194]. As a final example, we consider a method for generating the maximally entangled state which requires a fixed number of steps, independent of the number of ions. For simplicity, we illustrate the method for three ions. Starting with the state $\Psi(t=0) = |\downarrow\rangle|\downarrow\rangle|\downarrow\rangle|0\rangle$, we first selectively drive one of the ions (say ion 3) with a carrier π/2 pulse followed by a red sideband π pulse to give the sequence $|\downarrow\rangle|\downarrow\rangle|\downarrow\rangle|0\rangle \rightarrow 2^{-\frac{1}{2}}|\downarrow\rangle|\downarrow\rangle(|\downarrow\rangle + |\uparrow\rangle)|0\rangle \rightarrow 2^{-\frac{1}{2}}|\downarrow\rangle|\downarrow\rangle|\downarrow\rangle(|0\rangle + |1\rangle)$. We now use the Lamb-Dicke dependence of the carrier transition (Eqs. (56)) to make an odd (even) integer number of π flips correlated with the |n=0⟩ state and an even (odd) integer number of π flips correlated with the |n=1⟩ state (the laser beam intensity is assumed to be the same on all ions). We can now employ a transition to an auxiliary level. For example, if the state after the last step, is the state $2^{-\frac{1}{2}}(|\uparrow\rangle|\uparrow\rangle|\uparrow\rangle|1\rangle + |\downarrow\rangle|\downarrow\rangle|\downarrow\rangle|0\rangle)$, we could apply a blue sideband π pulse between states |↑⟩ and |aux⟩ of one of the ions (say the third) followed by a carrier π pulse on this transition to carry out the steps $2^{-\frac{1}{2}}(|\uparrow\rangle|\uparrow\rangle|\uparrow\rangle|1\rangle + |\downarrow\rangle|\downarrow\rangle|\downarrow\rangle|0\rangle) \rightarrow 2^{-\frac{1}{2}}(|\uparrow\rangle|\uparrow\rangle|aux\rangle + |\downarrow\rangle|\downarrow\rangle|\downarrow\rangle)|0\rangle \rightarrow 2^{-\frac{1}{2}}(|\uparrow\rangle|\uparrow\rangle|\uparrow\rangle + |\downarrow\rangle|\downarrow\rangle|\downarrow\rangle)|0\rangle$. (If the state after the previous step was $2^{-\frac{1}{2}}(|\uparrow\rangle|\uparrow\rangle|\uparrow\rangle|0\rangle + |\downarrow\rangle|\downarrow\rangle|\downarrow\rangle|1\rangle)$, we would sandwich the last operations between two π pulses on the $|\downarrow\rangle \rightarrow |\uparrow\rangle$ transition of the selected ion.)

Other correlated states can also be useful for spectroscopy. A strategy which essentially measures the variance of $\tilde{N}_\downarrow$ is discussed by Holland and Burnett [195] and Kim, *et al.* [196]. This method has also been incorporated into a proposed technique for spectroscopy of internal states of Bose-Einstein condensates [197].

In comparing the case for entangled vs. non-entangled states in spectroscopy, the above discussion has assumed that $T_R$ is fixed. This constraint would be valid if the ions were subject to a constant heating rate and we desired to maintain the second-order-Doppler (time dilation) shift below a certain value, for example. However, the use of entangled states may not be advantageous, given other conditions. For example, Huelga, *et al.* [198] assume that the ions are subject to a certain dephasing decoherence rate (decoherence time less than the total observation time). In this case, there is no advantage of using maximally entangled states over non-entangled states. The basic reason is that the maximally entangled state decoheres L times faster than the states of individual atoms. Therefore, when using the maximally entangled state, $T_R$ must be reduced by a factor of L for optimum performance. Because of this, the gain from use of the maximally entangled state is offset by the reduced value of $T_R$. (Huelga, *et al.* [198] actually show that a modest improvement can be obtained under these conditions by use of partially entangled states.) In appendix A, we compare entangled vs. nonentangled states in the context of a practical atomic clock application where a reference oscillator is locked to the atomic resonance.

4. Decoherence

The atomic motional and internal states, and the (logic) operations, were described above in an idealized fashion. In this section we consider some of the practical limitations to these idealizations. These limitations can generally be grouped under the heading of decoherence if, by decoherence, we mean any effect which limits the fidelity of these operations (see Sec. 4.3). This is a more general use of the term decoherence; in some treatments, decoherence refers only to dephasing of qubit states and does not include state changes. Although somewhat arbitrary, we also find it convenient to break decoherence into categories: (1) decoherence of the ion motion, (2) decoherence of the ion internal levels, and (3) decoherence caused by nonideal applied fields which are responsible for the logic operations.

4.1. Motional decoherence

For the trapped ion system discussed in this paper, decoherence may be dominated by that of the motional state. Scaling of decoherence will depend on the physical system being treated and the mechanism of decoherence [199,200]. For quantum computation with ions, motional decoherence is somewhat easier to characterize than for a general motional state since we are primarily interested in relaxation of the $|n=0\rangle$ and $|n=1\rangle$ motional states for a particular mode (for example, the center-of-mass (COM) mode along the axis of a linear trap).

4.1.1. Phase decoherence caused by unstable trap parameters

A simple form of motional decoherence is caused by fluctuations in trap parameters. Most likely, by employing electronic filtering, these parameters fluctuate slowly on the time scale of the basic operations ($\simeq 1/\Omega_{n',n}$), therefore, the motion is subject to dephasing due to the corresponding adiabatic changes in motion frequency. In the linear trap, if we assume $a_i \ll q_i^2$, $i \in \{x,y\}$ (Sec. 2.1), then for small fluctuations $\delta V_o$, $\delta U_o$, $\delta \Omega_T$, $\delta \kappa$, and $\delta R$, we have

$$\delta\omega_{x,y}/\omega_{x,y} = \delta V_o/V_o - \delta\Omega_T/\Omega_T - 2\delta R/R, \quad \delta\omega_z/\omega_z = \tfrac{1}{2}(\delta\kappa/\kappa + \delta U_o/U_o). \tag{61}$$

The relationship between these frequency fluctuations and phase fluctuations in a series of logic operations is discussed in Sec. 4.3.2. The effects of modulation of these parameters by high-frequency noise is considered in Sec. 4.1.3 and in Ref. [66]. Although an experimentally open question, it is expected that all of these parameters could be controlled sufficiently well that they should not be the primary cause of decoherence.

4.1.2. Radiative decoherence

Decoherence has received considerable attention in connection with quantum measurement [201] and has been put forth as a practical solution to the quantum measurement problem [55,56,57]. In quantum optics, a paradigm for decoherence has been to consider relaxation of the harmonic oscillator associated with a single mode of the radiation field by coupling to the environment [202- 206]. This kind of fundamental decoherence has recently been observed in the context of cavity QED by Brune, *et al.* [207]. An important result from these studies is that relaxation of superposition states occurs at a rate which increases with the separation of the states in Hilbert space and almost always precludes the existence of "large" Schrödinger-cat-like states of motion except on extremely short time scales.

A fundamental source of decoherence for the COM mode of ion motion is understood by considering that the dipole associated with the oscillating charged ion(s) is radiatively coupled to the thermal fields of the environment, at temperature T. The master equation describing the evolution of the density operator ρ for the motion (in the interaction picture) can be written [208,209,210]

$$\dot{\rho} = \frac{\gamma}{2}(\bar{n}+1)(2a\rho a^\dagger - a^\dagger a \rho - \rho a^\dagger a)$$
$$+ \frac{\gamma}{2}\bar{n}(2a^\dagger \rho a - aa^\dagger \rho - \rho a a^\dagger), \tag{62}$$

where $\bar{n}$ is the mean number of motional quanta when the ion is in equilibrium with the environment ($\bar{n} = [\exp(\hbar\omega_z/k_BT) - 1]^{-1}$), and γ is the relaxation rate of the energy to thermal equilibrium. Since ion trap experiments will typically operate in the situation where $\hbar\omega_z \ll k_BT$, then $\bar{n} \approx k_BT/\hbar\omega_z$. We will assume the ion(s) start in the subspace of density matrix elements $\rho_{00}$, $\rho_{01}$, $\rho_{10}$, and $\rho_{11}$ where $\rho_{ij} \equiv \langle i|\rho|j\rangle$. Equation (62) implies

$$\dot{\rho}_{01}(t=0) = \sqrt{2}\gamma(\bar{n}+1)\rho_{12} - \gamma(2\bar{n}+\tfrac{1}{2})\rho_{01} \approx -2\bar{n}\gamma\rho_{01},$$
$$\dot{\rho}_{00} = -\gamma[\bar{n}\rho_{00} - (\bar{n}+1)\rho_{11}], \tag{63}$$
$$\dot{\rho}_{11} = 2\gamma(\bar{n}+1)\rho_{22} - \gamma(3\bar{n}+1)\rho_{11} + \gamma\bar{n}\rho_{00}.$$

General expressions for dρ/dt and d⟨n̂⟩/dt are given in Appendix B. Based on these expressions, and in the limit that $\bar{n} \gg 1$, we will characterize the motional decoherence by the time $t^* = 1/(\bar{n}\gamma)$, which is approximately the time for the ion to make a transition from the ground state. This agrees with a classical estimate [103].

4.1.3. Radiative damping/heating

The electric dipole associated with the an ion's COM oscillatory motion will couple to thermal (black body) or ambient radiation in the environment. However, since the wavelength corresponding to ion oscillation frequencies will typically be much larger than the trap electrode spacings, this coupling can be described by lumped-circuit models [102,103]. In these models, we assume the ion's motion induces currents in the neighboring electrodes; these currents, in turn, couple to the resistance of the electrodes or circuit elements attached to the electrodes. In experiments where these resistances are purposely made high to maximize damping of the ion COM motion, observed time constants agree with the model [77,78,79,102,103,104]. In the two (single-ion) experiments which have been able to achieve cooling to the n=0 motional state, the measured value of $t^*$ was about 0.15 s for a $^{198}$Hg$^+$ ion [44] and about 1 ms for a single $^9$Be$^+$ [17,21,45,47,131,132,211]. In these experiments, $t^*$ was intended to be made as long as possible; however, the observed values of $t^*$ were considerably shorter than what we would calculate from the model, as shown below.

The model considers the electric-circuit equivalent shown in Fig. 6(a). Effectively, the electric-dipole oscillator formed by the ion COM motion can be considered to be confined in a cavity formed by the trap electrodes. A useful representation of this situation is to model the COM motion (in one direction) as a series inductive-capacitive ($\ell$c) circuit which is shunted by the capacitance of the trap electrodes as shown [9,103]. The resistance r is due to losses in the electrodes and conductors which connect the electrodes. The Johnson noise associated with this resistance can heat the ions. The equivalent inductance of the ion COM motion is given by $\ell_L \simeq md^2/L(\alpha q)^2$ where d is the characteristic internal dimension of the ion trap electrodes, L the number of ions, and $\alpha$ is a geometrical factor on the order of 1 which can be computed [212]. For traps with hyperbolic electrodes, if we consider motion in the z direction, $d = 2z_o$ (the separation of the endcap electrodes [67-70] and $\alpha \simeq 0.8$ [213,214,215]. For the trap used in the NIST experiments, where $z_o = 130$ μm, $\ell_1 \simeq 60\,000$ henries! The resistance r yields a time constant $\ell_L/r = 1/\gamma$. This implies [9]

$$t^* = \frac{1}{\gamma \bar{n}} = \frac{\hbar Q}{k_B T} = \frac{\hbar \omega_z \ell_L}{k_B T r} = \frac{4m\hbar \omega_z}{q^2 S_E(\omega_z)}, \tag{64}$$

where $Q = \omega_z \ell_L/r$ is the quality factor of the ion oscillator. The last expression in Eq. (64) shows $t^*$ in terms of the spectral density of electric field fluctuations at the site of the ion which can be written $S_E = 4k_B Tr(\alpha/d)^2$ where $4k_B Tr$ is the Johnson noise associated with the effective resistor r. The trap reported in the NIST experiments had the endcaps made of a single piece of molybdenum as shown schematically in Fig. 6(b) (ring electrode not shown). We assume the induced currents flow in the endcaps electrode as indicated in the shaded portion of Fig. 6(b), where $\delta_S$ is the skin depth. This seems to be a conservative estimate since currents will also flow in the sides of the endcap electrodes and will not be confined to the skin depth [216], thereby reducing the effective value of r. Taking the resistivity of molybdenum to be $\rho(Mo) = 5.7 \times 10^{-6}$ Ω-cm, $w_T = 125$ μm, and $x_T = 1$ mm, we find $r \simeq 2\rho x_T/\delta_S w_T = 0.0415$ Ω. If we assume $T \simeq 300$ K, $t^* \simeq 4.6$ s, considerably longer than the observed value of $t^*$ ($\simeq 1$ ms). An alternative model for dissipation of charges moving parallel to a nearby surface [200] predicts a much larger value of $t^*$. Lamoreaux [217] has derived an expression which agrees with Eq. (64), however he chooses a value of r higher than what we calculate.

Faster heating will occur if T >> 300 K. This can be expected at the relatively high powers delivered to the step-up transformer used to generate $V_o$ ($\simeq$ 1 W in the NIST experiments), but this alone cannot explain the difference between what the model predicts and the observed heating rate. Conversely, if the trap can be operated at cryogenic temperatures, this kind of heating should be substantially reduced.

In a linear trap, heating can occur in the axial and radial directions due to the interplay of a stray static field (e.g., from patch potentials on the electrodes) and a noise on $U_o$ or $V_o$ at one of the secular frequencies. Here, we explain what appears to be the most important case, a fluctuation of $U_o$ in the presence of a stray static field along the z direction. This and other cases are discussed in Ref. [66].

In equilibrium, the force on an ion from a stray static field $\vec{E} = E_s \hat{z}$ is balanced by the field from the trap given by Eq. (2). We have $E_s = 2\kappa q U_o z_{equil}$, where $z_{equil}$ is the equilibrium position of the ion (here, we assume $z_{equil} = 0$ in the ideal case). A fluctuation in $U_o$ therefore causes a fluctuation in the electric field seen by the ion. We can characterize the spectral density of these field fluctuations as $S_E(\omega) = (E_s/U_o)^2 S_{U_o}(\omega)$ where $S_{U_o}(\omega)$ is the spectral density of potential fluctuations. From Eq. (64), we have

$$t^* = \frac{4m\hbar\omega_z}{q^2 S_{U_o}(\omega_z)} \left[\frac{U_o}{E_s}\right]^2 .$$

For a very small linear trap where $\kappa \simeq (0.3 \text{ mm})^{-2}$, and for m = 9 u (e.g., $^9Be^+$) and $\omega_z/2\pi$ = 10 MHz, we have $U_o \simeq$ 17 V. For $E_s \simeq$ 100 V/m and $S_{U_o}$ = (1 nV)$^2$/Hz (the Johnson noise voltage from a 60 $\Omega$ resistor at room temperature), we have $t^* \simeq$ 430 s. Since $t^* \propto \omega_z^5 d^6$, we see there is a premium on having a relatively large trap with large values of $U_o$ to keep $\omega_z$ as large as possible.

In the above, we assumed that the ions' motion couples to the surroundings through the oscillating electric dipole due to the ion(s) COM motion. In situations where the extent of the ion sample is small compared to the distance to the electrodes, the induced currents result dominantly from the COM mode; therefore radiative decoherence from modes other than the COM mode can be substantially suppressed [1,103]. For example, for two trapped ions aligned along the z axis, we would expect electric fields from stray (fluctuating) potentials on one of the end electrodes to cause an excitation force on the z stretch mode which is suppressed by a factor equal to the ratio of the ion spacing to trap dimensions compared to the force on the COM mode.

Fluctuations in $V_o$ and $U_o$ can also cause heating of the ions. These sources, expected to be small in current experiments, are discussed in Ref. [66]. This heating might be caused by parametric processes. For example, heating could be induced if the trap pseudopotential is modulated (coherently or by noise) at twice the secular frequency. This problem has been treated by Savard, *et al.* [218] in the context of optical dipole traps for neutral atoms (a kind of Paul trap for an electron to which the atomic core is attached). For the conditions of the NIST $^9Be^+$, this kind of heating was estimated to be to small to account for the observed value of $t^*$ [66].

4.1.4. Injected noise

Noise from various ancillary electronic devices might be injected onto the electrodes; this additional electronic noise could then heat the ions. Added electronic noise can be modeled as a resistor r in Fig. 6 that has a temperature much higher than the ambient temperature. These sources of noise can be tested by injecting noise at a level equal to or above the ambient noise level and looking for a shortening of $t^*$. For this test to be valid, we must have a reliable means of sensing the noise at the trap electrodes. This may be difficult to achieve in practice, since, in the experiments, it usually desirable to filter the electrodes from the rest of the environment at the motional frequencies. This was the case in the NIST experiments, where electronic filtering at the motional frequencies precluded the direct observation of voltage noise on the electrodes.

4.1.5. Motional excitation from trap rf fields

The rf fields used for trapping in a Paul trap can lead to excitation of the ion motion. We will consider four types of effects in which the rf micromotion can, indirectly, cause heating. For the first type of effect considered, we will analyze heating of the axial motion of a single ion in a conventional spherical-quadrupole Paul trap; the results can be generalized to other cases such as the heating of radial modes in a linear trap. For a spherical quadrupole trap, motion in the z (axial) direction has the same form as Eq. (4). If we assume a potential $V_o \cos(\Omega_T t)$ is applied between the ring and endcap electrodes, we have

$$z(t) \simeq A_z \cos(\omega_z t + \phi_z) \left[1 + \frac{q_z}{2} \cos(\Omega_T t)\right], \qquad (66)$$

where $A_z$ and $\phi_z$ are set by initial conditions, $q_z \equiv 8qV_o/(m\Omega_T^2(r_o^2 + 2z_o^2))$, $\omega_z \simeq q_z \Omega_T/(8)^{1/2}$, $r_o$ is the inner radius of the ring electrode, $2z_o$ is the distance between endcaps, and we assume $q_z \ll 1$. In the radial direction, the motion will be similar with radial secular frequency $\omega_r = \omega_z/2$. From this equation, we see that the ion's motion in the z direction has components at frequencies $\omega = \Omega_T \pm \omega_z$. Since the rf voltage $V_o$ is typically applied through a resonant step-up transformer (shown schematically in Fig. 6(c)), the ion's motion at these frequencies might be expected to couple to the resistance $R_T$ between the ring and endcaps associated with this step-up transformer. At a frequency $\omega$ near $\Omega_T$, the impedance between the ring and endcaps electrode can be represented by a parallel tuned circuit as shown in Fig. 6(c). This impedance is given by

$$Z(\omega) = \frac{R_T}{1 + 2iQ\left(\frac{\omega - \Omega_T}{\Omega_T}\right)} \equiv r_s(\omega) + iX(\omega), \qquad (67)$$

where Q is the quality factor for the circuit ($Q \simeq R_T/(\Omega_T L_T) = R_T \Omega_T C_T$). Coupling to the effective series resistance $r_s(\omega)$ should not occur if the endcap electrodes are placed symmetrically around the ring electrode (as intended). However, since the relative electrode positions are difficult to control in small ion traps, a displacement of the ring electrode toward one of the endcaps will cause a net induced current from ion motion to flow between ring and endcaps at frequencies $\omega = \Omega_T \pm \omega_z$. We characterize this current (in the z direction) by $I = \beta q \dot{z}/(2z_o)$, where $\dot{z}$ is the ion velocity and $\beta$ is a geometrical parameter which expresses the coupling to the electrodes ($\beta = 0$ when the ring is placed symmetrically between endcaps). The effective inductance of the ions for this type of coupling is given by $\ell_L' = \ell_L/\beta^2$. Associated with $r_s(\omega)$ in a small bandwidth $\Delta \nu$ around $\omega$ is a series Johnson noise $\langle V_n^2 \rangle = 4k_B T \Delta \nu r_s$. The electric field associated with this noise at frequencies $\omega = \Omega_T \pm \omega_z$ can heat the ion motion in a way similar to the way in which the motion can be excited by a coherent excitation at these frequencies [68]. From Eq. (3.7) of Ref. [68], we see that an electric field $E_1 \hat{z}$ applied at a frequency $\Omega_T \pm \omega_z$ is equivalent to an electric field $[\omega_z/(\Omega_T \sqrt{2})]E_1 \hat{z}$ applied at frequency $\omega_z$. Therefore, the Johnson noise from the series resistance $r_s(\omega = \Omega_T \pm \omega_z)$ is equivalent to that from a series resistance $r_s' = [\omega_z^2/(2\Omega_T^2)]r_s(\omega = \Omega_T \pm \omega_z)$ at frequency $\omega_z$. The heating from this source is characterized by the heating time $t^{*\prime} = \ell_L'/(r_s' \bar{n})$. For the NIST single $^9Be^+$ ion experiments, this source of heating was estimated to be negligible.

A second type of rf heating can occur due to the Coulomb interaction between ions. In a collection of ions, such as a string of ions in an ideal linear trap, the Coulomb coupling between ions makes all of the motional modes, except the COM modes, anharmonic. This can lead to excitation of these modes in a Paul trap by the driving fields at frequency $\Omega_T$. This excitation and the resulting chaotic motion have been studied extensively for two ions trapped in a conventional Paul trap. Experiments have been performed at Munich and IBM; the results of the these experiments and comparisons to theory are included in a review by Walther [219]. Moreover, even for a single (harmonically bound) ion, nonlinear subharmonic excitation can occur if the exciting field is inhomogeneous [220]. Both types of heating can be made negligible when the mode frequencies are not submultiples of $\Omega_T$ and when all modes are sufficiently cooled and very harmonic.

Another type of rf heating occurs in some experiments when the condition $a_i, q_i^2 \ll 1$ is not rigorously satisfied and the trapped ions are fairly energetic. The motion of single ions (or multiple ions when the mutual Coulomb interaction, or "space charge," can be neglected) will be unstable when the condition $p\omega_z + m\omega_r = \Omega_T$ is satisfied in a spherical quadrupole trap or when $p\omega_x + m\omega_y = \Omega_T$ in a linear trap (p and m are integers). This type of heating has been observed in some beautiful experiments [221] and has been explained theoretically [222]. These "heating resonances" arise from terms in the trap potential which are higher order than quadratic. We briefly explain their origin for a single ion. For simplicity, we neglect the contributions of the static potentials; their inclusion will not change the analysis significantly.

In general, the potential of the trap can be expanded about the equilibrium position of the ion and written in spherical coordinates $(r,\theta,\phi)$ as

$$\Phi = V_o \cos\Omega_T t \sum_{l=0}^{\infty} \sum_{m=-l}^{l} C_{l,m} \left[\frac{r}{d}\right]^l Y_{l,m}(\theta,\phi), \quad \textbf{(68)}$$

where $Y_{l,m}$ are the spherical harmonics and d is a characteristic dimension of the trap. We take d

= R for the linear trap (and d = $(r_o^2 + 2z_o^2)^{1/2}$ for the spherical quadrupole trap [67,68,69,70]). For the ideal linear trap (Eq. (1)) with $U_r = 0$, and neglecting a constant term, only two terms in the expansion in Eq. (68) contribute, $C_{2,2} = C_{2,-2} = -(2\pi/15)^{1/2}$. (For the ideal spherical quadrupole trap, neglecting a constant term, only one term contributes, $C_{2,0} = 4(\pi/5)^{1/2}$.) For a nonideal linear trap, the resonant heating can be explained as due to terms in Eq. (68) which give rise to a Hamiltonian of the form

$$H_I = q\Phi = qV_{p,m}\left[\frac{x}{R}\right]^p\left[\frac{y}{R}\right]^m \cos\Omega_T t. \tag{69}$$

In the interaction picture for the motion, this becomes

$$H_I' \simeq \frac{qV_{p,m}}{2}\left[\frac{x_o}{R}\right]^p\left[\frac{y_o}{R}\right]^m \left((a_x^\dagger)^p(a_y^\dagger)^m + (a_x)^p(a_y)^m\right),$$

where this last expression is the leading term which satisfies the resonance condition $p\omega_x + m\omega_y = \Omega_T$. This interaction will be suppressed because of the inherent smallness of high-order anharmonic terms $V_{p,m}$ (for simple trap electrode shapes) and the smallness of the terms $(x_o/R)^p$ and $(y_o/R)^m$. Furthermore, if the x and y modes are cooled to near the zero-point energy, matrix elements of the motional operators will be near 1. For large amplitudes of motion, the mode frequencies are not well defined because of anharmonic terms and heating from this coupling would be expected to occur. In any case, it is easy to check for a resonant heating of this type by varying the resonant frequencies $\omega_x$ and $\omega_y$ relative to $\Omega_T$. It can also be checked by varying the initial amplitudes of motion in the modes. These tests were used for the NIST single $^9$Be$^+$ ion experiments; no change in the heating was observed.

Similarly, in a linear trap with many ions, we would expect resonances to occur when $\sum_{k=1}^{3L} m_k \omega_k = \Omega_T$ where $m_k$ are integers and $\omega_k$ are normal mode frequencies (see Sec. 2.3.2). If all modes are cooled to the point where only a few motional states are excited, then the mode absorption spectrum will consist of sharp features around the mode frequencies, and the resonances can be avoided by changing the trap parameters. Moreover, the coupling parameters (Sec. 4.1.8) will, in general, be very small.

A fourth type of heating due to rf trap fields is explained as follows. A common problem in ion trap experiments is the presence of stray static electric fields. These fields can give rise to coherent motion at frequency $\Omega_T$ and potentially to heating, which must be accounted for. Stray static electric fields can arise from potential variations on the electrode surfaces ("patch" fields) due, for example, to the finite crystalline grain size of the electrode material [223], or charge buildup on the trap electrodes. Charge buildup can occur because, typically, ions are created by electron impact ionization of neutral atoms which pass through the trap. Often, the ionizing electrons are also collected by the electrode surfaces. Electrode charging is particularly important at low temperatures where, apparently, adsorbed gases on the electrodes can provide an insulating surface upon which stray charge resides for long periods of time (hours).

If stray static electric fields are present, the equilibrium position of the ions is shifted to a place where the force from the stray field is counterbalanced by the force from the pseudopotential. We will analyze the effects of such stray fields using a classical treatment of the motion of a single trapped ion. In general, $\vec{E}_{stray} = E_{sx}\hat{x} + E_{sy}\hat{y} + E_{sz}\hat{z}$. Stray fields along the z direction in a linear trap merely shift the origin along this direction and can therefore be neglected. For balance in the x and y directions, we have $F_x = qE_{sx} - \partial(q\Phi_p)/\partial x = 0$ and $F_y = qE_{sy} - \partial(q\Phi_p)/\partial y = 0$ where $\Phi_p$ is given by Eq. (6). If the x,y coordinate system is chosen so that the equilibrium position of the ion is at the origin in the absence of stray fields, we find a new equilibrium position $\Delta x \hat{x} + \Delta y \hat{y}$ for the ion and a resultant motion which are, to first order in $q_x$ and $q_y$ given by

$$x(t) \simeq (\Delta x + A_x \cos(\omega t + \phi_x))\left[1 + \frac{q_x}{2}\cos(\Omega_T t)\right], \quad \Delta x = \frac{qE_{sx}}{m\omega^2},$$

$$y(t) \simeq (\Delta y + A_y \cos(\omega t + \phi_y))\left[1 - \frac{q_y}{2}\cos(\Omega_T t)\right], \quad \Delta y = \frac{qE_{sy}}{m\omega^2},$$

(71)

where, as in Eq. (4), $A_x$ and $A_y$ are the amplitudes of secular motion. The presence of the offsets $\Delta x$ and $\Delta y$ means that the ion motion has an additional component at frequency $\Omega_T$. This motion will effectively give rise to sidebands on the applied radiation as seen by the ion, thereby reducing the size of matrix elements between states. For example, for single-photon transitions driven by a traveling wave with wavevector $\vec{k} = k_x\hat{x} + k_y\hat{y}$, the electric field from this traveling wave at the site of the ion is proportional to $\exp(i[k_x x(t) + k_y y(t) - \omega t + \phi]) + c.c.$ Terms in the exponential like $k_x \Delta x$ are (constant) phase shifts which can be neglected. Terms like $A_x\cos(\omega t + \phi_x)[1 + (q_x/2)\cos(\Omega_T t)]$ are just the motion of the ions in the ponderomotive potential in the ideal case. The factor from the remaining term,

$$\exp(i\frac{q_x}{2}[k_x \Delta x - k_y \Delta y]\cos(\Omega_T t)) = \cos[\phi_\Omega \cos(\Omega_T t)] + i\sin[\phi_\Omega \cos(\Omega_T t)],$$

(72)

where $\phi_\Omega \equiv (q_x/2)[k_x\Delta x - k_y\Delta y]$ gives rise to frequency modulation sidebands on the spectrum which are spaced by $\Omega_T$. If we consider the carrier or central part of this spectrum (the first term in the expansion of $\cos[\phi_\Omega \cos(\Omega_T t)]$, we find that the matrix elements are reduced by the factor $J_0(\phi_\Omega)$ compared to the case where the static fields are absent. For $\phi_\Omega \ll 1$, $J_0(\phi_\Omega) \simeq 1 - (\phi_\Omega/2)^2 \simeq \exp(-(\phi_\Omega/2)^2)$. Therefore the effect of the micromotion looks like an additional Debye-Waller factor due to the smearing out of the atom's position over the exciting wave (see the discussion following Eq. (25) and Sec. 4.3.5).

To the extent that offset fields are constant, they should not cause heating unless the $\Omega_T$

sidebands give rise to unwanted spectral components that are close to transition frequencies of interest. However, the offsets $\Delta x$ and $\Delta y$ can lead to a problem if the trapping field has noise $V_n$ at frequencies $\Omega_T \pm \omega_r$, that is, $V_o \cos\Omega_T t \rightarrow V_o \cos\Omega_T t + V_n \cos(\Omega_T + \omega_r)t + V_n \cos(\Omega_T - \omega_r)t$. In this case (assuming $\Delta y = 0$ for simplicity), the ion experiences noise fields at frequencies $\Omega_T \pm \omega_r$ equal to $E_x = -\partial \Phi/\partial x = -V_n \Delta x/R^2$ (Eq. (1)). From the first part of this section, this is equivalent to noise fields of amplitude $V_n \Delta x q_x/(4R^2)$ applied at frequency $\omega_r$. For the NIST single $^9Be^+$ ion experiments, this effect was estimated to be negligible. Moreover, this source of heating was tested for by purposely applying a static field offset and seeing if the observed heating rate increased; a null result was obtained. Experimentally, it has been possible to reduce stray static electric fields by heating the electrodes [191,224] or cleaning the electrodes with electron bombardment [225]. Alternatively, they can be compensated for with the use of correction electrodes [226-228].

4.1.6. Fluctuating patch fields

Electrode patch fields might also vary in time; if the spectrum of these variations overlaps the mode frequencies, this could lead to ion heating. Investigations into patch fields have primarily been done for time independent or very slowly varying ($< 500$ Hz) components [223]. However, fluctuating patch fields caused by fluctuating adsorbate coverage has been studied in some cases [229,230]. These studies differ somewhat in the low frequency behavior at time scales comparable to diffusion times but at frequencies $\nu > \nu_c$, where $1/\nu_c$ is a time constant characteristic of surface diffusion, they predict $S(\Phi_n,\nu) \propto \nu^{-\alpha}$, where $S(\Phi_n,\nu)$ is the spectral density of rms potential fluctuations $\Phi_n$ (in units of $V^2$-$Hz^{-1}$) and $\alpha \simeq 3/2$ [230].

To estimate the effects of time-varying patch potentials on a single trapped ion, we assume the ion is sensitive only to the potential on a portion of a nearby electrode. We take the area of this portion equal to $\pi a_p^2$ where $a_p$ is the distance between the ion and the nearest part of the electrode surface. The effects of these potentials on the ion motion in one direction is then estimated by assuming the ion is centered between two capacitor plates of area $\pi a_p^2$ separated by a distance $2a_p$. The fluctuating potentials on these plates give rise to a fluctuating field at the site of the ion which can then excite its motion.

Patch potential fluctuations can be caused by the fluctuations in the surface coverage from adsorbed background gas molecules (or atoms). High frequency fluctuations appear to be dominated by surface diffusion rather than adsorption and desorption [230]. The number of adsorbed molecules in an area $\pi a_p^2$ can be approximated by $N(\theta) \simeq \theta(\pi a_p^2)/(\pi r_a^2)$ where $\theta$ is the fractional coverage and $r_a$ is the radius of the adsorbed molecule. For low coverages ($\theta \ll 1$), the number of molecules will fluctuate randomly, $\Delta N \simeq N^{1/2}$, which leads to fluctuations in coverage $\Delta \theta \simeq \theta^{1/2} r_a/a_p$. A simple model for changes in the surface potential due to adsorbed molecules is that the molecules are polarized by the surface and effectively screen the surface potential. We can relate the change in potential of the plate to the change in surface coverage by $\Delta \Phi \simeq \kappa \Delta \theta = \kappa \theta^{1/2} r_a/a_p$, where $\kappa$ is a proportionality constant.

If we take $S(\Phi_n,\nu) = S_o$ (constant) for $\nu < \nu_c$ and $S(\Phi_n,\nu) = S_o(\nu_c/\nu)^{3/2}$ for $\nu > \nu_c$, we have $\langle \Phi_n^2 \rangle = \int S(\Phi_n,\nu) d\nu = 3 S_o \nu_c$. Here, $\langle \Phi_n^2 \rangle^{1/2}$ is taken to be equal to the value of $\Delta \Phi$ estimated in the previous paragraph. The cutoff frequency $\nu_c$ is given by $\nu_c \simeq 1/t_{diff}$ where $t_{diff}$ can be approximated by $t_{diff} = l_d^2/(4D)$ where D is the diffusion constant and $l_d$ is the diffusion length [231]. Here, we take $l_d$ to be the radius of the effective patch ($l_d \simeq a_p$). We then find for $\nu > \nu_c$,

$$S(\Phi_n,\nu) = 4\theta D^{1/2} \left[\frac{(\kappa r_a)^2}{3a_p^3}\right] \nu^{-3/2}, \tag{73}$$

where an extra factor of 2 has been included to account for two capacitor plates which are placed on either side of the ion. To calculate the heating rate from these potential fluctuations, we first note that they will primarily act over a narrow bandwidth associated with the ion's motional frequency. In this case, we can represent the fluctuations as coming from the Johnson noise of a resistor r at temperature T connected between the capacitor plates, that is, $S(\Phi_n,\nu) = 4k_BTr$ (assuming the capacitive impedance is much greater than r). Therefore, we can rewrite Eq. (64) as

$$t^* = \frac{4\hbar\omega_z \ell_L}{S(\Phi_n,\nu)}, \tag{74}$$

where $\omega_z$ is the ion oscillation frequency.

Since the polarizability of molecules and atoms does not change dramatically for different species, we will estimate $\kappa$ from a measurement of change in surface potential for potassium atoms on tungsten. From Fig. 2 of Schmidt and Gomer [232], we find $\kappa \simeq 3$ V. For an estimate of $\theta$, we extrapolate the data presented in Fig. 6.6 of Tompkins [233] for $H_2$ on tungsten and find $\theta \simeq 0.13$ at a partial pressure of $10^{-8}$ Pa. The diffusion constant for $H_2$ on Mo is approximately equal to $10^{-11}$ cm$^2$-s$^{-1}$ [233]. To make a comparison with the heating observed on a single $^9Be^+$ ion in the NIST experiments, we take $l_d \simeq a_p \simeq 130$ μm, $\omega_z/2\pi = 11$ MHz, L = 1, $\ell_1 = 6.2 \times 10^4$ H, $r_a = 10$ nm (Sec. 4.1.3), and we find $\nu_c \simeq 2.4 \times 10^{-7}$ Hz and $t^* \simeq 30$ s.

This model is very sensitive to the high frequency dependence of $S(\Phi_n,\nu)$ on $\nu$, and because of the very low value of $\nu_c$ estimated here, the model should be refined. However, we note that if the value of $t^*$ ($\simeq 1$ ms) for $^9Be^+$ observed in the NIST experiments is caused by fluctuating potentials on the surfaces of the electrodes, this would correspond to $S(\Phi_n, 11$ MHz$) \simeq (1.3$ nV$)^2$/Hz. This should be detectable with a sensitive amplifier. Therefore, independent of the model, this type of noise may be detectable in a straightforward way. Conversely, we note that a single trapped ion in the experiments considered here will be an extremely sensitive detector of potential fluctuations on electrodes in vacuum.

### 4.1.7. Field emission

Field emission from sharp protrusions on the electrode surfaces can cause ion heating,

either from the direct electron-ion Coulomb coupling or from associated electronic noise on the electrodes. Field emission caused by the trap potentials is not unexpected, and field emission points have been observed to grow in a number of ion trap experiments. For typical values of $V_o$ and trap dimensions, the electron transit times from one electrode surface to another are much less than $1/\Omega_T$ so that field emission occurs as if the fields are quasi-static.

Since the onset of field emission varies exponentially with the applied voltage between electrodes, it is possible to check for field emission by varying the trap potentials by small factors, and monitoring the ion heating rate. This technique appears to have ruled out field-emission heating in the NIST experiments since the change in ion heating was much less than a factor of two when $V_o$ was reduced by a factor of two. This argument assumes the exponential variation of field emission with applied voltage; if the emission or some leakage current is less sensitive to voltage changes, this test may not be valid. If field emission points are formed, it is usually straightforward to remove them by momentarily applying a large negative potential to the electrode in question. The resulting high current is usually sufficient to "burn out" the field emission tip.

4.1.8. Mode cross-coupling from static electric field imperfections

According to the scheme of Cirac and Zoller [1], the operations which provide quantum entanglement of the internal states of L trapped ions involve the coherent manipulation of a single mode k of collective motion. In the quantum logic scheme discussed in Sec. 3.3, this mode is typically taken to be the COM mode along the axis of a linear trap. A potential source of motional decoherence is caused by the coupling of this kth mode to one or more of the 3L-1 other spectator modes of vibration in the trap. If the 3L-1 other modes of oscillation are not all laser-cooled to their zero-point energy, then energy can be transferred to the kth mode of interest. Even when the spectator modes are cooled to the zero-point state, they can act as a reservoir for energy from the COM mode. This does not lead to heating but can cause decoherence. Ideally, the ions are subjected to quadratic potentials as in Sec. 2.1. In practice, higher-order static potential terms are present; these terms can induce a coupling between the modes. Similar couplings are induced by the intentionally-applied time-varying fields necessary for providing entanglement; these are discussed in Sec. 4.4.7 below.

We will assume that the higher order field gradients act as a perturbation to the (harmonic) normal mode solution. Following the convention of Eq. (30), these fields will be specified by $E_i$ for $i \in \{1,2,...3L\}$ where the index i specifies both the ion and direction of $\vec{E}$. We write the electric field at the jth ion as

$$\vec{E}_j = E_j \hat{x} + E_{L+j} \hat{y} + E_{2L+j} \hat{z}, \quad j \in \{1, 2, ...L\}. \tag{75}$$

From Eq. (31), we can write the equation of the kth normal mode as [109,234]

$$\frac{\partial^2 q_k}{\partial t^2} + \omega_k^2 q_k = \frac{q}{m} \sum_{i=1}^{3L} D_k^i E_i. \tag{76}$$

In general, we can write

$$E_i = E_i(\{u_p\}) = E_i(\{q_j\}) =$$

$$E_i(\{q_j\}=0) + \sum_{m=1}^{3L} q_m \left[\frac{\partial E_i}{\partial q_m}\right]_{\{q_j\}=0} + \frac{1}{2} \sum_{l=1}^{3L} \sum_{m=1}^{3L} q_l q_m \left[\frac{\partial^2 E_i}{\partial q_l \partial q_m}\right]_{\{q_j\}=0} + \ldots \tag{77}$$

where the derivatives are evaluated at the equilibrium positions. The first term on the right side of this equation just gives rise to a shift of the equilibrium positions, and the second term can be absorbed into new definitions of the normal mode frequencies $\omega_i$. The second-order term (last term shown in this equation) can resonantly couple two modes of oscillation (l and m) to the normal mode of interest k. We find a possible resonant term:

$$\frac{\partial^2 q_k}{\partial t^2} + \omega_k^2 q_k = \frac{q}{m} \sum_{i=1}^{3L} D_k^i q_l q_m \left[\frac{\partial^2 E_i}{\partial q_l \partial q_m}\right]_{\{q_j\}=0}, \tag{78}$$

where the l and m mode frequencies satisfy $|\omega_l \pm \omega_m| = \omega_k$. This type of coupling can either add to or extract energy from mode k, depending on the relative phases of motion in the three modes. By substituting the free solution to modes l and m ($q_j(t) = Q_k \exp(\pm i(\omega_j t + \phi_j))$) into the last equation, we find that if $q_k(t=0) = (dq_k/dt)_{t=0} = 0$, the driven solution to the amplitude of mode k initially grows linearly with time:

$$|q_k(t)| = \left|\frac{qt}{2m\omega_k} \sum_{i=1}^{3L} D_k^i Q_l Q_m \left[\frac{\partial^2 E_i}{\partial q_l \partial q_m}\right]_{\{q_j\}=0}\right|. \tag{79}$$

We illustrate with an approximate numerical example which might have been expected to play a role in the heating that was observed in the NIST experiments. In those experiments, performed on single $^9Be^+$ ions, the heating that was observed was such that the ion made a transition from the n=0 to n=1 level in about 1 ms. For a single ion, the three normal modes are just the oscillation modes along the x, y, and z directions ($q_1 = x$, $q_2 = y$, $q_3 = z$; $D_k^i = \delta_{i,k}$). The mode frequencies were $(\omega_x, \omega_y, \omega_z)/2\pi \simeq (11.2, 18.2, 29.8)$ MHz, thus approximately satisfying the condition $\omega_x + \omega_y = \omega_z$. For sake of argument, we assume this resonance condition to be exactly satisfied. We consider heating of the x motion assuming both the y and z modes are excited. From Eq. (79), we find $|x(t)| = |qtA_yA_z[\partial^2E_x/\partial y\partial z]_{(y=z=0)}/(2m\omega_x)|$ where $A_y$ and $A_z$ are the amplitudes of motion in the y and z directions. For simplicity, we neglect the fact that energy is exchanged between the three modes; that is, we assume the amplitudes of the y and z motion remain fixed. In this approximation, if $A_y = A_z = \xi$, the time it takes to excite the x motion to the same amplitude ($\xi$) is given by $t = |2m\omega_x/(q\xi[\partial^2E_x/\partial y\partial z]_{(y=z=0)})|$. If $\xi \simeq 10$ nm (corresponding to $\langle\hat{n}_y\rangle \simeq \langle\hat{n}_z\rangle \simeq 1$ for the conditions of the single $^9$Be$^+$ ion NIST experiments, the field gradient required to drive the x motion to an amplitude of 10 nm ($\langle n_x\rangle \simeq 1$) in the observed time of 1 ms is approximately $\partial^2E_x/\partial y\partial z = 1000$ V/mm$^3$. It is highly unlikely the gradient was this large for the NIST experiments, and, furthermore, the resonance condition was only approximately satisfied. Moreover, this source of heating was easily tested by varying the initial values of $A_y$ and $A_z$ (by varying the Doppler-cooling minimum temperature through laser detuning) and studying the heating rate of the x motion which had previously been cooled to the zero point of motion. No dependence on the initial values of $A_y$ and $A_z$ was found. In any case, if all modes of motion are initially cooled to the zero-point state this source of heating vanishes because the assumed coupling only provides an exchange of energy between modes. This places a premium on cooling all modes to as low an energy as possible. Finally, it appears that this single-ion example gives a worst case scenario since, for large numbers of ions, the force on the generalized coordinates (right hand side of Eq. (76)) requires a high-order field gradient to be nonzero. These gradients are highly suppressed in the typical case where ion-ion separation is much smaller than the distance between the ions and the trap electrodes.

4.1.9. Collisions with background gas

Although trapped ion/quantum logic experiments will typically be carried out in a high-vacuum environment (P < $10^{-8}$ Pa), residual background gas collisions can be important. The effects of collisions can be broken up into two classes: (i) inelastic collisions, which alter the internal state of the trapped ion or even change the species of the ion, and (ii) elastic collisions, which only add kinetic energy to the ion. Both types of collisions will cause decoherence, although heating from elastic collisions is expected to be to be the chief concern.

The most troublesome inelastic processes are chemical reactions and charge exchange. A background gas atom or molecule can collide and chemically react with the trapped ion, creating a different ionic species which is no longer useful. For the reactions to occur, they must be energetically favorable (exothermic), and, if the background neutral is a molecule, the reactions almost always proceed since the internal degrees of freedom of the molecule can help satisfy energy and momentum conservation in the reaction. In the ion trap experiments discussed here, the ion can spend an appreciable amount of time in the (optically) excited state due to laser excitation; in this case the extra energy due to laser excitation can make an otherwise endothermic reaction become exothermic. For example, in experiments on laser-cooled Hg$^+$ ions [235], when ions were excited to the metastable 5d$^9$6s$^2$ $^2$D$_{5/2}$ level (approximately 4.4 eV above the ground state) they reacted with neutral Hg atoms in the background gas to cause loss of the Hg$^+$ ions (presumably due to radiative association causing Hg$_2^+$ dimer formation). As a second example, in experiments on $^9$Be$^+$ ions [47, 76, 211], the ions were converted to BeH$^+$ upon collision with an H$_2$ molecule when resonant light was applied to the $^2$S$_{1/2}$ → $^2$P$_{1/2,3/2}$ transitions.

As a final example, the formation of YbH$^+$ by a similar process has been carefully studied by Sugiyama and Yoda [236]. The second form of inelastic collision is charge exchange, where a neutral background atom gives up an electron and neutralizes the trapped ion. Both types of inelastic collision depend critically on the particular constituents involved.

Chemical reactions and charge exchange can occur only if the interparticle spacing of the two colliding partners approaches atomic dimensions. An upper limit on these rates is given by the Langevin rate, for which background neutrals penetrate the angular momentum barrier and undergo a spiraling-type collision into the ion [237]. In these collisions, the electric field from the trapped ion polarizes the background neutral (polarizability $\alpha$), resulting in an attractive interaction potential $U(r) = -\alpha q^2/(8\pi\epsilon_o r^4)$. Impact parameters less than a critical value $b_{crit} = (\alpha q^2/\pi\epsilon_o \mu v^2)^{1/4}$ will result in spiraling collisions, where $\mu$ and $v$ are the reduced mass and relative velocity of the pair. (Since the ions are assumed to be nearly at rest from laser cooling, $v$ is simply given by the velocity of the background constituent.) The velocity-independent Langevin rate constant $k_{Langevin} \equiv \sigma v = \pi b_{crit}^2 v$ leads to an overall reaction rate

$$\gamma_{Langevin} = n k_{Langevin} = nq\sqrt{\frac{\pi\alpha}{\epsilon_o \mu}}, \qquad (80)$$

where n is the density of the background gas. In a metal/glass room-temperature apparatus such as was used in the NIST $^9$Be$^+$ experiments [47,76,211], the dominant background gas constituent is usually H$_2$. For H$_2$, we obtain $k_{Langevin} = 1.64 \times 10^{-9}$ cm$^3$·s$^{-1}$. At a pressure of $10^{-8}$ Pa and a temperature of 300 K, we have $\gamma_{Langevin} \simeq 0.004$ s$^{-1}$. Other candidate background molecules and atoms have similar polarizabilities and values of $k_{Langevin}$. Experimentally, we observe lifetimes of several hours for trapped $^9$Be$^+$ and $^{24}$Mg$^+$ ions at pressures of around $\approx 10^{-8}$ Pa indicating that at least for ground state $^9$Be$^+$ and $^{24}$Mg$^+$ ions the probability of chemical reactions with the background gas constituents is small. In a cryogenic ion trap for $^{199}$Hg$^+$, the lifetime is many days [74].

Background gas can heat the trapped ions by transferring energy during an elastic collision. The Langevin rate above gives too low an estimate of the rate of collisions which transfer energy to the ion, since a "heating" collision need not penetrate the angular momentum barrier. A conservative estimate for the heating rate can be given from the more general expression for the total collision cross section $\sigma_{elastic}$ (in the quasi-classical limit) in a $C_4/r^4$ potential [238]. We take $C_4 = \alpha q^2/8\pi\epsilon_o$ and find

$$\sigma_{elastic} = \pi \Gamma(1/3)\left[\frac{\alpha q^2}{16\epsilon_o \hbar v}\right]^{2/3}. \qquad (81)$$

If we average over a thermal distribution of background H$_2$ velocities, this results in the rate constant $k_{elastic} = <\sigma_{elastic}v> = 1.23 \times 10^5 \alpha^{2/3}(\tilde{v})^{1/3}$ where $\tilde{v} \equiv (2k_BT/\mu)^{1/2}$. For H$_2$ at 300 K and a pressure of $10^{-8}$ Pa, we find $k_{elastic} \simeq 1.24 \times 10^{-8}$ cm$^3$·s$^{-1}$ and $\gamma_{elastic} \simeq 0.03$ s$^{-1}$. Although each collision on average transfers a large amount of energy to the trapped ion, we conclude that at typical UHV pressures, such collisions will also be rare. Collisional heating can be tested by raising the background gas pressure. A simple way to do this is simply turn off the vacuum pump; collisions as a source of heating in the NIST experiments was eliminated in this way. However, doing so does not insure that all partial pressures increase by the same factor. Preferably, the partial pressure of selected gases should be increased by leaking them into the vacuum system and looking for an increase in ion heating.

When ions are first loaded into a trap, their kinetic energy is in general comparable to the depth of the trap (typically greater than 1 eV). In this regime, elastic collisions with the background gas are actually beneficial, as the background gas can provide a viscous damping medium and bring the temperature of the trapped ions into thermal equilibrium with the surrounding gas [239]. This allows initial laser-cooling to proceed much faster.

4.1.10. Experimental studies of heating

Some experimental diagnostics for heating have been discussed above in Secs. 4.1.4 - 4.1.9. As discussed in Sec. 4.1.2, we are primarily interested in determining the heating from the $|n=0\rangle$ or $|n=1\rangle$ states. In the NIST single $^9$Be$^+$ ion experiments, heating could be estimated by first preparing the ion in the $|\downarrow\rangle|0\rangle$ state, waiting a certain delay time, and then measuring the strength and ratio of the first red and blue sidebands [44,45]. This method gives a simple indication of the heating rate, but a more complete method is described here.

More recently, we have determined that depletion of the $|n=0\rangle$ state was dominated by a nearly continuous and smooth heating which initially took the atom from n=0 to n=1. This was established in experiments where the distribution of n level populations was measured after a delay time using the technique of Ref. [21] and the tomographic technique of Leibfried, *et al.* [131,132]. For example, in the experiments of Ref.. [21], many measurements were repeated for each value of the time delay in order to extract the population distributions from $P_\downarrow(t)$ (Eq. (42)). In this way the time $t^*$ (Eq. (64)) was determined to be about 1 ms. In this same apparatus, if we make the assumption that the heating was caused by a coupling to the environment at 300 K, the overall time constant to reach equilibrium is $\tau_{equil} \simeq t^*(k_B T/\hbar\omega_x) \simeq 570$ s.

Interestingly, under typical operating conditions, these tests of heating are highly insensitive to heating caused by collisions. In the previous section, we saw that at the typical background pressure of $10^{-8}$ Pa for the experiments of the NIST single $^9$Be$^+$ ion experiments, collisions would be expected to occur at a maximum rate of about 0.03 s$^{-1}$, as estimated from total elastic scattering. Strong heating collisions, given by the Langevin rate, occur at a rate of about 0.004 s$^{-1}$. These rates indicate a time constant to reach equilibrium with the 300 K background gas of between 30 s and 250 s, shorter than the 570 s time estimated from the observed n=0 → n=1 heating rates (previous paragraph). However, since each heating-rate experiment takes only about 1 ms, at most, only one experiment in about 30 000 indicates a background gas collision; these events are simply lost in the experimental noise. In effect, the technique of measuring $t^*$ is able to detect the continuous, smooth heating occurring in between collisions, even though collisions are expected to be a stronger overall heating source. This effect was even more pronounced in the experiments of Ref. [44].

Although the source of heating in the NIST experiments is not understood at the present time, it should be possible to tell if the heating is radiative or from some other cause by comparing the heating rates of the axial-COM and axial-stretch modes of two ions in the trap.

The heating of the COM mode should be nearly the same as for a single ion since it can be excited by a spatially uniform (oscillating) field. However, since the stretch mode will be excited only by a field gradient, radiative heating of this mode by fields emanating from the trap electrodes should be significantly less.

4.1.11. Experimental studies of motional decoherence

It is desirable to have some methods to test for decoherence. Full characterization of decoherence in a motional state could be accomplished by reconstruction of the density matrix [131,132] coupled with a time delay between creation and measurement. This complete characterization may not be necessary and other methods have been used. One possibility is to create an interference signal between two states that depends on decoherence mechanisms and monitor the contrast of that signal in time. This type of measurement was used to characterize the purity of Schrödinger-cat states for trapped atoms [47]. Decoherence of Schrödinger cat states of the electromagnetic field, caused by radiative damping, has been monitored by studying a correlation between observed states of two successive atoms which probe the field [207]. One kind of Schrödinger cat for motional states has the form $(|\alpha\rangle + |-\alpha\rangle)$ where $|\alpha\rangle$ denotes a coherent state. A measurement of the value of the Wigner function at the origin $W(\alpha = 0)$ may be sufficient to characterize decoherence [131,240]. If $\alpha$ is sufficiently big ($\langle 0|\alpha\rangle \approx 0$), we would expect an initial nonzero value for $W(0)$ that is damped toward zero in the course of decoherence [240,241]. For quantum logic with ions, it appears that decoherence of the COM motional states in the submanifold of states $|0\rangle$ and $|1\rangle$ is of primary importance. It may therefore be sufficient to reconstruct the 2x2 matrix spanned by the $|0\rangle$ and $|1\rangle$ number states of the COM mode and characterize its evolution due to decoherence. One scheme is outlined here.

Suppose we initially create the state $\Psi(t=0) = |\downarrow\rangle(C_{\downarrow,0}(0)|0\rangle + C_{\downarrow,1}(0)|1\rangle)$. In a particular experiment, after a time $t_d$ the state becomes $\Psi(t_d) = |\downarrow\rangle(\Sigma_{i=0}^{\infty} C_{\downarrow,i}(t_d)|i\rangle)$ due to motional decoherence. We can then apply two analysis pulses of radiation to this state. We first apply a $\pi$ pulse on the $|\downarrow\rangle|1\rangle \rightarrow |\uparrow\rangle|0\rangle$ red sideband transition. We assume this pulse takes time $t_1$ ($\Omega_{0,1}t_1 = \pi/2$). We follow this with a $\pi/2$ pulse on the $|\downarrow\rangle \rightarrow |\uparrow\rangle$ carrier transition. Here, we simplify the discussion by assuming the time for both of these pulses is short enough that decoherence during the pulses can be neglected. The wavefunction is now given by

$$\Psi_{final} = \frac{1}{\sqrt{2}}\left(|\downarrow\rangle\psi_\downarrow(x) + |\uparrow\rangle\psi_\uparrow(x)\right),$$

$$\psi_\downarrow(x) = [C_{\downarrow,0}(t_d) - e^{i(\phi_1-\phi_2)}C_{\downarrow,1}(t_d)]|0\rangle - i\sin\Omega_{1,2}t_1 e^{i(\phi_1-\phi_2)}C_{\downarrow,2}(t_d)|1\rangle \quad \textbf{(82)}$$

$$+ \sum_{i=2}^{\infty}\left[\cos\Omega_{i-1,i}t_1 C_{\downarrow,i}(t_d) - i\sin\Omega_{i,i+1}t_1 e^{i(\phi_1-\phi_2)}C_{\downarrow,i+1}(t_d)\right]|i\rangle,$$

where $\phi_1$ and $\phi_2$ are the phases (Eq. (23)) of the analysis pulses relative to the pulses used to create the state $\Psi(0)$. We now measure the probability $P_\downarrow$ of finding the atom in the $|\downarrow\rangle$ state and obtain

$$P_\downarrow(\Delta\phi) = \tfrac{1}{2}\left[|C_{\downarrow,0}(t_d)|^2 + |C_{\downarrow,1}(t_d)|^2 - 2Re[iC_{\downarrow,0}^*(t_d)C_{\downarrow,1}(t_d)e^{i\Delta\phi}] + \sum_{i=2}^{\infty}|C_{\downarrow,i}(t_d)|^2\right], \quad \textbf{(83)}$$

where $\Delta\phi \equiv \phi_1 - \phi_2$. In this expression, we have assumed all cross terms of the form $C_{\downarrow,i}^* C_{\downarrow,i+1}^*$ for $i \geq 1$ average to zero over many measurements; this will be true if decoherence is caused by some process which is uncorrelated with the creation and analysis pulses. We find $\langle Re(C_{\downarrow,0}^* C_{\downarrow,1})\rangle = \tfrac{1}{2}\langle P_\downarrow(\pi) - P_\downarrow(0)\rangle$ and $\langle Im(C_{\downarrow,0}^* C_{\downarrow,1})\rangle = \tfrac{1}{2}\langle P_\downarrow(\pi/2) - P_\downarrow(-\pi/2)\rangle$. Therefore, from measurements of $P_\downarrow$ for four values of $\Delta\phi$, we measure the coherence $(C_{\downarrow,0}^* C_{\downarrow,1})$ (or $\rho_{0,1}(t_d)$, the off-diagonal matrix element of the motion after time $t_d$). If necessary, we can find the amplitudes $|C_{\downarrow,0}(t_d)|$ and $|C_{\downarrow,1}(t_d)|$ by applying the blue sideband to $\Psi(t_d)$ for a time $\tau$ and reducing $P_\downarrow(\tau)$ as described in Sec. 3.2.1.

4.2. Internal state decoherence

For many years, one of the principal applications of the stored-ion technique has been for high-resolution, high-accuracy studies of internal state structure of atomic ions. This capability can be applied for use in atomic clocks (for reviews of efforts from a number of laboratories, see Refs. [242] and [243]). High resolution and accuracy are obtained because ions can typically be stored for long times (many hours or days) with minimal perturbations to their internal structure from electric and magnetic fields.

Energy level shifts caused by electric fields (Stark shifts) are usually small and, in many cases, magnetic field level shifts can be controlled well enough; these properties lead to very weak decoherence between internal state superpositions. As an example, in Fig. 7, we show a spectrum taken of a particular hyperfine transition in $^9Be^+$ ions [76]. This resonance, obtained with the Ramsey method of separated fields [173], has a linewidth of less than 0.001 Hz (at a frequency $\omega_o/2\pi \simeq 303$ MHz) and corresponds to a coherence time between the two internal levels of the transition of more than 10 minutes. (We could independently establish that the noise apparent in Fig. 7 was primarily caused by the fluctuations in the oscillator driving the transitions and not due to decoherence of the $^9Be^+$ internal levels.) More recently, a slightly narrower resonance was reported by Fisk, *et al.* [244]. This resonance was observed on a much higher frequency hyperfine transition in $^{171}Yb^+$ ($\simeq 12.6$ GHz) which resulted in a Q factor of $1.5 \times 10^{13}$ (Q $\equiv$ transition frequency/linewidth). For levels separated by optical energies, very long coherence times are also possible because of the very long radiative lifetimes of particular optical levels. So far however, observed coherence times have been limited by the linewidth of the probing lasers to a few tens of hertz [235,245,246,247].

These relatively small rates of decoherence indicate that quantum states considered for clock transitions are also attractive as qubit levels for a quantum computer. However, the long coherence times obtained in the spectroscopy experiments were obtained under special conditions which may not always be compatible with their use in quantum logic. Therefore, in this section, we consider various sources of internal state decoherence and how they might be controlled.

4.2.1. Radiative decay

A fundamental limit to internal state coherence is given by radiative decay. In general, we must consider decay from both levels of a two-level system; for simplicity, and as is often the case, we will assume the lower state is stable.

For electric dipole radiation, the decay rate from upper state $|2\rangle$ to lower state $|1\rangle$ is given by $\gamma_{rad} = \omega_o^3 |\langle 2|\vec{\mu}|1\rangle|^2/(3\pi\epsilon_o \hbar c^3)$ where $\omega_o \equiv (E_2 - E_1)/\hbar$ and $\vec{\mu}$ denotes the atomic dipole operator. For magnetic dipole radiation, we have $\gamma_{rad} = \omega_o^3 |\langle 2|\vec{\mu}|1\rangle|^2/(3\pi\epsilon_o \hbar c^5)$. For hyperfine transitions, which decay by magnetic dipole radiation, we make the approximation $|\langle 2|\vec{\mu}|1\rangle| \simeq \mu_B$, the Bohr magneton. If we assume $\omega_o/2\pi = 10$ GHz, then $\gamma_{rad} \simeq 10^{-12}$ s$^{-1}$, clearly small enough for the discussion here. For optical electric dipole transitions, we make the approximation $|\langle 2|\vec{\mu}|1\rangle| \simeq qa_o$ where $-q$ is the electron charge and $a_o$ is the Bohr radius. If we assume $\omega_o/2\pi = 10^{15}$ Hz ($\lambda = 300$ nm), we find $\gamma_{rad} \simeq 7.5 \times 10^7$ s$^{-1}$, which is much too fast for our purposes. However, as is well known, a number of trapped-ion species have first excited optical levels which are metastable. A well-studied case is Ba$^+$ for which the first excited $^2D_{5/2}$ state radiatively decays by electric quadrupole radiation. It has lifetime measured to be about 35 s [95,248,249,250]. Other ions also have metastable optical levels which are interesting for the purposes here. Some of these are considered by Hughes, *et al.* [63,64] and Steane [60]. Lifetime measurements of various ions have been compiled and reviewed by Church [251].

4.2.2. Magnetic field fluctuations

In the absence of purposely applied electromagnetic fields which provide the logic operations, uncontrolled fluctuating external magnetic fields are expected to give the primary contribution to internal state decoherence. Decoherence results from the fact that the energy separation between two levels of interest depends on the external magnetic field $\vec{B}$. In most cases, it will be possible to express the transition frequency between two levels to sufficient accuracy by

$$\omega_o + \delta\omega = \omega_o + \left[\frac{\partial\omega}{\partial B}\right]_{\omega_o} (B - B_o) + \frac{1}{2}\left[\frac{\partial^2\omega}{\partial B^2}\right]_{\omega_o} (B - B_o)^2, \tag{84}$$

where $B_o$ is the average magnetic field and $\omega_o$ is the transition frequency for $B_o = 0$.

The effects of a fluctuating magnetic field depend on the spectrum of the fluctuations. We first assume the typical case that the fluctuations are slow enough to be considered quasi-static during the time of a single operation $\tau_{op}$. A common source of low frequency fluctuations which would typically fit into this category are sinusoidal field fluctuations due to unbalanced currents in AC power lines. Therefore, in the spin-½ analog, we assume

$$H_{internal} = \hbar\omega_o S_z(1 + \beta(t)), \tag{85}$$

where $\omega_o\beta \ll (\tau_{op})^{-2}$. For this Hamiltonian, Schrödinger's equation yields

$$C_\downarrow(t - t_o) = C_\downarrow(t_o)e^{i\phi(t)}, \quad C_\uparrow(t - t_o) = C_\uparrow(t_o)e^{-i\phi(t)}, \quad \phi(t) \equiv \frac{1}{2}\left[\omega_o t + \int_0^t \beta(t')dt'\right]. \tag{86}$$

In a sequence of logic operations, $\phi(t)$ must be small enough or be taken into account (see also Sec. 4.3.2).

As a second case, we assume the magnetic field varies rapidly; specifically, we assume B varies sinusoidally: $B - B_o = \Delta B \cos\omega_B t$ where $\omega_B \gg 1/\tau_{op}$. We find

$$\delta\omega = \left[\frac{\partial\omega}{\partial B}\right]_{\omega_o} \Delta B \cos\omega_B t + \frac{1}{4}\left[\frac{\partial^2\omega}{\partial B^2}\right]_{\omega_o}(\Delta B)^2(1 + \cos 2\omega_B t). \tag{87}$$

The term $[\partial^2\omega/\partial B^2]\Delta B^2/2$ on the right side of this expression is a frequency shift which can be absorbed into the definition of $\omega_o$ if $\Delta B$ remains constant. For simplicity, assume that one of the cosine terms in Eq. (87) dominates so that $\delta\omega$ sinusoidally oscillates at $\omega_m = \omega_B$ or $2\omega_B$ so that $\beta(t) = \beta_o\cos\omega_m t$ in Eq. (85). The effects of this fast modulation can be seen if we consider applying external radiation near the carrier frequency $\omega_o$ of the internal state transition. Schrödinger's equation leads to expressions similar to those of Eqs. (17) which take the form (for $\Delta = \delta = 0$)

$$\dot{C}_{\uparrow,n} = -ie^{i(\phi - \eta_m\sin\omega_m t)}\Omega_{n,n}C_{\downarrow,n}, \quad \dot{C}_{\downarrow,n} = -ie^{-i(\phi - \eta_m\sin\omega_m t)}\Omega_{n,n}C_{\uparrow,n}, \tag{88}$$

where $\eta_m \equiv \omega_o\beta/\omega_m$. We have

$$e^{\pm i\eta_m\sin\omega_m t} = J_o(\eta_m) + 2[J_2(\eta_m)\cos 2\omega_m t + J_4(\eta_m)\cos 4\omega_m t + ...] \pm 2i[J_1(\eta_m)\sin\omega_m t + ...]. \tag{89}$$

where $J_i(X)$ is the ith Bessel function with argument X. For $\omega_m \gg 1/\tau_{op} \simeq \Omega_{n,n}$, the sinusoidally varying terms on the right average to zero (as in the rotating wave approximation). Therefore the wavefunction evolves just as in Eq. (25) except we must replace $|\Omega_{n,n}|$ by $|\Omega_{n,n}|J_o(\eta_m)$. Since the $J_o$ factor can be absorbed into the definition of $\Omega$, this steady modulation should not cause a

problem. An important source of high-frequency magnetic fields is from the currents in the electrodes which oscillate at the trap drive frequency $\Omega_T$.

The gate demonstrated in Ref. [17] had a strong sensitivity to magnetic field fluctuations since the qubit frequency and the auxiliary transition frequency had a dependence $\partial\omega/\partial B \simeq \mu_B/\hbar \simeq 10^{10}$ Hz/T. If this scheme is used in future experiments, sufficient magnetic shielding must be provided. Alternatively, the magnetic field dependence can be minimized by operating at a magnetic field where the transition frequency becomes independent of magnetic field to first order (making the $\partial\omega/\partial B$ terms in Eq. (84) and (87) vanish). For example, most atomic clocks are based on hyperfine transitions in ground electronic $^2S_{1/2}$ states, where the nuclear spin $\vec{I}$ is half integral (½, 3/2, 5/2, ...). In these cases, where the total angular momentum ($\vec{F} = \vec{I} + \vec{S}$) is integer, the (F,$m_F$ = 0) ↔ (F',$m_F$ = 0) transition is nearly first-order field independent for $|\vec{B}| \simeq 0$. These particular transitions may not be useful in the applications discussed here since other transition frequencies (for example, (F,$m_F$ = 0) ↔ (F',$m_F$ = ±1) transitions) will be very close to that of the (F,$m_F$ = 0) ↔ (F',$m_F$ = 0) transition frequency as $|\vec{B}| \to 0$. This may cause unwanted couplings to these other levels. At optical frequencies, similar first-order field independent transitions in ions have been used in frequency standards [235]. However, other hyperfine and optical transitions become first-order field independent at nonzero magnetic field, and their frequencies are separated from competing transitions (see, for example, Refs. [76] and [189]).

For quantum logic, storing qubit information in two states whose energy separation is first-order field independent therefore appears to be attractive. However, the conditional dynamics which is central to quantum logic may necessitate a transition to an auxiliary state. In the notation of Sec. 3.3, a transition between the $|\uparrow\rangle$ and $|aux\rangle$ levels will have a transition frequency which will, in general, not be first-order field independent if the $|\downarrow\rangle \to |\uparrow\rangle$ transition frequency is field-independent. Therefore, if the external field fluctuates, the fidelity of this operation will be compromised. As an example, suppose quantum logic is performed on a collection of $^9$Be$^+$ ions using the 2s $^2S_{1/2}$ $|m_I = -3/2, m_J = 1/2\rangle \equiv$ "$|\downarrow\rangle$" and $|-1/2, 1/2\rangle \equiv$ "$|\uparrow\rangle$" levels as qubits (strong-field state representation). With these designations, level $|\downarrow\rangle$ has a higher energy than level $|\uparrow\rangle$. These are the same levels as those used for the clock reported in Refs. [75] and [76]; the transition frequency $\omega_0/2\pi \simeq 303$ MHz is first-order field independent at a field of $B_0 \simeq 0.8194$ T. At this magnetic field, the transition frequency has a second-order dependence of $\delta\omega_0/2\pi \simeq -5.2(\delta B/B_0)^2$ MHz. Therefore, a field fluctuation of $\delta B = 10^{-4}$ T from $B_0$ leads to a frequency offset of the transition of only about 80 mHz. For the first CN gate described in Sec. 3.3, we could use $2\pi$ transitions between the $|\uparrow\rangle$ and "$|aux\rangle$" $\equiv |1/2, 1/2\rangle$ level. However, this transition has a first-order dependence on magnetic field at 0.8194 T given by $(\partial\omega/2\pi)/\partial B \simeq -22$ MHz/T. Therefore if the field shifts by $\delta B = 10^{-4}$ T, this will cause a shift of this transition frequency of $\Delta/2\pi = -2200$ Hz. Similar considerations must be applied to quantum logic using other ions such as $^{201}$Hg$^+$ ions where the splittings between hyperfine transitions are much higher [252]. Since it will be difficult to find two field-independent transitions in the same ion, it may be advantageous to use logic gates which require only one internal state transition as in the second and third gates described in Sec. 3.3.

4.2.3. Electric field fluctuations

In Sec. 4.4, we treat the effects of electric fields from the applied laser beams; in this section we treat the effects of other electric fields. These electric field shifts are likely to be less important as a source of internal state decoherence than shifts due to external magnetic fields. First, electric-field shifts are second-order in the field, except for exceptional cases involving nearly-degenerate states of opposite parity. Second, the fields are relatively small at the site of the ion. In analogy to Eq. (84), we have

$$\omega_0 + \delta\omega_E = \omega_0 + \frac{1}{2}\left[\frac{\partial^2\omega}{\partial E^2}\right]_{E=0} E^2, \qquad (90)$$

where E is the magnitude of the electric field. Unlike Eq. (84), the expansion is made around E = 0, rather than a nonzero value $E_0$, since, as noted previously, the mean value of E is zero for an ion in a trap. If it were not zero, then the ion would have a mean acceleration and would not be spatially localized. Of course, the ion must experience an average electric field to counteract the force of gravity, but this field is extremely small. For an ideal rf trap, such as one in which the rf potential is described by Eq. (1), and in which there are no stray, static electric fields, the ions are attracted to the line along which the rf electric field is zero. In this case, the mean-squared electric fields due to confinement experienced by cooled ions are very small. If stray fields are present, the mean-squared electric field seen by the ions will be nonzero and can be significantly larger [228]. The stray fields can be reduced by applying potentials to compensation electrodes; to some extent this is necessary anyway for efficient laser cooling. It should be possible to reduce the rms electric fields from the electrodes and neighboring ions to less than 1 V/cm. We note that rms electric fields of about 8 V/cm are present at room temperature, due to the blackbody (thermal) electromagnetic field. Since the spectrum of this field is dominated by infrared frequencies, and because it is steady, this should lead only to a small AC Stark shift [253].

We briefly consider two cases that have been discussed for ion trap quantum computation: narrow optical transitions between the ground and a metastable state and transitions between ground-state hyperfine sublevels. For an optical transition between two energy levels, the energy shift, derived from perturbation theory, is roughly $(qa_0)^2 E^2/\Delta E$, where $qa_0$ is the product of the proton charge and the Bohr radius, that is, a typical electric-dipole matrix element, $\Delta E$ is a typical energy difference between one of the two levels and another which is connected by the electric-dipole operator, and E is the electric field. For the case of the $Ba^+$ 6s $^2S_{1/2}$-to-5d $^2D_{5/2}$ transition, which has a wavelength of 1.76 μm, this expression predicts a shift $\delta\omega_E/2\pi$ of approximately 3 mHz for E=1 V/cm, if $\Delta E$ is taken to be the 6s $^2S_{1/2}$-to-6p $^2P_{1/2}$ energy difference. This is in rough agreement with the experiment of Yu, *et al.* [254]. Similar values of $\delta\omega_E$ are estimated for the $5d^{10}6s\ ^2S_{1/2} \rightarrow 5d^96s^2\ ^2D_{5/2}$ transition in $^{199}Hg^+$ [228]. Transition frequencies between hyperfine levels, such as those used in the NIST single $^9Be^+$ ion experiments, are even less sensitive to electric fields, since, to a good approximation, an electric field shifts all sublevels of the hyperfine multiplet by the same amount. A differential shift between the sublevels arises in third-order perturbation theory [253]. For $^9Be^+$, for example, the fractional shift between hyperfine sublevels is approximately $-4\times10^{-18}$ for E = 1 V/cm [255].

So far, we have considered only electric dipole shifts, that is, the shifts that are due to a uniform electric field. An electric quadrupole shift, which is proportional to the product of the electric quadrupole moment of the state and the applied electric field gradient, may also be present. The shift $\delta\omega_Q$ is given approximately by

$$\delta\omega_Q \approx \frac{qQ_a}{\hbar}\frac{\partial E_i}{\partial r_j}, \qquad (91)$$

where $Q_a$ is the atomic quadrupole moment and $\partial E_i/\partial r_j$ is a typical component of the electric field gradient tensor, such as $\partial E_x/\partial z$. The precise value depends on the whole electric field gradient tensor and the details of the atomic state. There is no quadrupole shift for S-states or for other states with electronic angular momentum J<1, like the $^3P_0$ states in the Group IIIA ions, such as In$^+$ [256]. The trap rf electric fields have gradients, but, since they oscillate sinusoidally, they average to zero. Static electric field gradients due, for example, to patch effects, or the static potential to confine ions along the axis of a linear trap (Eq. (2)) would be more of a problem. For a typical value of $Q_a = 10\ a_0^2$, and an electric field gradient of 10 V/cm$^2$, $\delta\omega_Q/2\pi \approx$ 0.7 Hz. Fluctuations in this (small) shift will probably be small compared to those of magnetic field shifts. Electric field gradients due to neighboring trapped ions will also be present [257]. Consider an ion at position x = y = z = 0. The electric field gradient component $\partial E_z/\partial z$ due to ions having charge q, located at x = y = 0, z = ±d is -q/($\pi\epsilon_0 d^3$). For d=10 μm, this is 575 V/cm$^2$. However, this should be relatively constant, since the spacing between ions is determined by the trap fields, which must be kept nearly constant if motional decoherence is to be avoided. Thus, they may lead to energy shifts, but not to decoherence.

Garg [58,59] has considered the decoherence of internal states of ions due to fluctuating electric fields induced by the reservoir of vibrational modes of a string of trapped ions. Such decoherence is present even if the modes are in the ground state, because of zero-point motion. This form of intrinsic decoherence was found to be negligible, at least for the specific case in which the 6s $^2S_{1/2}$ and the 5d $^2D_{5/2}$ states of Ba$^+$ are used to store the qubits. For a string of 1000 ions, the decoherence time was found to be around $10^4$ - $10^8$ times longer than the natural lifetime of the upper state, which is 35 s.

4.3. Logic operation fidelity and rotation angle errors

The third and perhaps most important class of decoherence involves imperfections in the logic operations. Ideally, a quantum computation transforms an initialized state vector $\Psi(0)$ to $U_{ideal}\Psi(0)$, where $U_{ideal} = \exp[-iH_{ideal}t/\hbar]$ (with $H_{ideal}$ time independent) is the unitary evolution operator without errors. In principle, both $U_{ideal}$ and the corresponding interaction Hamiltonian $H_{ideal}$ operate on the Hilbert space formed by the $|\downarrow\rangle$ and $|\uparrow\rangle$ states of the L qubits. In practice, for the case of trapped ions, the physical implementation of logic gates requires that we must include at least the $|0\rangle$ and $|1\rangle$ states of the chosen motional mode and perhaps an internal auxiliary state (Sec. 3.3). If decoherence mechanisms cause other states to be populated, the Hilbert space must be expanded. Although more streamlined algorithms may be available, the operator $U_{ideal}$ can always be broken up into fundamental single and two-bit quantum gates [140,155,156,157]; for simplicity, this will be assumed in what follows.

In the previous sections, we have concentrated on the decoherence of the state vector $\Psi$ due to environmental coupling, in the absence of purposely applied additional fields. Here, we treat decoherence due to imperfections in the evolution operator $U_{ideal}$ which is implemented by these additional applied fields. We break this decoherence into two categories: (i) errors or noise in the (classical) gate parameters (that is, the rotation angles $\phi$ and $\Omega_{n',n}t$ in Eq. (23)) which result in undesired evolution and, (ii) coupling to the environment induced by the gates (for example, spontaneous emission induced by the applied fields). In some cases, we can model noisy logic gates by considering an ensemble of gates whose parameters are statistically distributed. This allows a simple characterization of several candidate noise sources.

The final step in an ideal quantum computation is a probabilistic measurement of all the quantum bits, which yields a given L-bit number k, or equivalently, the state k, with probability $P_k = |\langle k|U_{ideal}\Psi(0)\rangle|^2$ ($k = 0,1,2,...,2^L-1$). A particular quantum computation is thus completely characterized by the vector of probability amplitudes $A_k = \langle k|U_{ideal}\Psi(0)\rangle$. The usefulness of a quantum computer algorithm relies on the coherent interference of the probability amplitudes, resulting in only a small fraction of the $2^L$ possible numbers with appreciable measurement probability. If the actual evolution operator U contains imperfections, the probability amplitude vector is instead $B_k = \langle k|U\Psi(0)\rangle \neq A_k$. We characterize the fidelity of a computation by the expression

$$F = \left\langle |\langle U_{ideal}\Psi(0)|U\Psi(0)\rangle|^2 \right\rangle = \left\langle |\sum_{k=0}^{2^L-1} A_k^* B_k |^2 \right\rangle, \tag{92}$$

where the average is performed over any random variables affecting the operator U. Roughly speaking, F is the probability that errors in the operator U do not affect the result of a computation, with F=1 representing error-free computation. This characterization of the fidelity follows the approach of quantum trajectories [258,259]. A quantum computation can be interpreted as the evolution of a complicated path through Hilbert space. If the operations comprising the computation are imperfect, the path of computation bifurcates into a superpositions of the correct path and incorrect path [260]. In this model, the fidelity F is just the probability that the correct path has been followed. Schemes for complete characterization of a quantum process in terms of a input-output matrix have been proposed in Refs. [261] and [262]. This approach would be able to characterize the fidelity of any of the operations discussed here. It can also be used to test for decoherence of the internal and motional states, by use of time delays between state preparation and application of the diagnostic.

An adequate characterization of the effects of noise in a quantum computation requires knowing the type of noise processes present, <u>and</u> application of this noise to a specific computational algorithm on a specific input state. This is most straighforwardly accomplished using numerical simulations [1,263,264,265,266]. Here, we do not treat any specific algorithms, but will attempt to make some general observations on particular sources of logic operation errors and how these errors accumulate after many operations.

4.3.1. Accumulated errors

Each successive operation in a quantum computation based on the Cirac/Zoller trapped-ion scheme can be regarded as a rotation between two states $|A\rangle_j|n\rangle$ and $|A'\rangle_j|n'\rangle$ in a multidimensional Hilbert space where $|A\rangle_j$ and $|A'\rangle_j$ are from $|\downarrow\rangle_j$, $|\uparrow\rangle_j$, or $|aux\rangle_j$ and $|n\rangle$ and $|n'\rangle$ are from $|0\rangle$ or $|1\rangle$ of the selected motional mode. Before the kth operation, assumed to be applied to the jth ion, we can, to a good approximation, write the wavefunction as

$$\Psi_{k-1} = \alpha_{k-1}|\psi_{k-1,\alpha}\rangle_j|\Psi_{k-1,\alpha}\rangle + \beta_{k-1}|\psi_{k-1,\beta}\rangle_j|\Psi_{k-1,\beta}\rangle + \gamma_{k-1}|\psi_{k-1,\gamma}\rangle_j|\Psi_{k-1,\gamma}\rangle, \quad (93)$$

where $|\psi_{k-1,\alpha}\rangle_j$ and $|\psi_{k-1,\beta}\rangle_j$ are the basis states for the rotation (for example, the $|\uparrow\rangle_j|1\rangle$ and $|aux\rangle_j|0\rangle$ states required to carry out step 1b of Sec. 3.3). The state $|\psi_{k-1,\gamma}\rangle_j$ is the part of the wavefunction for the jth ion which does not include states involved in the rotation (for example, the $|\downarrow\rangle_j$ state). The states $|\Psi_{k-1,\alpha}\rangle$, $|\Psi_{k-1,\beta}\rangle$ and $|\Psi_{k-1,\gamma}\rangle$ are states which include other ions and, because of gate imperfections, may involve states outside the Hilbert space spanned by the $|\downarrow\rangle$, $|\uparrow\rangle$, and $|aux\rangle$ internal states, and $|0\rangle$ and $|1\rangle$ motional states. We choose $\langle\Psi_{k-1,\alpha}|\Psi_{k-1,\alpha}\rangle = \langle\Psi_{k-1,\beta}|\Psi_{k-1,\beta}\rangle = \langle\Psi_{k-1,\gamma}|\Psi_{k-1,\gamma}\rangle = 1$ so that $|\alpha_{k-1}|^2 + |\beta_{k-1}|^2 + |\gamma_{k-1}|^2 = 1$. From Eq. (23), the "actual" kth operation can be written as

$$R_{ka}(\theta_k+\zeta_k,\phi_k) = \begin{bmatrix} \cos(\theta_k+\zeta_k) & -ie^{i\phi_k}\sin(\theta_k+\zeta_k) \\ -ie^{-i\phi_k}\sin(\theta_k+\zeta_k) & \cos(\theta_k+\zeta_k) \end{bmatrix}, \quad \psi = \begin{bmatrix} \alpha_{k-1} \\ \beta_{k-1} \end{bmatrix}, \quad (94)$$

where, to simplify notation, we have chosen $\theta_k = (\Omega_{n',n}t)_k$ and $\zeta_k, \phi_k \ll \pi$ are the errors in the amplitude and phase of the rotation. We can write

$$R_{ka} = R_k + \Delta R_k, \quad (95)$$

where $R_k = R_{ko}\cos\zeta_k$, $R_{ko} \equiv R_k(\theta_k, 0)$ is the ideal kth operator and

$$\Delta R_k \equiv -\begin{bmatrix} \sin\theta_k\sin\zeta_k & i\left[(e^{i\phi_k}-1)\sin\theta_k\cos\zeta_k + e^{i\phi_k}\cos\theta_k\sin\zeta_k\right] \\ i\left[(e^{-i\phi_k}-1)\sin\theta_k\cos\zeta_k + e^{-i\phi_k}\cos\theta_k\sin\zeta_k\right] & \sin\theta_k\sin\zeta_k \end{bmatrix}. \quad (96)$$

$\Delta R_k$ has terms which are first order (and real) in $\zeta_k$ and $\phi_k$. Nevertheless, we might expect that

the overall fidelity has errors which depend only quadratically on $\zeta_k$ and $\phi_k$ since a general computation may approximate a sequence of nearly random rotations on a multidimensional Hilbert space. To see how this might come about we express a general computation consisting of M operations as

$$O = R_{Ma}R_{(M-1)a}...R_{ka}...R_{2a}R_{1a} \simeq O_1 + O_2 + O_3, \tag{97}$$

where

$$O_1 \equiv R_M R_{M-1}...R_2 R_1,$$

$$O_2 \equiv [R_M...R_2 \Delta R_1] + [R_M...\Delta R_2 R_1]...$$
$$+ ...[R_M \Delta R_{M-1}...R_1] + [\Delta R_M R_{M-1}...R_1], \tag{98}$$

$$O_3 \equiv R_M...R_3 \Delta R_2 \Delta R_1 + R_M...\Delta R_3 R_2 \Delta R_1 ...$$
$$+ ...\Delta R_M R_{M-1} \Delta R_{M-2}...R_1 + \Delta R_M \Delta R_{M-1} R_{M-3}...R_1,$$

and where we have neglected terms which are higher than second order in $\zeta_k$ and $\phi_k$. In this notation, the ideal computation is given by the operator

$$O_{ideal} = R_{Mo}...R_{2o}R_{1o}. \tag{99}$$

For the fidelity, we need to calculate $\langle\Psi(0)|O^\dagger_{ideal}O|\Psi(0)\rangle = \langle\Psi(0)|O^\dagger_{ideal}(O_1 + O_2 + O_3)|\Psi(0)\rangle$ where we will neglect terms of third order and higher in $\{\Delta R_i\}$. For the first term, we find

$$\langle\Psi(0)|O^\dagger_{ideal}O_1|\Psi(0)\rangle \simeq 1 - \sum_{i=1}^{M} \frac{\zeta_i^2}{2}. \tag{100}$$

For the second term, we have a sum of M terms; the kth of which is given by

$$\langle \Psi(0)|O^\dagger_{ideal}R_M R_{M-1}...\Delta R_k...R_2 R_1|\Psi(0)\rangle$$

$$\simeq \langle \Psi(0)|R^\dagger_{1o}R^\dagger_{2o}...R^\dagger_{ko}\Delta R_k R_{k-1}...R_1|\Psi(0)\rangle \tag{101}$$

$$= \langle \Psi_{k-1,o}|R^\dagger_{ko}\Delta R_k|\Psi_{k-1,o}\rangle$$

where $\Psi_{k-1,o}$ corresponds to the wavefunction represented in Eq. (93) for the ideal case. We find

$$\langle \psi_{k-1,o}|R^\dagger_{ko}\Delta R_k|\psi_{k-1,o}\rangle \simeq -\frac{\phi_k^2}{2}\sin^2\theta_k[|\alpha_{k-1}|^2 + |\beta_{k-1}|^2]$$

$$-i[(|\alpha_{k-1}|^2 - |\beta_{k-1}|^2)(\zeta_k\cos\theta_k + \sin\theta_k)\phi_k\sin\theta_k$$

$$+ 2Re[\alpha^*_{k-1}\beta_{k-1}](\frac{\phi_k^2}{2}\sin\theta_k\cos\theta_k - \zeta_k) \tag{102}$$

$$- 2Im[\alpha^*_{k-1}\beta_{k-1}](\sin\theta_k + \zeta_k\cos\theta_k)\phi_k\cos\theta_k].$$

The third term $\langle \Psi(0)|O^\dagger_{ideal}O_3|\Psi(0)\rangle$ is straightforward to evaluate and is clearly second order in $\zeta_k$ and $\phi_k$. Even though the imaginary term in Eq. (102) has terms linear in $\zeta_k$ and $\phi_k$, when the fidelity F is calculated, these terms add only in quadrature so the kth operation gives only a contribution of second order in $\zeta_k$ and $\phi_k$. Therefore, in general, we have

$$F \simeq 1 - \sum_{k=0}^{M}\left[C_{k1}\zeta_k^2 + C_{k2}\zeta_k\phi_k + C_{k3}\phi_k^2\right], \tag{103}$$

where the coefficients $C_{k1}$, $C_{k2}$, $C_{k3} \lesssim 1$. For general computations involving many operations M in a large Hilbert space, it might be expected that the errors in rotations are distributed in an approximately random fashion. In this case, we would expect the fidelity to be approximately given by $F \simeq 1 - M(F_{\zeta\zeta}\zeta^2 + F_{\phi\phi}\phi^2 + F_{\zeta\phi}\zeta\phi)$ where $\zeta$ and $\phi$ are characteristic of the errors for all operations and $F_{\zeta\zeta}$, $F_{\phi\phi}$, $F_{\zeta\phi} \lesssim 1$. In the case amplitude errors ($\zeta$) dominate, then the maximum number of operations before the fidelity drops appreciably below 1 is $M_{max} \simeq 1/\zeta^2$. For example, laser intensity fluctuations on the order of 1 part in $10^4$ would give $M_{max} \simeq 10^8$. At the other extreme, the errors might add coherently in some cases. These operations could conspire to give contributions to the fidelity that goes as $(\Sigma_k^M \zeta_k)^2$. We can illustrate this with a simple example. Assume that an intended rotation $\theta_k$ has an error $\zeta_k \ll \pi$. The state of the system before application of this rotation can be expressed by Eq. (93). From Eq. (92) we find F =

$\cos^2(\zeta_k/2)(|\alpha_{k-1}|^2 + |\beta_{k-1}|^2) + 4[\text{Re}\{\alpha_{k-1}^*\beta_{k-1}\exp(i\phi_k)\}]^2 \sin^2(\zeta_k/2)$. The existence of the second term only serves to improve the fidelity, so at worst (when $\alpha_j = 1$ or $\beta_j = 1$), the fidelity of this single operation is given by $\cos^2(\zeta_k/2) \simeq 1 - (\zeta_k)^2/8$. Now suppose that a second rotation $\theta_m$ about the same axis has error $\zeta_m$. A similar analysis shows that when these two operations are applied in succession, a worst case for the fidelity is given by $F = \cos(\zeta_k/2 + \zeta_m/2) \simeq 1 - (\zeta_k + \zeta_m)^2/8$. Similarly, for rotations on M bits, the worst case fidelity is given by $F = \cos(\Sigma_i^M \zeta_i/2)$ so that the errors could accumulate linearly. Most likely, this case will correspond to rather uninteresting computations and one can generally expect that the errors will accumulate in a way which is between random walk and linear.

The effects of operation errors ($\zeta_i$, $\phi_i$) have been observed in simulations of quantum computations to factorize small numbers [263,266]. If the errors fluctuate randomly about the correct value, the fidelity is given by a simple expression which is essentially the same as Eq. (103) for $F \simeq 1$ [266]. However, for constant errors the fidelity indicated by the simulations is somewhat worse indicating that the computational algorithms cannot be assumed to be rotations about axes chosen pseudo randomly. Therefore, the simulations indicate that requirements on systematic offsets are more stringent than the requirements on random fluctuations from operation to operation.

4.3.2. Pulse area and phase fluctuations

As in the last section, we will consider two sources of noise: (1) those which result in offsets from, and fluctuations in, the ideal Rabi frequencies (for example, offsets and fluctuations of $\Omega_{n',n}t$ in Eq. (21)) and (2) those caused by phase fluctuations between applications of separate operations to a particular ion (fluctuations from some constant value of $\phi$ in Eq. (21)).

We first make some remarks regarding the coherent evolution between two states of the system which comprise the basis states for a single operation. First, there are a number of methods of coherently transferring population between a two level system with radiation, including constant amplitude Rabi pulses as represented by Eqs. (21) and (23), adiabatic rapid passage [267], and adiabatic transfer via dark states [268-270]. The advantage of the last two techniques is that for $\pi$ pulses, their efficiency does not depend critically upon the pulse area (Eq. (104) below) of the applied field. However, for rotations other than $\pi$, this advantage is lost. Moreover, for these techniques to work with high efficiency, particularly simple atomic level structures are required and coupling to any off-resonant levels can potentially contaminate the transfer. For example, adiabatic transfer can be extremely efficient between ground state hyperfine levels if the transfer occurs only through a single excited (electronic) state. However, in practice, other excited states are driven by the coupling (laser fields) thereby affecting the fidelity of the transfer [271]. Furthermore, the high powers required for adiabatic rapid passage and adiabatic transfer aggravate decoherence from spontaneous Raman scattering (Sec. 4.4.6.4). Therefore, we will only consider the case of Rabi pulses, where population is transferred by applying a resonant field with a well-controlled amplitude envelope and duration.

When a resonant radiation field is applied to a two-level system, the resultant unitary transformation is given by Eq. (23). We can further generalize this expression replacing $\Omega_{n',n}t$ by $\Theta_{n',n}(t)/2$ where

$$\Theta_{n',n}(t) = 2 \int_{t'=0}^{t} \Omega_{n',n}(t')\, dt' \qquad (\textit{"pulse area"}). \tag{104}$$

The factor of two in this equation is introduced so that the condition $\int \Omega_{n',n}(t')dt' = \pi/2$ corresponds to a $\pi$ pulse of the effective Bloch vector. Here, as opposed to the situation assumed in Sec. 2.3.1 (where the coupling radiation was assumed to be turned on instantaneously, left at constant amplitude, and then turned off instantaneously), we now allow the coupling radiation to be turned on and off gradually. This more closely approximates a real experimental situation and can be advantageous since the spectrum of $\Omega_{n',n}(t)$ can be tailored to suppress off-resonant transitions (Sec. 4.4.6). Here, we call $\Theta_{n',n}$ the pulse area of the applied radiation and $2\Omega_{n',n}(t)$ the Rabi frequency envelope. If the applied pulse area is noisy or set incorrectly, then the output state will fluctuate from it's ideal value as discussed in the previous section.

We must also consider detunings between the applied frequencies and resonant frequencies of the ions ($\Delta$ in Eq. (21)) and phase fluctuations between successive operations on the same ion (fluctuations in $\phi$ in Eq. (21) from operation to operation). This could occur from frequency or phase fluctuations in the applied fields or fluctuations in the ion internal or motional frequencies. For frequency fluctuations, we consider that for a given operation on ion j, $\Delta_j(t)$ is slowly varying over the time of a single operation (Sec. 4.2.2). In the limit that detunings are small, the dominant effect can be characterized by assuming that $\Delta = 0$ for each operation, but that different phases

$$\phi_k = \int_{t_o}^{t_k} \Delta_j(t')\, dt', \tag{105}$$

are present for successive applications of the same operation. In this expression, $t_o$ is the time that the first operation is applied to ion j and $t_k$ is the time of the kth operation (applied to the jth ion).

4.4. Sources of induced decoherence

Below, we consider various sources of fluctuations and decoherence caused by the logic operations and how these might be evaluated and controlled in experiments. First, we are interested in controlling the accuracy and stability of the Rabi pulse area $\Theta_{n',n}$ given by Eq. (104) and the accumulated phase as expressed in Eq. (105). Since gates will most likely be implemented with laser transitions we will examine laser intensity and timing fluctuations.

4.4.1. Applied field amplitude and timing fluctuations

Fluctuations in the laser intensity at the site of a given ion can arise from both fluctuations in the relative position of the beam with respect to the ion and fluctuations in laser power. Laser/ion position stability is particularly important since the Cirac/Zoller scheme of quantum logic assumes that ions in an array be selectively addressed, thereby requiring a high degree of control of the laser beam spatial profile (Sec. 4.4.4). Of course, the simplest method for minimizing position fluctuations is to employ mechanically stiff mounts for the optics and ion trap electrodes, and have the laser source as close as possible to the ions. A quadrant detector indexed to the trap electrodes and placed near the ion may also be used to actively stabilize the beam position by feeding back to a galvanometer or acousto-optic modulator. If optical fibers are used to deliver laser beams to the ions, position fluctuations between the fiber and the ions could be made small; however, we must also consider position fluctuations between the laser source and the input to the fiber. If the position tolerances can be adequately controlled, the dominant source of intensity noise at the ion would likely be given by fluctuations in optical power and laser mode. Here, we estimate limits on laser amplitude noise.

If we assume the laser fields responsible for quantum logic operations are coherent states, the fundamental noise floor is photon shot noise. For a laser beam of average power $P_0$, the fractional level of shot noise is

$$\frac{\delta P}{P_0} = \sqrt{\frac{\hbar\omega}{P_0 \tau_{op}}} \quad , \tag{106}$$

where $\omega$ is the (optical) photon frequency, $\tau_{op}$ is the time the radiation is applied, and, for simplicity, we assume square pulse envelopes as in Eq. (23)). Of course, lasers seldom produce amplitude noise at the photon shot-noise limit. Almost all laser sources have significant amplitude noise in the 10 Hz-10 kHz range due to acoustic vibrations which, for example, affect the laser cavity resonators. Much of this noise can be removed by active power stabilization, where a beamsplitter directs a portion of the laser power to a photodetector, and an error signal is derived and fed back to an upstream modulator or, in the case of a diode laser, directly to the amplitude of the laser source [272,273]. The limiting noise of this stabilization scheme is degraded slightly by the imperfect quantum efficiency of the photodetector as well as the beamsplitter. If the beamsplitter directs a fraction $\epsilon$ of the input optical power to the stabilizer (which then gives an optical power $P_u \simeq (1-\epsilon)P_0$ directed to the ion), and the quantum efficiency of the photo detector is $\eta_{det}$, the limit of fractional power noise in the logic pulse is (assuming no added electronic noise in the feedback loop)

$$\frac{\delta P_u}{P_u} \geq \sqrt{\frac{\hbar\omega}{P_u \tau_{op} \eta_{det} \epsilon(1-\epsilon)}} \quad . \tag{107}$$

For a laser wavelength of 313 nm, and assuming $\epsilon = 0.5$ and $\eta_{det} = 0.5$, we have $\delta P_u/P_u \geq 2.3\times 10^-$

$^9$ $(P_\mu \tau_{op})^{-\frac{1}{2}}$. For 1 W of usable laser power and $\tau_{op}$ = 1 μs, this corresponds to a fractional power fluctuation of $\geq 2.3 \times 10^{-6}$.

This estimate applies only to the laser power fluctuations at the beamsplitter and assumes no additional noise is introduced between the beamsplitter and the photo detector or the beam splitter and the ions. Typically, the usable part of the laser beam must be directed further through optics, the air, and a window to the vacuum envelope enclosing the ion trap. Fluctuating etalon effects in the optics and air currents may therefore seriously increase the power fluctuations beyond Eq. (107).

The effects of (Gaussian) noise on laser intensity have been treated by Schneider and Milburn [274]. These effects show up in a well-characterized way for transitions involving Fock states.

Fluctuations in timing errors may also be important. Clearly, fractional fluctuations in the duration of laser pulses will correspond directly to the same fractional fluctuations in the desired value of $\Theta_{n',n}$. If we require fractional fluctuations of $10^{-6}$ on $\Theta_{n',n}$, then we require timing precision of 1 ps on a 1 μs pulse. Similar considerations apply to the stability of pulse envelope shapes.

For both amplitude and timing fluctuations, it may be possible to sample a portion of the beam used for logic and apply it to a "check bit" ion. The response of this ion could then be used to monitor and control the amplitude and timing of the pulses.

4.4.2. Characterization of amplitude and timing fluctuations

In practice, it may be easiest to characterize the amplitude and timing fluctuations with a power detector placed as close as possible to the position of the ions. However, it would be useful to characterize these effects using the ions themselves. One possibility is to observe the sinusoidal Rabi oscillations predicted by Eq. (21). For example, for a single ion initially prepared in the $|\downarrow\rangle$ state, we can record the probability $P_\downarrow(\tau)$ of detecting the ion in the $|\downarrow\rangle$ state after the laser is applied to the ion for time $\tau$. From Eq. (23), if we first prepare the ion in the $|\downarrow,n\rangle$ state and if the power is constant, $P_\downarrow(\tau) = |C_{\downarrow,n}|^2 = \cos^2|\Omega_{n',n}t| = \frac{1}{2}[1 + \cos 2\Omega_{n',n}t]$. In Fig. 2, we show this type of curve for n=0 and n' =1. This plot includes a decay due to decoherence. Here, we consider decoherence caused by specific types of noise. We will cast the results in terms of fluctuations in $\Omega_{n',n}$, however, they may easily be converted to corresponding time fluctuations since we are really interested in fluctuations in the net rotation angle given by Eq. (104). For simplicity of notation, in this section, we write $\Omega_{n',n} = \Omega$.

First assume that Ω fluctuates slowly so that it is constant over the time $\tau$ for an individual measurement of $P_\downarrow(\tau)$ but fluctuates over the time taken to make an average measurement of $\langle P_\downarrow(\tau) \rangle$. We characterize power fluctuations by a spectrum $D(\Omega-\Omega_o)$ of Ω values around the desired value $\Omega_o$. We have

$$\langle P_\downarrow(\tau) \rangle = \frac{1}{2} \int_{-\infty}^{\infty} [1 + cos2\Omega\tau] D(\Omega - \Omega_o) d(\Omega - \Omega_o) . \qquad \textbf{(108)}$$

We illustrate with two possibilities for D, finding

$$\langle P_\downarrow(\tau)\rangle = \frac{1}{2}\left[1 + cos2\Omega\tau\, e^{-2(\Delta\Omega_{rms}\tau)^2}\right] \quad for \quad D(\Omega - \Omega_o) = \frac{1}{\sqrt{2\pi}\Delta\Omega_{rms}} e^{-\frac{1}{2}\left(\frac{\Omega - \Omega_o}{\Delta\Omega_{rms}}\right)^2},$$

$$\langle P_\downarrow(\tau)\rangle = \frac{1}{2}\left[1 + cos2\Omega\tau\, \frac{1}{1 + 2(\Delta\Omega_{rms}\tau)^2}\right] \quad for \quad D(\Omega - \Omega_o) = \frac{1}{\Delta\Omega_{rms}} e^{-\frac{\sqrt{2}|\Omega - \Omega_o|}{\Delta\Omega_{rms}}}$$

(109)

where $(\Delta\Omega_{rms})^2 \equiv \langle(\Omega - \Omega_o)^2\rangle$. In both cases the contrast of the signal drops about a factor of two for $\tau \simeq 1/\Delta\Omega_{rms}$. Although the second choice of $D(\Omega - \Omega_o)$ seems less physical, it indicates the exact spectrum of the noise may not be important for a qualitative test. If we need to detect fluctuations of order $\Delta\Omega_{rms}/\Omega$, then we must measure $P_\downarrow$ after approximately $\Omega\tau/2\pi \simeq (2\pi\Delta\Omega_{rms}/\Omega)^{-1}$ Rabi cycles. For example, if we require sensitivity $\Delta\Omega_{rms}/\Omega \simeq 10^{-4}$, we need to make measurements of $P_\downarrow(\tau)$ after about 1500 Rabi cycles.

If, instead, the intensity fluctuates slowly compared to $\tau_{op}$, but fast compared to the time $\tau$, $\langle P_\downarrow(\tau)\rangle$ is a less sensitive test of the fluctuations since their effects tend to average out. In order to show the effects of high-frequency fluctuations of $\Omega$ for a simple case, we consider a sinusoidal time variation:

$$\Omega(t) = \Omega_0 + \Delta\Omega\sin(\omega_{amp}t + \varphi),$$

(110)

where $\omega_{amp}$ is the modulation frequency of the Rabi amplitude $\Omega(t)$. This analysis should, at least, yield the dependence on the amplitude and frequency of the fluctuations. The measured quantity $P_\downarrow(\tau)$ is given by

$$P_\downarrow(\tau) = \frac{1}{2}[1 + \cos\theta(\tau)],$$

(111)

where the rotation angle $\theta(\tau)$ is given by Eq. (104) which, here, takes the form

$$\frac{\theta(\tau)}{2} = \int_0^\tau \Omega(t)dt = \Omega_0\tau + \Delta\Omega\int_0^\tau \sin(\omega_{amp}t + \varphi)dt$$
$$= \Omega_0\tau + \frac{\Delta\Omega}{\omega_{amp}}[\cos\varphi(1 - \cos\omega_{amp}\tau) + \sin\varphi\sin\omega_{amp}\tau].$$

(112)

This yields

$$P_\downarrow(\tau) = \frac{1}{2}\left[1 + \cos\left(2\Omega_0\tau + \frac{2\Delta\Omega}{\omega_{amp}}[\cos\varphi(1-\cos\omega_{amp}\tau) + \sin\varphi\sin\omega_{amp}\tau]\right)\right]$$

$$= \frac{1}{2}\left[1 + \cos(2\Omega_0\tau)\cos\left(\frac{2\Delta\Omega}{\omega_{amp}}[\cos\varphi(1-\cos\omega_{amp}\tau) + \sin\varphi\sin\omega_{amp}\tau]\right)\right. \quad \textbf{(113)}$$

$$\left. - \sin(2\Omega_0\tau)\sin\left(\frac{2\Delta\Omega}{\omega_{amp}}[\cos\varphi(1-\cos\omega_{amp}\tau) + \sin\varphi\sin\omega_{amp}\tau]\right)\right].$$

Since $P_\downarrow(\tau)$ is obtained by repeating the measurements, and since the phase $\varphi$ will be random from one measurement to the next, we average over $\varphi$:

$$\langle P_\downarrow(\tau)\rangle_\varphi = \frac{1}{2} + \frac{\cos(2\Omega_0\tau)}{4\pi}\int_0^{2\pi}\cos\left(\frac{2\Delta\Omega}{\omega_{amp}}[\cos\varphi(1-\cos\omega_{amp}\tau) + \sin\varphi\sin\omega_{amp}\tau]\right)d\varphi$$

$$- \frac{\sin(2\Omega_0\tau)}{4\pi}\int_0^{2\pi}\sin\left(\frac{2\Delta\Omega}{\omega_{amp}}[\cos\varphi(1-\cos\omega_{amp}\tau) + \sin\varphi\sin\omega_{amp}\tau]\right)d\varphi. \quad \textbf{(114)}$$

If we assume that $|\Delta\Omega/\omega_{amp}| \ll 1$, which is valid for a small, high-frequency, modulation, then the integrands can be expanded to second order in $(\Delta\Omega/\omega_{amp})$:

$$\langle P_\downarrow(\tau)\rangle_\varphi \approx \frac{1}{2} + \frac{\cos(2\Omega_0\tau)}{4\pi}\int_0^{2\pi}\left[1 - \frac{1}{2}\left(\frac{2\Delta\Omega}{\omega_{amp}}\right)^2[\cos\varphi(1-\cos\omega_{amp}\tau) + \sin\varphi\sin\omega_{amp}\tau]^2\right]d\varphi$$

$$- \frac{\sin(2\Omega_0\tau)}{4\pi}\int_0^{2\pi}\left(\frac{2\Delta\Omega}{\omega_{amp}}\right)[\cos\varphi(1-\cos\omega_{amp}\tau) + \sin\varphi\sin\omega_{amp}\tau]\,d\varphi \quad \textbf{(115)}$$

$$= \frac{1}{2} + \frac{\cos(2\Omega_0\tau)}{2}\left[1 - 2\left(\frac{\Delta\Omega}{\omega_{amp}}\right)^2(1-\cos\omega_{amp}\tau)\right].$$

Aside from the fast oscillations due to the $\cos\omega_{amp}\tau$ term, which might be difficult to observe, the effect of the high-frequency fluctuations of $\Omega$ is to reduce the visibility of the signal by $2(\Delta\Omega/\omega_{amp})^2$.

4.4.3. Applied field frequency and phase fluctuations

A simple source of fluctuations is due to frequency or phase fluctuations in the radiation that is used to carry out the basic operations. Schneider and Milburn [274] have considered a specific model of phase fluctuations for ion experiments and calculate the corresponding decoherence for operations of the type used in quantum logic. If qubit energy levels are separated by optical energies, the lasers that drive qubit transitions must have the required frequency and phase stability. Given the performance of current stable lasers [235,245,246,247], this may be a problem for long computations. As discussed in Sec. 2.3.3, optical stimulated-Raman transitions provide the required strong field gradients whereas the overall frequency sensitivity in the transitions is dominated by the difference frequency of the lasers, not the frequency of each laser. Since the two Raman beams can be derived from one laser beam with the use of frequency modulators, the frequency fluctuations are dominated by those of the oscillator which drives the modulator. The phase stability of these sources can be high and does not appear to pose problems. For example, the data shown in Fig. 7 can be regarded as one test of the phase sensitivity of the oscillator used to drive the indicated internal state transition. Stimulated-Raman transitions have the disadvantage that they give rise to AC Stark frequency shifts as indicated by the $|g_{1,2}|^2/\Delta_R$ terms on the right side of Eqs. (40). If these shifts are equal for both qubit levels, the net shift is zero. If they are different, the effective qubit frequency is shifted during the operation. This must be measured and accounted for. Moreover, the overall shift can be tuned to zero by appropriately adjusting the relative intensities of the Raman beams. A problem still exists if the Stark shifts fluctuate; we examine the consequences of those fluctuations here.

From Eqs. (40), the AC Stark frequency shift of the qubit transition due to stimulated-Raman transitions is

$$\omega_S = \omega_{S2} - \omega_{S1} = -(|g_2|^2 - |g_1|^2)/\Delta_R, \qquad (116)$$

so that the shifted qubit frequency is given by $\omega'_o = \omega_o + \omega_s$. We first consider the effects of frequency fluctuations of the Raman beams. If the two beams are derived from the same beam with the use of a modulator, the frequency shift of the two laser beams will be the same. Therefore, to a good approximation, we find $\delta\omega'_o \simeq -(\delta\Delta_R/\Delta_R)\omega_s$ where $\delta\Delta_R$ represents the frequency shift of both laser beams. Typically, $\Delta_R > 10$ GHz, and frequency fluctuations of lasers can be controlled to less than 1 kHz, so this source of error should be small. Moreover, it is essentially absent if we tune $\omega_s$ to be zero.

More serious are fluctuations in laser intensity. We will characterize these fluctuations by the corresponding fluctuations in $g_i$ ($\propto$ square root of laser intensity). From Eq. (116), we have

$$\delta\omega_S = 2\omega_{S2}\frac{\delta g_2}{g_2} - 2\omega_{S1}\frac{\delta g_1}{g_1}. \qquad (117)$$

As a worst case, we will assume fluctuations in $g_1$ and $g_2$ are random and uncorrelated and, for simplicity, we assume $\langle(\delta g_1/g_1)^2\rangle \simeq \langle(\delta g_2/g_2)^2\rangle \equiv \xi^2$, where $\langle\rangle$ denotes an average over the spectrum of the fluctuations. Therefore

$$\langle(\delta\omega_S)^2\rangle = 4\xi^2(\omega_{S1}^2 + \omega_{S2}^2). \tag{118}$$

As discussed in Sec. 4.3.2, we assume the fluctuations are slow compared to $\tau_{op}$, where $\tau_{op}$ is the time of the operation. In a particular operation, a frequency offset $\delta\omega_S$ gives rise to a phase fluctuation $\delta\phi_S = \delta\omega_S\tau_{op}$. It is useful to characterize the effects of these phase fluctuations by comparing their size relative to fluctuations in rotation angles caused by the corresponding fluctuations in the Rabi rates $\Omega_{n',n}$. With the same assumptions regarding $\delta g_1$ and $\delta g_2$, we have $\langle(\delta\Omega_{n',n}/\Omega_{n',n})^2\rangle = 2\xi^2$. Therefore, fluctuations in the overall rotation angle $\Theta_{n',n} = 2\Omega_{n',n}\tau_{op}$ (Eq. (104)) are given by $\langle(\delta\Theta_{n',n}/\Theta_{n',n})^2\rangle = 2\xi^2$. From these expressions we find

$$\frac{\langle(\delta\phi_S)^2\rangle}{\langle(\delta\Theta_{n',n})^2\rangle} = \frac{(\omega_{S1}^2 + \omega_{S2}^2)}{2\Omega_{n',n}^2}. \tag{119}$$

For quantum logic, a worst case appears to be for sideband excitation $n' = n \pm 1$. If we assume the Lamb-Dicke limit, we find

$$\frac{\langle(\delta\phi_S)^2\rangle}{\langle(\delta\Theta_{n',n})^2\rangle} = \frac{(g_1^4 + g_2^4)}{2\eta^2 g_1^2 g_2^2}. \tag{120}$$

From this expression, it is desirable to have $g_1 \simeq g_2$ to minimize the effects of $\delta\phi_s$ relative to $\delta\Theta_{n',n}$. Even in this case, rms phase fluctuations caused by Stark shifts are worse than those caused by Rabi frequency fluctuations by a factor of $\simeq 1/\eta$. Most likely, however, fluctuations in intensity will be dominated by fluctuations from the primary laser from which both beams are generated. Therefore we expect fluctuations to be correlated $\delta g_1/g_1 = \delta g_2/g_2$. When, in addition, $\omega_{S1} \simeq \omega_{S2}$, the phase fluctuations caused by Stark shifts will be less than those caused by Rabi frequency fluctuations. If these conditions hold true, it appears that stimulated Raman transitions between two qubit levels separated by fairly low (for example, hyperfine) frequencies are superior to single-photon transitions between qubit levels separated by optical frequencies. This situation might change as laser oscillators become more stable.

In the above discussion, we have neglected the alteration of the ion motional frequencies caused by the superimposed dipole force potentials of the focussed laser beams. However, if we assume the Raman beams have waists $w_o$ which are approximately equal and that the magnitudes of the projections of their $\vec{k}$ vectors along the z axis are the same, it is straightforward to show that the relative shift of (single trapped ion) secular frequencies for states $|\downarrow\rangle$ and $|\uparrow\rangle$ is approximately equal to $\delta\omega_S(z_o/w_o)^2$. Therefore, since $z_o \ll w_o$, the frequency shifting effects will be dominated by the shifts of the internal states.

Phase fluctuations of the laser beams themselves will also directly affect the fidelity of the operations. In the stimulated-Raman case, where both beams are derived from the same beam, phase fluctuations between Raman beams will be very nearly canceled. However, another source of laser phase fluctuations will come from path length fluctuations between the (laser) source and the trapped ions. Path length fluctuations are expected to be dominated by mechanical vibrations; these vibrations are typically restricted to low frequencies (< 1 kHz). They could result from a number of causes such as fluctuating mirror mounts or trap mounting hardware. For single photon laser transitions, the overall path length between the laser and ions is important; for stimulated-Raman transitions, the primary problem will be caused by path length differences between the two Raman beams after the frequency modulator. Therefore, for stimulated-Raman transitions using overlapping, copropagating beams (which can be used to drive carrier n' = n transitions) the paths are the same and there should not be a problem. However, for stimulated Raman sideband transitions, we require $\vec{k}_1$ and $\vec{k}_2$ along different paths, so that the path length problem is analogous to the problem for single photon transitions. For brevity, we will treat the problem of single-photon transitions; other cases (including the stimulated-Raman case) are easily generalized from this.

We assume the laser electric field at the exit of the laser oscillates as $\cos(\omega_L t)$. At the position of the ion, the field is given by the same expression with the time replaced by the retarded time $t - d_L/c$ where $d_L$ is the distance between the laser and ion. Therefore, the overall phase difference between the field at the laser and ion is $\omega_L d_L/c$. We will assume the fluctuations in $d_L$ are slow enough that over the time of a single operation the phase can be considered constant. If $d_L$ fluctuates by an amount $\Delta d_L$, the phase of the field fluctuates by $\Delta d_L \omega_L/c = 2\pi \Delta d_L/\lambda$ where $\lambda$ is the laser wavelength. In the NIST single $^9Be^+$ ion experiments, all of the operations have typically taken less than 1 ms; therefore vibrations have not been a problem. However, in longer computations, the requirements on $\Delta d_L$ are stringent. Some form of active stabilization, such as the method described by Bergquist, *et al.* [235], will probably have to be used. That technique can be viewed as a Doppler shift cancellation scheme and was based on a Doppler shift cancellation scheme used in spacecraft tracking [275]. To illustrate what is, in principle, possible, we note that optical cavities can be made to track the frequency of lasers to precision much smaller than 1 Hz [235,245,246,247]. For a laser frequency of $5 \times 10^{14}$ Hz ($\lambda$ = 600 nm), and a cavity composed of two mirrors separated by a distance of 50 cm, this corresponds to holding relative positions of the mirror to within $10^{-13}$ cm. Therefore, although they add additional complications to the experiments, such schemes for length stabilization can be used to effectively null the effects of path length variations.

4.4.4. Individual ion addressing and applied field position sensitivity

The scheme of Cirac and Zoller for trapped ion quantum logic requires that ions (along the axis of a linear trap) be addressable individually with laser beams for logic operations. This may be difficult, because the high vibrational frequencies desired for efficient laser cooling and suppression of decoherence also results in closely spaced ions. As discussed in section II.A., the minimum separation of adjacent ions in a linear trap is between the center ions and is approximately $s_{min} \simeq 2sL^{-0.56}$, with $s = (q^2/4\pi\epsilon_o m\omega_z^2)^{1/3}$ where L is the number of ions. For $^9Be^+$ ions with an axial COM frequency of $\omega_z/2\pi = 1$ MHz, this separation is about 10 μm for 2 ions, and 4 μm for 10 ions.

The most straightforward method for individual optical addressing is to tightly focus laser beams on the selected ion [1]. The transverse intensity distribution of a Gaussian optical beam of power P is

$$I(r) = \frac{2P}{\pi w_o^2} \exp\left(-\frac{2r^2}{w_o^2}\right), \qquad (121)$$

where $w_0 \simeq \lambda/(\pi \cdot NA)$ is the beam waist, $\lambda$ the radiation wavelength, and $NA = \tan\theta$ is the numerical aperture of the beam with cone half-angle $\theta$ (the formula for $w_0$ in the paraxial ray approximation is valid only for NA < 1) [276]. For large numerical apertures (NA ≈ 0.5), it appears that laser beams can thus be focussed down to a spot on the order of a wavelength, but this is difficult to realize in the laboratory. If we can realize $w_o = 5$ μm in a Gaussian beam, at a distance 10 μm from the center of the beam, this would imply a relative intensity of about $3\times10^{-4}$ or a electric field amplitude (proportional to Rabi frequency) of 1.8 % relative to the center of the beam - clearly a problem. If $w_o = 2$ μm could be obtained, the intensity (electric field) would be down by a factor of $1.3\times10^{-14}$ ($1.1\times10^{-7}$). However, these estimates are too optimistic because they assume the laser beam is normal to the axis of the linear trap; to address the axial modes by the methods described in Secs. 2 and 3, we need a component of the laser beam $\vec{k}$ vector along the axis. Therefore, if the angle of the laser beam relative to the trap axis is $\phi_k$, we must replace r in Eq. (121) by $s_{min}\sin\phi_k$. In addition, imperfections in the surfaces of the intervening vacuum port window, multiple reflections from these windows, and diffraction typically distribute laser intensity outside of the theoretical waist of the beam. The degree to which this occurs depends on the details of window surfaces, etc. and must be resolved experimentally.

Alternatively, we could accomplish all of the operations described in Secs. 2 and 3 by using the *transverse* gradient of the field associated with a focussed beam [125]. Therefore to accomplish sideband transtitions, we would displace the laser beam laterally, with respect to its direction, from the position of the ion. This would give rise to a coupling to the ion's motion which is in the direction of the transverse intensity gradient of the laser beam. In this case, we can make the direction of the beam normal to the trap axis. This method has the disadvantage that the field gradient would be reduced relative to the case treated above thereby leading to smaller Lamb-Dicke parameters and correspondingly reduced values of $\Omega_{n',n}$ for n' ≠ n.

The transverse intensity gradients of focused laser beams can also cause significant intensity fluctuations at the selected ion if the relative position of the beam with respect to the ion is not stable on the time scale of the computation. An alternative to using tightly focused Gaussian laser beams is to first feed the (expanded) laser beam through a sharply defined aperture, and use a lens to image the aperture at the position of the ions. With this technique, the beam intensity can be distributed more smoothly around the selected ion and have very steep intensity edges (on the order of the original aperture sharpness) away from the ion, thus suppressing beam vibration problems and confining the radiation to a single ion. This technique has been used to make relatively "hard" walls for an optical dipole trap [277]. For this technique to work well, the imaging lens must collect a large fraction of the light transmitted through the aperture or else diffraction effects will result in light intensity outside the image of the aperture. To address individual ions, we require very small aperture images, which gives rise to a design tradeoff. If a one-to-one relay lens is used to image a small object aperture, effects of diffraction are enhanced. If a demagnifying lens is used to reduce a large object aperture, then the aperture must be placed a large distance from the lens, requiring a relatively large lens. For two ions, imaging a sharp edge such as a razor blade at the space between the ions may be sufficient. We might also consider having every other ion in a string be a "garbage" ion which is not used in the computation, thereby increasing the spacing between qubit ions by a factor of two (or more, if more garbage ions are used between each qubit ion). This has the disadvantage that total number of ions (and spectator modes) increases, aggravating the problems associated with large quantum registers. If sufficiently good addressing on one ion in a string can be accomplished, it may be simpler to adiabatically shift the position of the ions, rather than the laser beams, in order to address different ions. This could be accomplished by applying different static potentials $U_o(t)$ and $U_o'(t)$ to the end segments of the rods in Fig. 1. However, changes in $U_o(t)$ and $U_o'(t)$ would have to be coordinated to keep the COM axial frequency constant or else additional phase shifts would be introduced. Stimulated-Raman transitions have the advantage that $\Delta \vec{k}$ can be made parallel to the axis of the trap even though each beam is at an angle with respect to the trap axis to facilitate selection of a particular ion. This is important, since coupling to transverse modes is eliminated.

    Another method of optically addressing individual ions is to cause a destructive optical interference at the position of a specific ion, with a net coupling at the other ion(s). For instance, if ion j is positioned at the node of a resonant standing wave laser field, the coupling between states $|\downarrow\rangle_j|n_k\rangle$ and $|\uparrow\rangle_j|n_k'\rangle$ is proportional to $\langle n_k'|\sin[\eta_k^j(a_k+a_k^\dagger)]|n_k\rangle$. In this case, the coupling of the standing wave to ion j vanishes when the laser frequency is tuned to an even order sideband such as the carrier ($n_k'=n_k$) (also see Sec. 4.4.6.2). If, instead, the ion is positioned at an antinode, the coupling is proportional to $\langle n_k'|\cos[\eta_k^j(a_k+a_k^\dagger)]|n_k\rangle$; thus, the coupling vanishes when the laser frequency is tuned to an odd order sideband, such as the first blue or red sideband ($n_k' = n_k \pm 1$). By appropriately choosing the angles of focussed laser beams relative to the trap axis or the spacing between ions, it should be possible to position an antinode (node) at ion j while positioning nodes (antinodes) at the ions adjacent to ion j (for equally spaced ions). In the case of two-photon stimulated Raman transitions, we desire to place ion j at common nodes or antinodes of two standing waves. Although this interference technique should allow individual access to each of two trapped ions, it appears technically difficult to extend this technique to more than three ions.

    Finally, we consider the application of external field gradients which shift the internal energy levels of ions depending on their position. For a magnetic field gradient to give this selectivity, we require the Zeeman splitting between adjacent ions to be much larger than the Rabi frequency, or $\Delta\mu(\partial|\vec{B}|/\partial z)s/\hbar \gg \Omega$, where $\Delta\mu$ is the difference in $\langle\vec{\mu}\cdot\vec{B}\rangle/|\vec{B}|$ between the two levels of interest, and s is the ion separation along the z direction. For $\Delta\mu = \mu_B$, s = 10 μm, and a Rabi frequency of $\Omega/2\pi \approx 1$ MHz, this requires $\partial|\vec{B}|/\partial z \gg 0.1$ T/cm. Field gradients of this

magnitude can be achieved; however, they would introduce large, and not-easily-controlled phase shifts for the other ions in a quantum register.

The laser beam itself can provide ion selectivity by employing the transverse gradient in the optical field intensity. For instance, if we desire to perform a $\theta$-pulse on ion j without affecting neighboring ion k, the intensity profile of the laser beam can be set so that the ratio of field strengths (intensities for the case of two-photon stimulated-Raman transitions) at ion j vs. ion k is $\theta/2\pi m$, where m is an integer. Now if the pulse duration is set so that ion j is rotated by $\theta$, ion k receives a rotation of $2\pi m$ and hence returns back to its initial state (with an extra phase factor of $(-1)^m$).

For the case of two-photon stimulated-Raman transitions, the laser beam can provide ion frequency selectivity by employing the Stark shift and the transverse gradient of the optical field. Here, for example, we could assume that the two counterpropagating Raman beams of equal intensities and spatial profiles are offset so that beam 1 is centered on ion j, and beam 2 is centered on adjacent ion k as depicted in Fig. 8. We will assume the coupling scheme of Fig. 3. Let $\epsilon$ be the fraction of peak intensity seen by the offset ions (that is, the intensity of beam 2 at ion j and beam 1 at ion k). Assume that when both beams are positioned on either ion, $g_1 = g_2 = g$. When the beams are offset, the two-photon resonant Rabi frequency at each ion is $\Omega = \epsilon^{1/2}(g^2/\Delta_R)$, where $g^2/\Delta_R$ is the Rabi frequency expected if both beams were centered on a given ion. The Stark shifts of the two ions are in opposite directions: $\delta_j = +\delta_o$, $\delta_k = -\delta_o$, where $\delta_o = \Omega(1-\epsilon)/\epsilon^{1/2}$. If we make $\delta_o \gg \Omega$ ($\epsilon \ll 1$), then by appropriately tuning the difference frequency of the laser beams, we can selectively drive transitions on either ion j or k. Alternatively, if, for example, we desire to perform a $\theta$-pulse to ion j without affecting ion k in an "unrepairable" way, $\epsilon$ can be adjusted to a particular value which results in ion k making the transformation $\psi_k = \alpha|\downarrow\rangle_k + \beta|\uparrow\rangle_k \rightarrow \psi_k' = e^{i\zeta}(\alpha|\downarrow\rangle_k + e^{i\xi}\beta|\uparrow\rangle_k)$. For example, if we use Rabi pulses which have a square pulse shape, Eq. (21) applies. For ion j, the laser must be tuned to the shifted frequency $\omega_{jo}' = \omega_{jo} + \delta_o$; Eq. (21) applies with $\Delta = 0$ and $|\Omega_{n',n}|t = \theta/2$ where t is the pulse time. For ion k, $\omega_{ko}' = \omega_{ko} - \delta_o$; Eq. (21) applies with $\Delta = 2\delta_o$. To achieve the desired form of $\psi_k'$, we want $\theta/2(1+\delta^2/\Omega^2)^{1/2} = m\pi$, or $\epsilon^2 - [1+(2n\pi/\theta)^2]\epsilon + 1 = 0$, where m is an integer. For m=1 and $\theta = \pi$ (a $\pi$-pulse on ion j), we find $\epsilon = 0.208$, and $\xi = 1.74\pi$. The phase shift $\xi$ on ion k and the corresponding phase shift on ion j must be kept track of in subsequent operations on these ions. Generalizing this to more than two ions becomes difficult if the laser beams also overlap other qubit ions. This scheme places an additional premium on laser power stability, since the light shifts are bigger than the Rabi frequencies by $1/\epsilon^{1/2}$ for $\epsilon \ll 1$. In addition, in both of the above schemes, employing the laser beams to differentially affect neighboring ions, one major drawback is that the positions and profiles of the laser beams must be accurately controlled.

Many of the above individual addressing schemes are improved greatly when dealing with only two ions instead of a string of many. This leads us to seriously consider systems where quantum logic operations are performed on accumulators consisting of only two ions, with the other ions located somewhere else (Sec. 5.1). A proposal has also been made for transferring quantum information from one register to another by optical means [151,278].

4.4.5. Effects of ion motion (Debye-Waller factors)

The Rabi frequency $\Omega_{n',n}$ describes the transition rate between states $|\downarrow\rangle|n\rangle$ and $|\uparrow\rangle|n'\rangle$. To realize the conditional dynamics desired for quantum logic, we want $\Omega_{n',n}$ to depend on the motional quantum numbers n and n' of a particular vibrational mode and be independent of the state of motion of other modes. In addition, for simple rotations on internal states, we want $\Omega_{n',n}$ to be independent of the motional state for all modes. It is not possible, in practice, to rigorously satisfy both of these requirements. For instance, the conventional controlled-not gate employs two carrier pulses (steps (1a) and (1c) in Sec. 3.3) which are intended to not depend on the state of motion; this requires the Lamb-Dicke parameter $\eta$ to be small (see Eq. (56)). In the Raman configuration, $\eta$ is proportional to the difference in two wavevectors and can be made negligible by using co-propagating beams ($\Delta\vec{k} \simeq 0$). On the other hand, with single-photon optical transitions, the Rabi frequencies depend on the motion of all modes which have a component of motion along the direction of $\vec{k}$. As discussed in Sec. 2.3, we can take advantage of the motional dependence of the carrier to construct a logic gate, but in this case also, the Rabi frequency will depend on the motion in the other modes along the direction of $\vec{k}$ or $\Delta\vec{k}$. Similarly, for sideband operations, such as step (1b) in Sec. 3.3, it will, in general, be impossible to have $\Omega_{n',n}$ depend on only one mode of motion. In this section, we examine the influence of extraneous modes on the Rabi frequencies $\Omega_{n',n}$.

In a collection of L ions, there will be motion in the 3L-1 spectator modes of vibration. The motion in these other modes reduces the Rabi frequency in much the same way as lattice vibrations affect a single emitter or scatterer embedded in a crystal, as described by the Debye-Waller effect [279,280]. From Eq. (36), we have the Rabi frequency for the jth ion (assuming all modes are in specific Fock states)

$$\Omega^j_{n_{k'},n_k} = \Omega^{j(o)}_{n_{k'},n_k} \left| \langle \{n_{p\neq k}\} | \prod_{p\neq k} e^{i\eta^j_p(a_p+a_p^\dagger)} | \{n_{p\neq k}\} \rangle \right| \\
= \Omega^{j(o)}_{n_{k'},n_k} \prod_{p\neq k} e^{-\frac{1}{2}(\eta^j_p)^2} L_{n_p}((\eta^j_p)^2) , \quad (122)$$

where $\Omega^{j(o)}_{n_{k'},n_k} \equiv \Omega |\langle n_k'|\exp(i\eta^j_k(a_k+a_k^\dagger))|n_k\rangle|$ is the Rabi frequency of the kth mode (selected for logic), in the absence of the 3L-1 spectator modes labeled by index p and vibrational number $n_p$. We assume the vibrational states of spectator modes are independently thermally distributed with mean vibrational number $\bar{n}_p$:

$$P_{n_p} = \frac{\bar{n}_p^{n_p}}{(1+\bar{n}_p)^{n_p+1}} . \quad (123)$$

If we average over this distribution, the mean and mean-squared Rabi frequencies are given by

$$\overline{\Omega_{n_k',n_k}^j} = \Omega_{n_k',n_k}^{j\,(0)} \prod_{p \neq k} \left[ \sum_{n_p=1}^{\infty} P_{n_p} e^{-\frac{1}{2}(\eta_p^j)^2} L_{n_p}\!\left((\eta_p^j)^2\right) \right] = \Omega_{n_k',n_k}^{j\,(0)} \prod_{p \neq k} e^{-(\eta_p^j)^2 (\bar{n}_p + \frac{1}{2})}$$

$$\overline{\left(\Omega_{n_k',n_k}^j\right)^2} = \left(\Omega_{n_k',n_k}^{j\,(0)}\right)^2 \prod_{p \neq k} \left[ \sum_{n_p=1}^{\infty} P_{n_p} e^{-(\eta_p^j)^2} L_{n_p}\!\left((\eta_p^j)^2\right)^2 \right] = \left(\Omega_{n_k',n_k}^{j\,(0)}\right)^2 \prod_{p \neq k} e^{-2\eta_p^2 (\bar{n}_p + \frac{1}{2})} I_0\!\left(2(\eta_p^j)^2 \sqrt{\bar{n}(\bar{n}+1)}\right)$$

(124)

where we have used the Laguerre polynomial sum identities

$$\sum_{n=0}^{\infty} z^n L_n(x) = \frac{e^{-\frac{zx}{1-z}}}{1-z} \quad , \quad \sum_{n=0}^{\infty} z^n L_n(x)^2 = \frac{e^{-\frac{2zx}{1-z}}}{1-z} I_0\!\left(\frac{2x\sqrt{z}}{1-z}\right) \quad (125)$$

In these expressions, $|z| < 1$ and $I_0(\epsilon)$ is the zeroth modified Bessel function with argument $\epsilon$.

The first line in Eq. (124) shows the exponential reduction of the Rabi frequency due to motion in modes p (Debye-Waller factor). For each mode p, we can write $(\eta_p^j)^2(\bar{n}_p + \frac{1}{2}) = \langle \frac{1}{2} k_{eff}^2 (x_p^j)^2 \rangle$ where $\vec{k}_{eff}$ is the effective k vector of the radiation and $x_p^j$ is the amplitude of the component of motion parallel to $\vec{k}_{eff}$ for the jth ion. For L large, the product term on the right side of the first expression is approximately equal to $\exp(-\frac{1}{2} k_{eff}^2 (x_{rms}^j)^2)$ where $x_{rms}^j$ is the total rms amplitude of motion parallel to $\vec{k}_{eff}$. This is the Debye-Waller reduction factor due to the thermal energy of the jth ion.

The second line in Eq. (124) allows us to determine the fractional fluctuations in the Rabi frequency from experiment to experiment

$$\frac{\Delta\Omega_{n_k',n_k}^{j\,(rms)}}{\overline{\Omega_{n_k',n_k}^j}} \equiv \frac{\sqrt{\overline{\left(\Omega_{n_k',n_k}^j\right)^2} - \left(\overline{\Omega_{n_k',n_k}^j}\right)^2}}{\overline{\Omega_{n_k',n_k}^j}} = \sqrt{\left[\prod_{p \neq k} I_0\!\left(2(\eta_p^j)^2 \sqrt{\bar{n}_p(\bar{n}_p+1)}\right)\right] - 1} \approx \sqrt{\sum_{p \neq k} (\eta_p^j)^4 \bar{n}_p(\bar{n}_p+1)}. \quad (126)$$

In this last approximation, the Bessel function is expanded to lowest order $[I_0(\epsilon) = 1 + \epsilon^2/4 + \cdots]$ which is appropriate if the 3L-1 arguments of the Bessel function in Eq. (126) are all small compared to 1. This is expected to be the case if all modes are cooled to the Lamb-Dicke limit $[(\eta_p^j)^2 \bar{n}_p \ll 1]$. Eq. (126) describes the fractional rms fluctuations in the Rabi frequency due to thermal motion in the spectator modes of vibration.

Typically, the thermal motion is determined by initial conditions and the reservoir is "turned off" once the experiment begins. This is expected to be the case for imperfect initial laser-cooling, resulting in a probability distribution of stable Rabi frequencies with the above mean and rms values. In this case, the Rabi frequencies for each ion maintain a constant value during a single run of the experiment, but deviate from the mean as indicated in the previous equation. Stating the results in another way, if the total number of modes is large, then the distribution of Rabi frequencies is nearly Gaussian, and the probability that a given run of the experiment results in a Rabi frequency which is fractionally within $\epsilon$ of the mean Rabi frequency $\overline{\Omega}^j_{n_{k'},n_k}$ is

$$Pr\left(\left|\frac{\Omega^j_{n_{k'},n_k} - \overline{\Omega}^j_{n_{k'},n_k}}{\overline{\Omega}^j_{n_{k'},n_k}}\right| < \epsilon\right) \approx erf\left(\frac{\epsilon}{\sqrt{2\sum_{p \neq k}(\eta^j_p)^4 \bar{n}_p(\bar{n}_p + 1)}}\right). \tag{127}$$

For example, if each of 100 spectator modes is laser-cooled to $\bar{n}_p = 0.1$ and has Lamb-Dicke parameter $\eta^j_p = 0.01$, then the probability that the Rabi frequency is within $10^{-4}$ of the mean value is approximately 0.23.

To see how Eq. (126) scales with the number of ions, assume that $\vec{k}_{eff}$ is aligned with the axis of a linear trap, where the axial motion is described by L modes. If the frequencies and amplitudes of all modes contributing to the axial motion of ion j are assumed to be about the same, we can write $\eta^j_p \simeq \eta_1/L^{1/2}$ and $\bar{n}^j_p \simeq \bar{n}$ where $\eta_1$ and $\bar{n}$ are the Lamb-Dicke parameter and mean occupation for the axial motion of a single (thermalized) trapped ion. In this case, Eq. (126) becomes

$$\frac{\Delta\Omega^{j,rms}_{n_{k'},n_k}}{\overline{\Omega}^j_{n_{k'},n_k}} \simeq \eta_1^2 \left[\frac{\bar{n}(\bar{n}+1)}{L}\right]^{1/2}. \tag{128}$$

This expression indicates that a large number of ions is beneficial because it tends to average out the effects of motion in the (L - 1) extraneous modes. Eq. (128) is an overestimate of the fluctuations since the L-1 extraneous modes will have higher frequency than the COM mode, leading to smaller amplitudes of motion than assumed in this crude estimate.

The Rabi frequency fluctuations discussed here translate to fractional phase offsets ($\zeta_k/\Theta_k$ of Eq. (94)) for each gate operation by the same expressions given in Eqs. (124) and (126). In practice, the mean Rabi frequency $\overline{\Omega}_{n_{k'},n_k}$ can be measured by averaging over many experiments. If we assume the COM mode is used for logic, then $\eta_k = \eta_1/L^{1/2}$. Therefore, we cannot choose $\eta_1$ to be too small, or else the entangling operations for quantum logic become too slow (Eq. (24)). To suppress the effects of Debye-Waller factor fluctuations, it is therefore desirable to cool all modes (whose motion is parallel to $\vec{k}_{eff}$) to the zero-point state.

4.4.6. Coupling to spectator levels

We have assumed that each time an external field is applied to form part of a logic operation, only two quantum states take part in the interaction. This assumption is valid when any other state is far from resonance. To explain this in a simple example, we refer to Fig. 9. Suppose we want to drive a transition between states $|\downarrow\rangle$ to $|\uparrow\rangle$ with radiation resonant with this transition. If level $|\downarrow\rangle$ is coupled to level $|s\rangle$ with this radiation, the wavefunction will have a small admixture of state $|s\rangle$ after the operation; the amplitude of state $|s\rangle$ will become larger as $\Delta$ becomes smaller. After a sequence of many operations, the amplitudes of these "spectator" states can build up and cause errors in the computation.

4.4.6.1 Polarization discrimination of internal states

In an array of trapped ions, two internal states of each ion comprise a qubit of information. In addition, a third auxiliary internal level $|aux\rangle$ may be required (transiently) for the operation of a CN gate as described in Sec. 3.3. This state might be a particular Zeeman sublevel of a hyperfine multiplet [17] or an optical metastable state [1]. By employing suitable polarizations of the driving field, particular internal state transitions can be selected with high discrimination. Furthermore, the Zeeman sublevels can be spectrally resolved (assuming the Zeeman splitting is much larger than the Rabi frequency), by applying a magnetic field. This combination will help to isolate the two internal levels of interest from other internal levels. Here, we give an example of how polarization selection can be used to suppress coupling to internal "spectator" states.

We consider the case of Ref. [17], which uses two 2s $^2S_{1/2}$ ground state $|F,m_F\rangle$ hyperfine levels of $^9Be^+$ ions as qubit levels. These levels are designated $|\downarrow\rangle \equiv |2,2\rangle$ and $|\uparrow\rangle \equiv |1,1\rangle$ (Fig. 5). We also make the identification $|aux\rangle \equiv |2,0\rangle$. They are driven using stimulated-Raman transitions by detuning the lasers from the $^2P_{1/2}$ state. By choosing the Raman beam coupling between the $^2P_{1/2}$ and $|\downarrow\rangle$ states to be $\sigma^-$ polarized and the coupling between the $^2P_{1/2}$ and $|\uparrow\rangle$ states to be $\pi$ polarized, the $|\downarrow\rangle$ and $|\uparrow\rangle$ states form a closed family; that is, neither the $|\downarrow\rangle$ or $|\uparrow\rangle$ state can be driven to other Zeeman sublevels. In this case, the nearest off-resonant transitions to consider are detuned by at least the hyperfine frequency ($\simeq$ 1.25 GHz in $^9Be^+$). Similarly, by choosing the beam coupling between the $^2P_{1/2}$ and $|aux\rangle$ states to be $\sigma^+$ polarized (and the coupling between the $^2P_{1/2}$ and $|\uparrow\rangle$ states to be $\pi$ polarized), the $|aux\rangle$ level is coupled only to the $|\uparrow\rangle$ level.

4.4.6.2 Spectral discrimination of states

Spectator states can include both motional and internal states. For example, a single CN gate uses pulses which drive on the carrier as well as first motional sidebands of the COM mode. Discrimination between these two types of transitions cannot be made with polarization selection since the relative strengths of the matrix elements are independent of polarization. Since, in the Lamb-Dicke limit, the carrier operation will have a higher resonant Rabi frequency than the sideband operations, the largest source of contamination may be due to an off-resonant excitation of the carrier during the sideband operations. The basic problem can be illustrated by considering the simple example illustrated in Fig. 9. Here, we assume decay from the excited states is negligible, $\Gamma_s = \Gamma_\uparrow = 0$. We assume we want to carry out an operation which coherently drives the $|\downarrow\rangle \leftrightarrow |\uparrow\rangle$ transition (resonance frequency $\omega_o$) and avoid the presence of any amplitude in the "spectator" state $|s\rangle$ after the operation. We assume radiation of frequency $\omega \simeq \omega_o$ is applied which couples levels $|\downarrow\rangle$ and $|\uparrow\rangle$ with strength $\Omega$. We assume this radiation also couples levels $|\downarrow\rangle$ and $|s\rangle$ with (resonant) strength $\Omega'$, but, for simplicity, does not couple levels $|\uparrow\rangle$ and $|s\rangle$. In the context of quantum logic, $\Omega$ might correspond to the Rabi frequency for a sideband transition and Ω' the Rabi frequency for the carrier transition. If we take the zero of energy in Fig. 9 to be the energy of the $|\downarrow\rangle$ level, we can write the Hamiltonian as $H = H_o + H_I$, where

$$H_o = \hbar\omega_o |\uparrow\rangle\langle\uparrow| + \hbar(\omega_o + \Delta)|s\rangle\langle s|,$$

$$H_I = 2\hbar\cos\omega t(\Omega[|\downarrow\rangle\langle\uparrow| + |\uparrow\rangle\langle\downarrow|] + \Omega'[|\downarrow\rangle\langle s| + |s\rangle\langle\downarrow|]), \qquad (129)$$

where the expression for $H_I$ is equivalent to that in Eq. (14). In the Schrödinger picture, we write $\Psi = C_\downarrow|\downarrow\rangle + C_\uparrow \exp(-i\omega_o t)|\uparrow\rangle + C_s \exp(-i[\omega_o + \Delta]t)|s\rangle$. This leads to equations for the amplitudes

$$\dot{C}_\downarrow = -i\Omega e^{i\delta t} C_\uparrow - i\Omega' e^{i(\delta - \Delta)t} C_s,$$

$$\dot{C}_\uparrow = -i\Omega e^{-i\delta t} C_\downarrow, \qquad (130)$$

$$\dot{C}_s = -i\Omega' e^{-i(\delta - \Delta)t} C_\downarrow,$$

where $\delta \equiv \omega - \omega_o$ and we have neglected rapidly varying terms proportional to $\exp(\pm i(\omega + \omega_o)t)$ (the usual rotating wave approximation). If we make the substitution $C_s = C_s' e^{i\Delta t}$, the equation for $\dot{C}_s$ can be written

$$\dot{C}_s' + i\Delta C_s' = -i\Omega' e^{-i\delta t} C_\downarrow.$$

for $|\Omega'|$ small enough relative to $|\Delta|$, we can "adiabatically eliminate" level s, by assuming $|dC_s'/dt| \ll |\Delta C_s'|$. In this case, Eqs. (130) become

$$\dot{C}_\downarrow \simeq -i\Omega e^{i\delta t} C_\uparrow + i\frac{(\Omega')^2}{\Delta} C_\downarrow, \qquad \dot{C}_\uparrow = -i\Omega e^{-i\delta t} C_\downarrow, \qquad C_s' \simeq -\frac{\Omega'}{\Delta} e^{-i\delta t} C_\downarrow. \qquad (132)$$

The Stark shift term $\delta_s \equiv (\Omega')^2/\Delta$ is a downward energy shift of the $|\downarrow\rangle$ state. It can be suppressed in the previous equation by including it into a shift of $\omega_o$. That is, by making the substitution $C_\downarrow = C_\downarrow' \exp(i\delta_s t)$ and choosing $\omega - \omega_o = \delta_s$, we find

$$\dot{C}_\downarrow' = -i\Omega C_\uparrow, \quad \dot{C}_\uparrow = -i\Omega C_\downarrow', \quad C_s' = -\frac{\Omega'}{\Delta} C_\downarrow'. \qquad (133)$$

In this case, $C_\downarrow'$ and $C_\uparrow$ are given by Eqs. (17), leading to a the desired evolution between the $|\downarrow\rangle$ and $|\uparrow\rangle$ states. Moreover, if the couplings $\Omega$ (and $\Omega'$) are turned on and off slowly compared to $1/\Delta$ (the condition of adiabaticity), then the amplitude in state $|s\rangle$ adiabatically grows during the operation and reduces to zero upon completion of the operation, since it is proportional to $\Omega'$.

The general problem, however, is that we desire to make all transitions as fast as possible. This means that we cannot satisfy the adiabatic condition; therefore $C_s'$ will not be simply proportional to $\Omega'$ and will have some nonzero value after $\Omega'$ is reduced to zero. For example, if $\Omega'$ is turned on adiabatically, but turned off diabatically, $C_s$ will have a final amplitude with magnitude $|\Omega' C_\downarrow'/\Delta|$. Estimates of these effects in specific contexts are given by Poyatos, *et al.* [261] and James [61].

Since it should be possible to separate internal states by relatively large frequencies and further discriminate from internal spectator states with polarization, the most important task will be to suppress transitions to motional spectator levels. The fundamental problem is that in order to discriminate between carrier and sideband transitions, the ion must be able to tell that it is moving. If the time of the operation $\tau_{op}$ is small compared to the inverse of the frequency of the motional mode, this is impossible. Therefore, the fundamental limit on $\tau_{op}$ is given approximately by $1/\omega_{COM}$. To maximize computational speed, we desire to approach this condition as closely as possible. This can be aided by using Rabi envelopes which can suppress certain spectral components [281,282]. Since the general problem will involve more than one spectator level (more motional states), a more general approach might be to apply optimal control theory [283-286].

4.4.6.3 Tailoring of laser fields

As discussed in the last section, the speed of an ion-trap quantum computer will be fundamentally limited by excitation of motional sidebands other that the desired one. To see this in a specific case, the conditional dynamics for the first CN gate described in Sec. 3.3 occurs in step (1b). That step uses a $2\pi$ pulse on 1st blue sideband of the $|\uparrow\rangle|1\rangle \leftrightarrow |aux\rangle|0\rangle$ transition. To make this step fast, we would like to increase the field intensity and decrease the interaction time accordingly. However, as the pulse becomes shorter, the spectrum of the pulse becomes wider. Eventually, the spectral width of this pulse is larger than $\omega_z$ so that we drive the relatively strong and unwanted $|\uparrow\rangle|1\rangle \leftrightarrow |aux\rangle|1\rangle$ carrier transition. In general, we also drive $|\uparrow\rangle|1\rangle \leftrightarrow |aux\rangle|n'\rangle$ transitions, where n' > 1. Most of these unwanted transitions can be suppressed by making all terms in the expansion in of Eq. (28) equal to zero except the term proportional to $\partial B/\partial z$ (or $\partial E/\partial z$ for electric dipole transitions). In this case, we suppress coupling to all unwanted $|aux\rangle$ levels except the $|aux\rangle|2\rangle$ state whose resonant frequency is detuned from the $|\uparrow\rangle|1\rangle \leftrightarrow |aux\rangle|0\rangle$ transition by $2\omega_z$.

Suppression of even or odd order terms in Eq. (28) can be accomplished by appropriate positioning of standing waves relative to the ion [40,61,287,288]. To illustrate with an example, we first examine the transition $|\downarrow\rangle|0\rangle \leftrightarrow |\uparrow\rangle|1\rangle$. For simplicity, we assume single photon electric-dipole (rather than stimulated-Raman) transitions and consider only motion along the z direction. We want to suppress all terms in the expansion of the field (re., Eq. (28)) except the $\partial E/\partial z$ term. We need to synthesize, using M separate standing wave laser beams, a wave which has the form

$$E = E_o \left[ \sum_{m=1}^{M} C_m \sin(k_m z) \right] \cos(\omega t), \tag{134}$$

where $k_m \equiv \vec{k}_m \cdot \hat{z}$, and where $z = 0$ corresponds to the equilibrium position of the ion in question. The case of $M = 1$ has already been suggested in various contexts to suppress coupling to the carrier ($\Delta n = 0$) transition (see Ref. [40] and Sec. 4.4.4 above). For $M > 1$, we want to satisfy

$$C_1 k_1^n + C_2 k_2^n + ... + C_M k_M^n = 0, \quad n = 3, 5, ... 2M - 1. \tag{135}$$

In this case, the next sideband spectator level which has nonzero coupling is one where $|n' - n| = 2M + 1$. This transition is detuned by $2M\omega_z$ from the transition of interest and the coupling is suppressed by approximately $\eta^{2M} = (k_1 z_o)^{2M}$ compared to $\Omega_{1,0}$ or $\Omega_{0,1}$.

Second, we consider carrier transitions $|\downarrow\rangle|n\rangle \leftrightarrow |\uparrow\rangle|n\rangle$, where $n = 0,1$. With the stimulated-Raman technique on hyperfine transitions, transitions of the type $|\downarrow\rangle|n\rangle \leftrightarrow |\uparrow\rangle|n'\rangle$, $n \neq n'$, are highly suppressed by the use of copropagating beams. For single photon transitions, we need to synthesize, using separate laser beams, a wave which has the form

$$E = E_o \left[ \sum_{m=1}^{M} C_m \cos(k_m z) \right] \cos(\omega t). \tag{136}$$

For $M > 1$, we want

$$C_1 k_1^n + C_2 k_2^n + ... + C_M k_M^n = 0, \quad n = 2, 4, ... 2(M - 1). \tag{137}$$

In this case, the next spectator level which has nonzero coupling is one where $|n' - n| = 2M$. This transition is detuned by $2M\omega_z$ from the carrier and the coupling is suppressed by approximately $\eta^{2M} = (k_1 z_o)^{2M}$ compared to $\Omega_{1,1}$ or $\Omega_{0,0}$. For the carrier transitions, we do not have the problem with the remaining spectator level as noted for the first sideband case above. On the other hand, for logic operations which use the carrier transition (Sec. 2.3 and Ref. [17]), we do not want $\eta$ to be too small, or else the gates take to long to implement. Therefore the suppression of the higher-order sidebands may not be as great as in the case of logic using first sidebands.

For qubits coupled by single-photon transitions, it may be difficult to suppress couplings to motional modes in the x and y directions. This will be true if we carry out logic on a string of ions in a linear trap and use the axial COM mode. The laser beams must be at an angle with respect to the z axis to be able to select particular ions. This kind of problem can be eliminated by using the x or y COM mode for logic and making $\vec{k}$ orthogonal to the other directions. Finally, we remark that the use of standing waves will give rise to optical dipole potentials which can shift the ion motional frequencies. These shifts should be small, and they can be incorporated into the definitions of the motional oscillation frequencies.

4.4.6.4 Spontaneous emission

The internal atomic states of trapped ions, which store quantum bits of information, must be protected from spontaneous emission, at least for the duration of the computation. This excludes the possibility of "error correction," (See Sec. 3.3) which can tolerate a certain level of errors due to spontaneous emission. As discussed in Sec. 4.2.1, for qubit levels coupled by single photon optical transitions, this may be accomplished by employing long-lived energy levels which do not have an allowed electric dipole coupling, such as metastable electronic levels with a quadrupole or intercombination coupling to the ground state. Although the interaction of these states with the vacuum (causing spontaneous emission) is reduced, their interaction with an external field for use in quantum logic operations is also reduced. This results in a fundamental limit on the accuracy of each operation by roughly the ratio of the spontaneous emission rate to the Rabi frequency $\xi = \Gamma/\Omega$. In the case of optical transitions, $\Omega$ cannot be increased indefinitely, since at optical intensities beyond about $10^{14}$ W/cm$^2$, the atom is quickly photoionized. This amounts to inaccuracies due to spontaneous emission on the order of $\xi = 10^{-6} - 10^{-7}$ [289]. Even this limit may be too optimistic, as the two-level approximation breaks down before photoionization occurs, and the coupling to other electronic levels must also be considered [188,289]. This results in inaccuracies due to spontaneous emission on the order of $\xi = 10^{-5} - 10^{-6}$, depending on the particular ion used. To understand the nature of the problem, we examine a simplified system. More detailed treatments are given in Refs. [188] and [289].

Consider the situation shown in Fig. 9. We assume that levels $|\downarrow\rangle$ and $|\uparrow\rangle$ comprise the qubit states. We want to drive coherent transitions between these two levels but, in general, we must consider spontaneous Raman scattering from other non resonant spectator levels. For this approximate treatment, we will consider only the effects of the nearest, most strongly coupled state which is designated as state $|s\rangle$ in the figure. For simplicity, we assume that levels $|s\rangle$ and $|\uparrow\rangle$ are coupled only to the ground state $|\downarrow\rangle$ (and not each other) by the applied radiation, and that they decay by the same coupling process to the ground state with rates $\Gamma_s$ and $\Gamma_\uparrow$. To allow for different coupling processes for the $|s\rangle$ and $|\uparrow\rangle$ states, we let $\Gamma_\uparrow = \kappa \Gamma_s$ and we neglect $\omega^3$ factors in the differences of the lifetimes (this would be valid if the $|\downarrow\rangle \rightarrow |\uparrow\rangle$ transition frequency is much larger than $\Delta$). The worst case is given by considering sideband transitions between levels $|\downarrow\rangle$ and $|\uparrow\rangle$. From Sec. 2.3.1, the resonant Rabi frequency is given by $\Omega_1 \simeq \eta \Omega_{\uparrow,\downarrow} \simeq \eta \kappa^{1/2} \Omega_{s,\downarrow}$. We assume that during the operation, the ion has a probability of about 0.5 of being in the $|\downarrow\rangle$ and $|\uparrow\rangle$ state. With these assumptions, the ratio $\xi$ of spontaneous emission (from both upper levels) to Rabi frequency is

$$\xi = \frac{R_s}{\Omega_1} \simeq \frac{1}{2}\Gamma_\uparrow + \frac{1}{2}\Gamma_s\left(\frac{\Omega_{s,\downarrow}}{\Delta}\right)^2 = \frac{\Gamma_s}{2\Omega_1}\left(\kappa + \frac{\zeta}{\kappa}\right), \tag{138}$$

where $\zeta \equiv \Omega_I^2/(\eta\Delta)^2$, and the factors of ½ come from the probabilities the ion is in the $|\downarrow\rangle$ or $|\uparrow\rangle$ state. The most optimistic answer is given by minimizing $\xi$ with respect to $\kappa$ which leads to $\xi \simeq \Gamma_s/(\eta\Delta)$. As an approximate "best case" we take $\Gamma_s/2\pi = 25$ MHz, $\Delta/2\pi = 10^{14}$ Hz, $\Omega_I/2\pi = 10$ MHz, $\eta = 0.1$, giving $\xi \simeq 2.5 \times 10^{-6}$ and requiring $\kappa \simeq 0.4\xi$ and $\Gamma_\Uparrow \simeq 10^{-6}\Gamma_s$. Therefore, weakly-allowed transitions are desirable if single photon optical transitions are used for qubit transitions.

      In the case of two-photon stimulated-Raman transitions between stable electronic ground states, the ratio of spontaneous emission rate to Rabi frequency is approximately $\xi_{SR} = \gamma_{se}/(g^2/\Delta_R)$, where $\gamma_{se} \approx \Gamma g^2/\Delta_R^2$ is the off-resonant spontaneous emission rate, g is the resonant single-photon Rabi frequency of each laser beam, and $\Delta_R$ is the detuning of the Raman beams from the excited state (Sec. 2.3.3). This results in an inaccuracy $\Gamma/\Delta_R$ due to spontaneous emission, which is independent of optical intensity. Since Raman transitions between S electronic ground states are effective only when the detuning $\Delta_R$ is not much greater than the fine structure splitting of the atom [290]; this results in an inaccuracy $\xi_{SR}$ due to spontaneous emission in range from about $10^{-4}$ ($^9$Be$^+$) to $10^{-7}$ ($^{199}$Hg$^+$), depending on the particular ion used. Spontaneous emission from spectator electronic levels should not significantly affect this limit, provided that their splitting from the virtual excited state significantly exceeds $\Delta_R$ and that the single photon resonant Rabi frequencies coupling the ground states to the spectator levels are not much bigger than g [188]. These appear to be reasonable assumptions for most candidate ions.

      The decohering effects of spontaneous emission can be overcome by error correction schemes. Error correction is complicated by the fact that when spontaneous emission occurs, the atoms may decay to states outside the original set of computational basis states. However, this situation can, in principle, be detected by optically pumping the ions back to the computational basis and applying the error correction schemes [291,292].

      Spontaneous emission decoherence could, in principle, be nearly eliminated by driving single-photon transitions between ground-state-hyperfine or Zeeman levels with rf or microwave radiation since spontaneous emission from these levels is negligible. Refs. [3] and [293] discuss this possibility, where inhomogeneous magnetic fields couple the internal and motional states. The speed of sideband operations is limited by the size of the field inhomogeneity one can obtain. From Eq. (28), we want a coupling Hamiltonian $H_I = -\mu_x(\partial B_x/\partial z)z$. Assume $\partial B_x/\partial z = (\partial B_x/\partial z)_o \cos\omega t$, $\mu_x = \mu_M S_x$, and $z = z_o(a + a^\dagger)$ as in Sec. 2.3.1. For resonance at the first red sideband ($\omega = \omega_o - \omega_z$), $H_I$ is given by Eq. (27) with $\eta\Omega = -\mu_M(\partial B_x/\partial z)_o z_o/(4\hbar)$. If we take $\mu_M = \mu_B$, $z_o = 10$ nm, then to achieve $\Omega_I/2\pi = 1$ MHz, we require $(\partial B_x/\partial z)_o \simeq 2.9 \times 10^2$ T-cm$^{-1}$; a difficult task. Moreover, it would be difficult to address selected ions because of the long wavelength of the radiation relative to typical ion spacings.

4.4.7. Mode cross coupling during logic operations

      In the preceding sections, we have assumed that when transitions are driven between $|\downarrow\rangle_j|n_k\rangle$ and $|\uparrow\rangle_j|n_k'\rangle$ involving a single mode of motion k, the other 3L-1 spectator modes of motion are not affected because coupling to them is nonresonant. However, when the sum or difference frequency of two or more spectator modes is near the frequency of the desired mode-k transition ($\omega_k|n_k' - n_k|$), higher order couplings can entangle the $|\downarrow\rangle_j|n_k\rangle$ and $|\uparrow\rangle_j|n_k'\rangle$ states with the spectator mode states.

Equation (33) describes the general interaction Hamiltonian between the internal levels of ion j and all 3L modes of motion. By expanding the exponential in Eq. (33) to all orders, we find

$$H'_{Ij} = \hbar\Omega'\left[S_{+j}\left(\prod_{l=1}^{3L}\sum_{b_l,d_l=0}^{\infty}\frac{(i\eta_l^j a_l^\dagger)^{b_l}(i\eta_l^j a_l)^{d_l}}{b_l!d_l!}e^{i(b_l-d_l)\omega_l t}\right)e^{-i(\delta t - \phi_j)} + h.c.\right], \quad (139)$$

where $\Omega' = \Omega\exp[-\frac{1}{2}\sum_l(\eta_l^j)^2]$. This equation describes the processes of each mode l gaining or losing ($b_l$-$d_l$) vibrational quanta, accompanied by the raising or lowering of the internal electronic levels of ion j. In general, we must account for all terms in Eq. (139) which do not vary rapidly in time, or terms in which the resonance condition is nearly met: $\sum_l(b_l-d_l)\omega_l \simeq \delta = \omega_k(n_k'-n_k)$. Although detailed treatment of this problem is beyond our intent, a couple of comments may be made.

In general, we must account for all the terms in Eq. (139) which cause significant errors in the overall computation we are trying to carry out. For two or more trapped ions, some combination of modes will nearly always satisfy the resonance condition. However, this may occur only for high orders of $b_l$ and $d_l$, and if the Lamb-Dicke criterion is met, the contributions are vanishingly small. The terms that will cause problems are the ones that are close to satisfying the resonance condition and are relatively low order in $b_l$ and $d_l$. If the Lamb-Dicke criterion is satisfied, it will always be possible to avoid these spurious couplings, but it may be at the expense of making the Rabi frequency so small (in order to avoid coupling to relatively nearby off-resonant terms) that the operations will become too slow.

To understand this problem in the context of a simple example, we assume that a cross-mode coupling of this type occurs when two modes, p and q, have frequencies which satisfy the condition $n_p\omega_p - n_q\omega_q \simeq 0$, $\omega_k$, or $-\omega_k$ corresponding to possible extraneous mode coupling on the carrier, first blue sideband, or first red sideband of the logic operations (assumed to utilize mode k). This additional resonance condition yields, to lowest order in the Lamb-Dicke parameters, the effective Hamiltonian

$$H'_{Ij} = \hbar\Omega'S_{+j}\left\{1 + i\eta_k^j(a_k e^{-i\omega_k t} + a_k^\dagger e^{i\omega_k t}) + \frac{(i\eta_p a_p^\dagger)^{n_p}(i\eta_q a_q)^{n_q}}{n_p!n_q!}e^{i(n_p\omega_p - n_q\omega_q)t}\right\}e^{-i(\delta t - \phi_j)} + h.c., \quad (140)$$

where the resonance conditions are given by $\delta \equiv \omega - \omega_o = 0$, $+\omega_k$, or $-\omega_k$. A specific example is relevant to the NIST single $^9$Be$^+$ ion experiments. Here, mode k was the x oscillation, and modes p and q are identified with the z and y oscillations of the single ion in the trap. In this experiment $\eta_x = \vec{k}\cdot\hat{x}x_o$, $\eta_y = \vec{k}\cdot\hat{y}y_o$, $\eta_z = \vec{k}\cdot\hat{z}z_o$, and $\omega_x \simeq \omega_z - \omega_y$. (The frequency relationship $\omega_x = \omega_z - \omega_y$ is a consequence of Maxwell's equations for a quadrupole rf trap in the absence of static potentials applied to the electrodes [211].) We assume that the desired transition is the first blue sideband of mode x ($\delta = \omega_x$). In this case, the resonant part of Eq. (140) becomes

$$H_I' \simeq \hbar\Omega' \left[ S_+ \left\{ i\eta_x a_x^\dagger - \eta_z\eta_y a_z^\dagger a_y + O(\eta^3) \right\} + h.c. \right]. \tag{141}$$

The term proportional to $a_x^\dagger$ is the desired anti-Jaynes-Cummings operator, and the term proportional to $a_z^\dagger a_y$ can entangle the internal state with the other spectator modes (z and y), leading to errors.

For logic operations on a string of ions in a linear trap, we will assume that all other mode frequencies are higher. With the use of stimulated-Raman transitions, we can make $\Delta\vec{\mathbf{k}} \parallel \hat{\mathbf{z}}$ and restrict our attention to spectator modes along the z axis. Nevertheless, as L becomes large, nearby resonances of the type shown in Eq. (140) will become harder to avoid. These coupling terms always scale as products of Lamb-Dicke parameters. Thus if the spectator mode Lamb-Dicke parameters are small enough, or if at least one Lamb-Dicke parameter is approximately zero, the higher order unwanted resonances may be sufficiently suppressed. Furthermore, if the spectator modes are cooled to near the zero-point energy ($\langle \hat{n} \rangle \ll 1$), then any couplings in Eq. (140) with powers of the annihilation operator $a_q$ will be absent most of the time. Hence, in large registers, it will probably be important to cool all modes to near the zero-point energy.

5. Variations

5.1. Few-ion accumulators

Many of the problems anticipated above, such as unwanted coupling to adjacent ions or spectator modes, will be aggravated by a large number of ion qubits in a single ion trap "register." Moreover, implementation of quantum error correction schemes will require highly parallel processing [294]. Therefore, a multiplexing scheme for ion qubit registers is desirable. One possibility is to perform all logic in minimal accumulators which hold one or two ions at a time [65]. Ions would be shuffled around in a "super-register" and into and out of (multiple) accumulators which are well shielded from the other ions. The shuffling could be accomplished with interconnected linear traps with segmented electrodes as shown schematically in Fig. 10; construction of such traps appears to be possible with the use of lithographic techniques [295]. Single-bit rotations on the mth ion would be accomplished by moving that ion into an accumulator. Logic operations between ions m and k would be accomplished by first moving these ions into an accumulator. An accumulator could also hold a second species of ion (say Mg$^+$) which could be used to provide laser cooling to the $|n=0\rangle$ state of the motional mode used for the logic operations, if necessary. Therefore, for logic operations, an accumulator would hold two computational ions and the auxiliary ion. This scheme should make it easier to select ions with laser beams because it should be possible to address one ion while nulling the laser intensity on the other ion, even with very high trap frequencies (see Sec. 4.4.4). The very small number of logic ions in an accumulator (1 or 2) would make extraneous mode coupling much easier to avoid. The main disadvantage appears to be that computational speed is reduced because of the time required to shuffle ions in and out of the accumulator and provide laser cooling with the auxiliary ion, if required. However, energy shifts of the ion's internal structure, due to the electric fields required to move the ion, need not be severe. For example, to move a

$^9$Be$^+$ from rest to a location 1 cm away (and back to rest) in 1 μs would require a field of less than 50 V/cm. Electric fields of this order should give negligible phase shifts in qubits based on hyperfine structure (Sec. 4.2.3). The accompanying phase shift caused by time dilation for a transition frequency $\omega_o/2\pi$ = 1.25 GHz (hyperfine structure in $^9$Be$^+$) would be less than 1 μrad.

5.2. Multiplexing with internal states

In principle, it should be possible to multiplex quantum information into all 3L modes of ion motion. This will probably be useful in experiments which use a small number of ions. However, because of the higher potential for decoherence of motional modes, it might prove more useful to multiplex quantum information in multiple internal states. This would give more work space with a smaller number of ions. Quantum logic within multiple internal states would most likely follow the ideas of NMR quantum computing [165,166]. A simple example is given by the eight ground state hyperfine levels in $^9$Be$^+$ which can be labeled like states of three coupled spin-½ particles $|\epsilon_1,\epsilon_2,\epsilon_3\rangle$ where $\epsilon_i \in \{0,1\}$. For instance, a Toffoli gate [140] is realized by driving a π pulse on the $|F,M_F\rangle = |2,2\rangle \to |1,1\rangle$ transition in the ground state hyperfine manifold of $^9$Be$^+$ if we label the states $|2,2\rangle \equiv |1,1,0\rangle$ and $|1,1\rangle \equiv |1,1,1\rangle$, in the notation $|F,M_F\rangle \equiv |\epsilon_1,\epsilon_2,\epsilon_3\rangle$. Technically, the readout of specific internal quantum states would be more complicated since the methods employed in the experiments so far distinguish between only two possibilities, seeing fluorescence or not seeing fluorescence, on any given ion. However, by appropriate mappings, superpositions of any two internal states could be mapped onto two states of "readout" ions and the detection accomplished as described above. To multiplex information into internal states, we need a way to map qubits from any ion into the internal levels of other ions. To give a specific example of how this can be accomplished, we consider the case of $^9$Be$^+$ where we might wish to multiplex into the $|F,M_F\rangle = |2,2\rangle, |1,1\rangle, |2,1\rangle$, and $|1,0\rangle$ states shown in Fig. 11. This figure is essentially the same as Fig. 5, but for simplicity of notation, we label these hyperfine states as $|0\rangle, |1\rangle, |0'\rangle$, and $|1'\rangle$ respectively. We want to accomplish the transformation

$$\Psi_i = \alpha|0\rangle_A|0\rangle_B + \beta|0\rangle_A|1\rangle_B + \gamma|1\rangle_A|0\rangle_B + \delta|1\rangle_A|1\rangle_B$$
$$\Rightarrow \left(\alpha|0\rangle_A + \beta|1\rangle_A + \gamma|0'\rangle_A + \delta|1'\rangle_A\right)|0\rangle_B, \qquad (142)$$

between ions A and B. This can realized with the transformation

$$\hat{M} = \hat{C}_{A',B} \cdot \hat{C}_{A,B} \cdot \hat{C}_{B,A'} \cdot \hat{C}_{B,A} \cdot \hat{\Pi}_{A,1,0'}, \qquad (143)$$

where the operator $\hat{\Pi}_{A,1,0'}$ denotes a π transition between the states $|1\rangle_A$ and $|0'\rangle_A$ and the operators $\hat{C}_{c,t}$ are defined in Sec. 3.3. For example, $\hat{C}_{A',B}$ denotes a controlled-not operation with the $|0'\rangle_A$ and $|1'\rangle_A$ states acting as the control bit states and the $|0\rangle_B$ and $|1\rangle_B$ states acting as the target bit states.

## 5.3. High-Z hyperfine transitions

The possibility of using highly charged ions as qubits poses some interesting possibilities. For simplicity, we consider using hyperfine states in high-Z hydrogenic ions as qubit levels where Z is the nuclear charge. More complicated high-Z ions appear to show similar features. If Z is high enough, hyperfine transitions occur in the optical region of the spectrum. Therefore, the required high field gradients necessary for quantum logic would be provided by the laser fields used to drive the hyperfine transitions. The potential advantages are: (1) Off-resonant spontaneous scattering, which is a source of decoherence in other cases (Refs. [188], [289], and Sec. 4.4.6.4), is essentially eliminated since the first excited 2p level is at much higher energy above the ground state. (2) The trap binding of high-Z hydrogenic ions can typically be stronger that singly-ionized atoms, thereby increasing motional frequencies and reducing the time of fundamental operations. The disadvantages are (1) High-Z ions are hard to produce. (2) The lifetimes of the upper hyperfine states are short enough that spontaneous emission decoherence from these states cannot be neglected. (3) Detection of internal states may be difficult.

The ion production problem is not intractable; for example, recent storage ring measurements have been able to determine the transition wavelengths and lifetimes of hydrogenic $^{209}$Bi$^{82+}$ [296] and $^{165}$Ho$^{66+}$ [297]. Effort in some laboratories is directed to transferring similar high-Z ions to low energy ion traps. The hyperfine frequency of a hydrogenic ion can be estimated to be $\omega_{hfs}/2\pi \simeq 2.54 \times 10^8 \, Z^3 F |g_n|$ Hz, where F and F-1 are the values of angular momentum in the hyperfine states and $g_n$ is the nuclear g-factor ($\mu = \mu_n I g_n$, $\mu_n$ = nuclear magneton, I = nuclear spin) [298]. The spontaneous decay rate from the excited state is estimated to be $\gamma_{rad} = 1/\tau_{rad} \simeq 1.01 \times 10^{-42} [M_{ab}/(2F_b+1)](\omega_{hfs}/2\pi)^3$ s$^{-1}$, where the term in square brackets [298] depends on matrix elements and is on the order of 1. For example, the hyperfine transition in $^{209}$Bi$^{82+}$ has a wavelength of 244 nm and the upper level decays with a lifetime of 351 μs [296]. For $^{165}$Ho$^{66+}$, the transition wavelength is 573 nm and the decay time is 2 to 3 ms [297]. These experimental numbers agree roughly with the formulas above. These wavelengths are reasonable for quantum logic, but the lifetimes are somewhat short. However since the lifetime scales as $1/Z^9$, it might be possible to use lower Z, longer wavelength transitions. The readout of the qubits appears difficult, because coupling to a higher electronic level and observing fluorescence is impractical due to the short wavelengths involved. In principle, however, the qubit information from these ions could be transferred to other ions whose states are easier to read, perhaps using small accumulators as described in the previous section and in Sec. 4.3.

## 6. Other applications

### 6.1. Quantum correlations

In a classic paper, Einstein, Podolsky, and Rosen (EPR) [299] argued that quantum mechanics provides an incomplete description of physical reality and speculated that it might be an approximation to some theory which would provide a complete description. Today such theories, which supplement quantum mechanics with additional, as-yet unobserved parameters, are called hidden-variables theories. If such a theory obeys some physically reasonable conditions forbidding action at a distance, it is called a "local hidden-variables theory."

Simple quantum logic gates performed on small numbers of trapped ions can lead to interesting experiments which may shed light on the viability of local hidden-variables theories. For example, as described by Cirac and Zoller [1] and Sec. 3.4, using controlled-not gates, we can generate the state

$$\Psi = \frac{1}{\sqrt{2}}(|\downarrow\rangle_1 |\downarrow\rangle_2 + e^{i\phi}|\uparrow\rangle_1 |\uparrow\rangle_2). \tag{144}$$

In the context of EPR, this is called a Bell state. According to Bell's theorem [300], such a state allows us to distinguish between quantum mechanics and all local hidden-variables theories. If the states of the two entangled particles are detected outside of each other's light cones, then, for particular sets of measurements, we may derive inequalities which all local hidden-variables theories must obey, but which quantum mechanics violates. The experiments performed by Clauser and Shimony [301] and Aspect *et al.* [302] provide strong evidence against local hidden-variables theories. Their work used polarization measurements on entangled pairs of photons. In their experiments, the detection of the photons' polarization states occurred outside each others' light cones. Thus, the measurement on one photon could not possibly have affected the other measurement, which closed possible "loopholes" in the proof of quantum mechanics over other explanations.

However, some loopholes still remain open. Since the photon detection in the Aspect, et al. experiments was not 100% efficient, the group had to make assumptions that the photons they measured were a "fair" sample of the whole population of events. Thus, their experiments do not rule out the (seemingly implausible) possibility of local hidden-variables theories in which the hidden variables cause some sub-ensemble of the photon pairs to preferentially interact with the measurement apparatus.

In the system of two ions, we may detect the state of either ion with nearly 100% efficiency through the use of "electron-shelving" (Sec. 2.2.1). On the contrary, it will be difficult to perform measurements on two ions outside each other's light cone. Such a measurement would require separating the ions by a distance larger than the speed of light times the measurement time or transferring quantum information over large distances [151]. (In principle, the ions could be first entangled and then placed in different traps which could be separated by large distances before measurements were performed.) Nonetheless, an experiment with two trapped ions could be viewed as complementary: the photon experiments definitively close loopholes of causality, and the ion experiments could close loopholes due to detection inefficiency. Such experiments have the additional appeal of studying EPR on massive particles [303]. EPR states of atoms have recently been created in an atomic beam using the methods of cavity QED [304]; if detection efficiency can be improved, these experiments could also close loopholes due to detection inefficiency. Moreover, even though measurements of quantum correlations between entangled ions cannot be easily performed outside each other's light cone, one can argue strongly that the ions cannot transfer information by any known mechanism. Therefore, if the observed correlations violate Bell's inequalities, the correlations are established by some new force of nature or are, in fact, inherent in the structure of quantum mechanics.

An intriguing possibility for ions is the possibility of making "GHZ states" [54,305,306]. For three ions, the GHZ state is

$$\Psi = \frac{1}{\sqrt{2}}(|\downarrow\rangle_1|\downarrow\rangle_2|\downarrow\rangle_3 + e^{i\phi}|\uparrow\rangle_1|\uparrow\rangle_2|\uparrow\rangle_3). \tag{145}$$

For such a state, a single measurement can distinguish between the predictions of quantum mechanics and those of any local hidden-variables theory [54,305].

Aside from this, Bell states, GHZ states, and Schrödinger-cat states are highly entangled, and are thus of inherent interest for the study of uniquely quantum behavior. As the experiments improve, it will be interesting to push the size of entangled states to be as large as possible. The question is not whether we can make states which have the attributes of Schrödinger cats, but how big can we make the cats? Certain theories which address the measurement problem will be amenable to experimental tests, for example, quantitative limits on spontaneous wavefunction collapse theories [307,308] can be established. The isolation from the environment exhibited by trapped ions, coupled with the control possible over their quantum state and high detection efficiency make them an interesting laboratory for the study of fundamental issues in quantum mechanics.

6.2. Simulations

The nonlinearities with respect to motional raising and lowering operators inherent in the coupling Hamiltonians of Eqs. (14), (37) and (48) can lead to a rich variety of dynamics between the ion motion and internal levels. Some of these dynamics for a single ion have been discussed in Sec. 3.2. An interesting system which can be simulated with these couplings is a "phonon maser" which provides vibrational amplification by stimulated emission of radiation [309]. This would be an analog of the micromaser with some interesting differences such as the effects of recoil in the pumping process.

The Hamiltonians of Eqs. (14), (37) and (48) for a single ion can be consolidated into the general interaction Hamiltonian

$$H_I = \hbar\Omega S_+^\epsilon e^{i[\vec{k}\cdot\vec{x} - \delta t + \phi]} + h.c., \tag{146}$$

where $\delta = \omega - \omega_o$, where we have used the rotating-wave approximation, and where we have neglected phase factors of the field. The parameter $\epsilon$ is equal to 1 when internal state transitions are involved and $\epsilon = 0$ when the internal state is unchanged. In Eq. (146), $\vec{k}$ is the wavevector of the field for single photon transitions or $\vec{k} \to \vec{k}_1 - \vec{k}_2$ when stimulated-Raman transitions are used. Similarly, $\omega$ is the frequency of the applied field for single-photon transitions and $\omega \to \omega_{L1} - \omega_{L2}$ when stimulated-Raman transitions are used. In an interaction picture of the ion's motion, this Hamiltonian becomes

$$H_I' = \hbar\Omega\left[S_+^\epsilon e^{-i(\delta t - \phi)} \prod_{j=x,y,z} \exp(i[\eta_j(a_j e^{-i\omega_j t} + a_j^\dagger e^{i\omega_j t}) + h.c.\right], \quad (147)$$

where $\eta_x \equiv \vec{k}\cdot\hat{x}x_o$, $a_x$ is the lowering operator for the x motion of frequency $\omega_x$, etc. Now, assume that $\Omega$ is small enough, and that, in general, $\omega_x$, $\omega_y$, and $\omega_z$ are incommensurate that we can excite only one spectral component of the possible transitions induced by this interaction. For a particular resonance condition $\delta = -\ell_x\omega_x - \ell_y\omega_y - \ell_z\omega_z$, and in the Lamb-Dicke limit, we find

$$H_I' \simeq \hbar\Omega e^{i\phi} S_+^\epsilon \prod_{j=x,y,z} \left[\delta_{\ell_j,|\ell_j|}\frac{(i\eta_j a_j)^{|\ell_j|}}{|\ell_j|!} + (1 - \delta_{\ell_j,|\ell_j|})\frac{(i\eta_j a_j^\dagger)^{|\ell_j|}}{|\ell_i|!}\right] + h.c. \quad (148)$$

The two mode case where $\epsilon = \ell_z = 0$ is considered by Drobný and Hladký [310], and in a different excitation scheme in Ref. [110]. If the Lamb-Dicke limit is not rigorously satisfied, we must consider higher order nonlinear corrections to this expression; specific examples are discussed in Refs. [28] and [110]. These nonlinear terms are the origin of the high-order corrections to the Rabi frequencies (Eq. (18)). The case of carrier, and first red and blue sidebands on internal state transitions (e.g., $\epsilon = 1$, $\ell_x = \ell_y = 0$, $\ell_z = 0, \pm 1$) are used extensively for quantum logic and generation of nonclassical motional states and are discussed above. The case $\epsilon = 0$, $\ell_x = \ell_y = 0$, $|\ell_z| = 1$ has been used to create coherent [21] and Schrödinger cat [47] states of motion and is discussed in Sec. 3.1. The case $\epsilon = 0$, $\ell_x = \ell_y = 0$, $|\ell_z| = 2$ has been used to create squeezed states; this is discussed in Ref. [21]. A realization of the Hamiltonian $H_I' \propto S_+(a_x^\dagger)^2 +$ h.c. ($\epsilon = 1$, $\ell_x = \ell_y = 0$, $\ell_z = -2$) has been reported by Leibfried, *et al.* [132]. This is similar to the case of two-photon excitation in cavity QED analyzed by Buck and Sukumar [311] and Knight [312]. Interactions proportional to $S_+a_x^\dagger a_y +$ h.c., $S_+a_x a_y^\dagger +$ h.c., and $a_y^\dagger a_z +$ h.c. might be used to generate the maximally entangled state of Eq. (60) without the need to address individual ions [194]. An example of an interesting new case would perhaps be the realization of three-phonon downconversion (e.g., $\epsilon = 0$, $\ell_x = 3$, $\ell_y = -1$, $\ell_z = 0$). This case is analogous to three-photon downconversion in quantum optics (see Refs. [110],[313], and references therein). Here, it corresponds to driving a two-mode resonance using stimulated-Raman transitions where $\omega_{L1} - \omega_{L2} = \omega_y - 3\omega_x$. A suggestion to realize a Hamiltonian $\propto a_x^2 a_y^\dagger +$ h.c. ($\epsilon = 0$, $\ell_z = 0$, $\ell_x = 2$, $\ell_y = -1$) is discussed by Agarwal and Banerji [314]. Clearly, a large number of possibilities could, in principle, be realized just for a single ion; moreover, the number of possibilities increases dramatically if we consider all modes of motion for multiple trapped ions. The only limitation on how high $|\ell_j|$ in Eq. (148) can be is that $\Omega$ be chosen sufficiently small that couplings to other (unwanted) resonances are avoided. This will require that decoherence be small enough to see the desired dynamical behavior before coherence is lost.

Various forms of interactions which satisfy the requirements of quantum nondemolition (QND) measurements [20, 25,315] of ion motion or quantum feedback [316] can be extracted from Eq. (148). QND experiments employing dispersive interactions are considered by Retamal and Zagury [36]. These schemes rely on measuring the ion's internal state however, which almost always involves recoil heating, thereby destroying the state we wish to preserve. This could be circumvented by coupling the ion's motion to a cavity field which then serves as the probe [317].

6.2.1. Mach-Zehnder boson interferometer with entangled states

Realization of the various Hamiltonians indicated in Eq. (148) can lead to simulation of various devices of practical interest. As an example, consider a Mach-Zehnder interferometer which acts on two modes of oscillation of a single trapped ion; to be specific, we will consider the x and y modes of motion. The analogy with a Mach-Zehnder interferometer for bosons is that the two input modes to the boson interferometer are replaced by the x and y modes of ion oscillation. The (50/50) beamsplitters in the boson interferometer are replaced by an operator $B_\pm$ = $\exp[\pm i\pi(a_x^\dagger a_y + a_x a_y^\dagger)/4]$ [193,318,319]. This operator can be realized by applying the interaction in Eq. (148) with $\epsilon = 0$, and $\ell_x = -\ell_y = 1$ for a time given by $\Omega\eta_x\eta_y t = \pi/4$. A differential phase shift between the two arms of the interferometer can be simulated by shifting the relative phases of the fields in Eq. (148) between successive applications of $B_\pm$. In a particle (e.g., boson) interferometer, one typically measures the number of particles in either one or both output modes. In a single ion experiment, we have only one convenient observable, the internal state of the ion (either $|\downarrow\rangle$ or $|\uparrow\rangle$). Nevertheless, we can characterize the action of the phonon interferometer by repeating the experiment many times and measuring the density matrix of the output state [132].

It will be most interesting to characterize the action of the interferometer for various nonclassical input states. One example is the two-mode Fock state $|n_x\rangle_x|n_y\rangle_y$ [195]. This state could be prepared by applying the techniques described in Sec. 3.2 sequentially to the ion's x and y modes. This state is interesting because it has been shown that one could approach the Heisenberg uncertainty limit in a Mach-Zehnder interferometer by measuring the distribution of bosons in the output modes [195,196,197]. An alternative technique for studying the action of a beamsplitter on the two-mode Fock states has been suggested by Gou and Knight [23] when $\omega_x = \omega_y$. Here, a beamsplitter could be simulated by first preparing $|n_x\rangle_x|n_y\rangle_y$ along two orthogonal axes and then probing along two other axes (x' and y') which are rotated (in the xy plane) with respect to the first. This technique could also be used to analyze, for example, the $(|0\rangle_{x'}|2\rangle_{y'} + |2\rangle_{x'}|0\rangle_{y'})/2^{1/2}$ state from an initially prepared $|1\rangle_x|1\rangle_y$ state [23].

Another interesting state to consider for the phonon interferometer is the $(|N\rangle_x|0\rangle_y + |0\rangle_x|N\rangle_y)/2^{1/2}$ state (which is the desired state after the first beam splitter). This state has been shown to yield exactly the Heisenberg uncertainty limit for an interferometer for any value of N [194], if after the second beamsplitter, we measure the number of particles N(x) in the x output port. The result of this measurement is assigned the value $(-1)^{N(x)}$. This state could be prepared

from the $|\downarrow\rangle|0\rangle_x|0\rangle_y$ state by the following two steps:

(1) Apply a π/2 pulse on the Nth blue sideband of mode x ($\epsilon = 1$, $\ell_x = -N$, $\ell_y = 0$ in Eq. (148));
this creates the state $(|\downarrow\rangle|0\rangle_x + |\uparrow\rangle|N\rangle_x)|0\rangle_y/2^{1/2}$.
(2) Apply a π pulse on the Nth blue sideband of mode y ($\epsilon = 1$, $\ell_x = 0$ $\ell_y = -N$ in Eq. (148)); this creates the state $|\uparrow\rangle(|N\rangle_x|0\rangle_y + |0\rangle_x|N\rangle_y)/2^{1/2}$.

After the second beamsplitter, we have a state which can be written as

$$\Psi_{final} = |\uparrow\rangle \sum_{n_x=0}^{N} C_{n_x} |n_x\rangle_x |N-n_x\rangle_y. \tag{149}$$

In principle, we would like to measure $n_x$, record the value N(x), and assign the value $(-1)^{N(x)}$. Effectively, this assignment can be accomplished if we can find an interaction M which provides the mapping

$$M\Psi_{final} = |\uparrow\rangle \sum_{n_x \text{ even}}^{N} C_{n_x} e^{i\phi(n_x)} |n_x\rangle_x |N-n_x\rangle_y + |\downarrow\rangle \sum_{n_x \text{ odd}}^{N} C_{n_x} e^{i\phi(n_x)} |n_x\rangle_x |N-n_x\rangle_y. \tag{150}$$

After this mapping, we need only measure the internal state; if the ion is found in the $|\uparrow\rangle$ state we assign the value +1 to the measurement; if the ion is found in the $|\downarrow\rangle$ state, we assign the value -1. The mapping M can be achieved by applying radiation with $\vec{k} \parallel \hat{x}$ at the carrier frequency ($\epsilon = 1$, $\ell_x = 0$) and insuring $\Omega_{n_x,n_x} t = 2\pi m \pm n_x \pi$ where m is an integer. From Eq. (18), we have

$$\Omega_{n_x,n_x} t \simeq \Omega t e^{-\frac{\eta_x^2}{2}} \left[ 1 - n_x \eta_x^2 \left( 1 + \frac{\eta_x^2}{4} - n_x \frac{\eta_x^2}{4} \right) \right]. \tag{151}$$

Therefore, if we make $\Omega \exp(-\eta_x^2/2)t = 2m\pi$ and $\eta_x^2(1 + \eta_x^2/4) = (2m)^{-1}$, we achieve the desired mapping as long as the contribution to the phase from the term proportional to $n_x^2$ in this equation is small compared to π. Therefore we require $m \gg N^2/8$ or, equivalently, $\eta_x \ll 2/N$.

One final example of a two-mode phonon interferometer which directly yields Heisenberg 1/N phase sensitivity is a "beamsplitter" which creates the state $|\downarrow\rangle|0\rangle_x|N\rangle_y + |\uparrow\rangle|N\rangle_x|0\rangle_y/\sqrt{2}$ This state can be created by starting with an initial $(|\downarrow\rangle + |\uparrow\rangle)|N/2\rangle_x|N/2\rangle_y/\sqrt{2}$ dual Fock state (Sec. 3.2). Next, we apply N/2 π-pulses alternating between the two interaction Hamiltonians $H_1 = \Omega\eta_x\eta_y(S_+ a_x^\dagger a_y + h.c.)$ and $H_2 = \Omega\eta_x\eta_y(S_+ a_x a_y^\dagger + h.c.)$ which can be realized in the Lamb-Dicke regime. In this way, the state of motion of the ion is stepped through

$$|\downarrow\rangle|\frac{N}{2}\rangle_x|\frac{N}{2}\rangle_y + |\uparrow\rangle|\frac{N}{2}\rangle_x|\frac{N}{2}\rangle_y \quad \overset{H_1}{\rightarrow}$$

$$|\uparrow\rangle|\frac{N}{2}+1\rangle_x|\frac{N}{2}-1\rangle_y + |\downarrow\rangle|\frac{N}{2}-1\rangle_x|\frac{N}{2}+1\rangle_y \quad \overset{H_2}{\rightarrow}$$

$$|\downarrow\rangle|\frac{N}{2}+2\rangle_x|\frac{N}{2}-2\rangle_y + |\uparrow\rangle|\frac{N}{2}-2\rangle_x|\frac{N}{2}+2\rangle_y \quad \overset{H_1}{\rightarrow} \qquad (152)$$

$$\vdots$$

$$|\downarrow\rangle|N\rangle_x|0\rangle_y + |\uparrow\rangle|0\rangle_x|N\rangle_y$$

where, in this example, we assume $N/2$ is even. The interactions $H_1$ and $H_2$ follow from Eq. (148) with $\epsilon=1$, $\ell_x=-1$, $\ell_y=1$, $\ell_z=0$ and $\epsilon=1$, $\ell_x=1$, $\ell_y=-1$, $\ell_z=0$, respectively. The kth pulse has Rabi frequency $\Omega\eta_x\eta_y[(N/2+k)(N/2-k+1)]^{1/2}$ in the Lamb-Dicke regime. After a relative phase is accumulated in the two "paths" of the interferometer (simulated by adjusting the phase of the laser pulses as discussed above), we reverse the above steps and apply a final $\pi/2$ pulse on the carrier. Upon measuring the probability of occupation in state $|\downarrow\rangle$ or $|\uparrow\rangle$, the interference fringes exhibit $1/N$ phase sensitivity.

If the Lamb-Dicke criterion is not satisfied, the two components of the wavefunction superposition may experience different Rabi frequencies during each pulse, leading to undesired evolution. The exact Rabi frequencies of the two components of the wavefunction upon application of the kth pulse (interaction $H_1$ or $H_2$) follow from Eq. (18):

$$\Omega_A = \frac{\Omega\eta_x\eta_y \, e^{-\frac{1}{2}(\eta_x^2+\eta_y^2)} \, L_{N/2+k-1}^1(\eta_x^2) \, L_{N/2-k}^1(\eta_y^2)}{\sqrt{(N/2+k)(N/2-k+1)}} \,,$$

$$\Omega_B = \frac{\Omega\eta_x\eta_y \, e^{-\frac{1}{2}(\eta_x^2+\eta_y^2)} \, L_{N/2+k-1}^1(\eta_y^2) \, L_{N/2-k}^1(\eta_x^2)}{\sqrt{(N/2+k)(N/2-k+1)}} \,, \qquad (153)$$

where $\Omega_A$ is the Rabi frequency between states $|N/2 + k -1\rangle_x|N/2 - k +1\rangle_y$ and $|N/2 + k\rangle_x|N/2 - k\rangle_y$ and $\Omega_B$ is the Rabi frequency between states $|N/2 - k +1\rangle_x|N/2 + k -1\rangle_y$ and $|N/2 - k\rangle_x|N/2 + k\rangle_y$. The only differences between $\Omega_A$ and $\Omega_B$ are the arguments of the associated Laguerre polynomials. Thus, as long as $\eta_x = \eta_y$, the system will evolve as in Eq. (152), even when the Lamb-Dicke criterion is not satisfied.

6.2.2. Squeezed-spin states

As an example which demonstrates the advantages of using entangled states in spectroscopy, we discuss the following simple experiment which can be carried out on a single ion. Let us suppose we are interested in measuring, with maximum signal-to-noise ratio, the Zeeman frequency between states of a J = 1 manifold in an atom. To be specific, consider that we are interested in measuring the Zeeman frequency of the $\Delta M_F = \pm 1$ transitions in the $^2S_{1/2}$ (F = 1) ground-state hyperfine multiplet of a $^9Be^+$ ion (Figs. 5 and 11). This manifold is composed of the three levels $|F,M_F\rangle \in \{|1,0,\rangle, |1,\pm 1\rangle\}$. We will assume that the applied field is small enough that the frequency of the $|1,-1\rangle \rightarrow |1,0\rangle$ transition is equal to the frequency of the $|1,0\rangle \rightarrow |1,1\rangle$ transition. A straightforward way to measure the Zeeman frequency would be to prepare the atom in the $|1,1\rangle$ (or $|1,-1\rangle$) state, drive the Zeeman transition (using the Ramsey method), and then measure $J_z$. We will assume this detection can be accomplished with 100% detection efficiency, so the noise in the measurement is limited by the quantum statistics in the measurement process [100]. After many measurements, requiring a total averaging time $\tau$, a certain precision in the Zeeman frequency would be obtained. However an alternative measurement strategy, described below, would require an averaging time $\tau/2$ to reach the same measurement precision.

The basic idea is outlined in Ref. [9]; for J = 1, it is equivalent to the more general technique described by [194]. We prepare the atom in the state $\psi(0) = |1,0\rangle$ and then apply the Ramsey fields. Subsequently, we measure the probability of the ion to remain in the $|1,0\rangle$ state. (This measurement scheme has been used in the experiments of Abdullah, *et al.* [320]; the noise in these experiments, however, was not limited by projection noise.) This measurement is equivalent to measuring the operator $\mathbb{I} - J_z^2$, where $\mathbb{I}$ is the identity operator. This is equivalent to measuring the operator $J_z^2$, as discussed by Wineland, *et al.* [9], or the operator $\Pi_i \sigma_{zi}$, as discussed by Bollinger, *et al.* [194]. After application of the first Ramsey pulse to the $|1,0\rangle$ state, the ion is in the state $(|1,1\rangle + |1,-1\rangle)/\sqrt{2}$; this is equivalent to the maximally entangled state of Eq. (60) which could be formed by two spin-½ particles. As described in Sec. 3.3, we would expect the time required to reach a certain measurement precision to be reduced by a factor of 2 (L = 2) over the case of uncorrelated particles, represented by starting the ion in the state $\psi(0) = |1,1\rangle$ or $|1,-1\rangle$ state and finally measuring $J_z$.

We could carry out this experiment in the following way. We first optically pump the ion to the $|2,2\rangle$ hyperfine state. We then apply two successive $\pi$ pulses which carry out the transformations $|2,2\rangle \rightarrow |2,1\rangle \rightarrow |1,0\rangle$ with stimulated-Raman transitions. We then perform Ramsey spectroscopy on the $|1,0\rangle$ state. We can measure the probability of subsequently finding the atom in the $|1,0\rangle$ state by first reversing the order of the two $\pi$ pulses and then measuring the probability of finding the ion in the $|2,2\rangle$ state.

6.3. Mass spectroscopy and NMR at the single quantum level

Quantum logic operations may be useful in precision measurements other than spectroscopy (Sec. 3.4). For example, Ref. [48] discusses a method to measure cyclotron resonance frequencies of single ions at the quantum level; this technique essentially employs quantum logic to distinguish between motional quantum states. This capability would provide mass spectroscopy at such low energies that anharmonic shifts (electric potential and relativistic) would be very small. (At present, however, high-resolution mass spectrometry experiments are limited by other effects such as magnetic field drifts [77,78,79]. In mass spectroscopy, the basic problem is to measure a "test" ion's cyclotron frequency and compare it to the cyclotron frequency of a "measurement" ion (in the same magnetic field) which serves as the transfer standard.

One idea [48] is to store two ions in separate Penning traps which are stacked along their symmetry axis and share a common endcap and a common, axial magnetic field. The axial frequencies of the ions are adjusted to be the same, in which case the axial harmonic oscillators are coupled through the charge in the common endcap. The technique might work as follows: The axial motion of the "measurement" ion (for example, $^9Be^+$) is first laser cooled to the n=0 level. When the ions' axial modes are resonantly coupled, this cooling can be transferred to the test ion's axial motion; if the coupling is left on for a certain amount of time, the energies in the ions' axial oscillations are exchanged. Subsequently, the $^9Be^+$ ion's axial oscillation is recooled so that both ions are cooled to the zero-point energy. This axial cooling is then transferred to the test ion's cyclotron mode by parametric coupling, after which the axial motion of the test and measurement ion must be recooled to the zero-point energy. An external field is then applied to weakly excite the test ion's cyclotron motion. When the amplitude of this field is adjusted appropriately and the resonance condition is met, the n=0 to n=1 transition in the test ion is driven with high probability. The steps above are then reversed so that if the test ion's cyclotron motion was excited to the n=1 level, the $^9Be^+$ ion's axial motion is now in the n=1 level. When the resonance condition for the test ion's cyclotron frequency is not met, the $^9Be^+$ ion's motion remains in the n=0 level. Discrimination between the n=0 and n=1 axial level is then performed using quantum logic operations applied to the $^9Be^+$ ion as discussed in Sec. 3. For example, after the steps above, an axial red sideband $\pi$ pulse will excite a $|\downarrow\rangle \rightarrow |\uparrow\rangle$ transition in the $^9Be^+$ ion conditioned on whether or not the $^9Be^+$ ion's axial mode was in the n = 0 or n = 1 state. The test ion's cyclotron frequency can be referenced to a spin flip frequency in $^9Be^+$ (either electron or nuclear) which then acts as a transfer standard. Finally, by performing a cyclotron resonance measurement on a second test ion in the same fashion we can find the ratio of the two test ions' cyclotron frequencies and therefore derive their mass ratio.

Ref. [48] also suggested that these ideas could be applied to measure magnetic moments of test ions. In this section, we describe a variation on the method they discussed. The basic idea is to perform NMR on an unknown magnetic moment (the test ion) at the single spin level. Using quantum logic operations, the spin flip is detected in a measurement ion to which the test ion is coupled. We will describe the ideas in the context of a specific example; but the techniques are easily generalized to measuring the magnetic moments of other ions.

Suppose we want to measure the ratio of proton and antiproton magnetic moment. Such a measurement is of high current interest and can provide a second test of CPT on baryons in addition to the precise mass comparisons already performed [79]. Precision measurements of the magnetic moment ratios coupled with the mass ratios should test much of the same physics in the context of CPT as precision spectroscopy of antihydrogen [321]. A measurement scheme for the proton/antiproton moment ratio has been suggested previously by Quint and Gabrielse [322]. This method would employ the same basic ideas as in the electron g-2 experiments of Dehmelt and coworkers [108], where the proton's (or antiproton's) magnetic moment energy is transferred parametrically into its cyclotron energy (which is then detected) by use of applied inhomogeneous (oscillating) magnetic fields. The apparent drawback to this scheme is that it is slow because of the weakness of the parametric coupling and the difficulty of detecting small changes in the cyclotron energy. The method suggested here incorporates the same kind of parametric coupling but is potentially more efficient because it is sensitive to transfer at the single quantum level through the use of quantum logic techniques.

The basic idea of the method we propose here is to first store a proton and a $^9$Be$^+$ ion in trapping potentials which are close and in a common magnetic field (Fig. 12). The proton spin flip frequency is then compared to the $^9$Be$^+$ spin flip frequency (electron or nuclear), which acts as a transfer standard, effectively calibrating the (common) magnetic field. We then perform the same kind of measurement on a simultaneously trapped antiproton and $^9$Be$^+$ ion. By combining the two measurements, we determine the proton/antiproton magnetic moment ratio.

First consider the measurement of the proton to $^9$Be$^+$ spin flip frequency ratio. We simultaneously store a single proton and $^9$Be$^+$ ion in a double-well potential as indicated in Fig. 12. We assume that the ions are confined in a direction perpendicular to the z axis by a linear rf trap combined with a superimposed static magnetic field $\vec{B} = B_o \hat{z}$. (One particular geometry for a Paul trap in a strong magnetic fields is described by Bate, *et al.* [323].) The quadrupole electrodes are segmented to provide a static double-well potential along the z direction. We assume the electrode segments are capacitively coupled together so the rf electrode potentials are independent of z. A vertical wire (shown in cross section in the figure) is at one end of the trap; oscillating currents in this wire generate an oscillating magnetic field which can be used to drive the spin flip transitions of the proton and nuclear spin flip transition in the $^9$Be$^+$ ion (the electron spin flip transition in $^9$Be$^+$ could be driven with injected microwave radiation). Oscillating currents in this wire will also provide a parametric coupling between the proton spin and axial motion (below).

The experiment could proceed as follows: We first tune the static potentials so that the proton and $^9$Be$^+$ axial frequencies are the same. By performing Doppler laser cooling on the $^9$Be$^+$ ion's axial and radial modes, the proton's axial motion is cooled. The proton's radial modes will also likely be cooled through the Coulomb coupling. If not, we must apply an inhomogeneous rf field which parametrically couples the proton's radial and axial frequencies. We now uncouple the proton and $^9$Be$^+$ ion by adjusting their axial frequencies to be different. We then cool the $^9$Be$^+$ axial motion to the ground state (Sec. 3.1) and follow this by switching the axial modes back into resonance. The Coulomb interaction between the ions gives rise to a coupling $q_1 q_2 / (4\pi\epsilon_o |Z_2 - Z_1|)$, where $Z_1$ and $Z_2$ are the respective axial positions of the ions and $q_1$ and $q_2$ are, here, the charges of the proton and $^9$Be$^+$ ion. We have neglected the effects of induced charges in the trap electrodes, this will not be a large effect if the distance between ions is on the order of or less than the distance of each ion to the trap electrodes. For small amplitudes of oscillation we can write $Z_i = Z_{io} + z_i$, $i \in \{1,2\}$ where $Z_{io}$ are the equilibrium positions of each ion. In the limit of weak coupling and to lowest order in $z_i$, the interaction between ions is given by

$$H_{q_1,q_2} = \frac{q_1 q_2 z_1 z_2}{2\pi \epsilon_o d^3}, \tag{154}$$

where $d \equiv |Z_{1o} - Z_{2o}|$. This gives rise to a frequency splitting between axial normal modes of [48]

$$\delta\omega_{1,2} = \pi/t_{exch} = \frac{q_1 q_2}{2\pi \epsilon_o d^3 \omega_z \sqrt{m_1 m_2}}, \tag{155}$$

where $m_i$ are the ion masses and $t_{exch}$ is the time to exchange the energy of the two ions' axial energies after they are coupled together. At the time $t_{exch}$, the ions are again decoupled, leaving the proton in its axial ground state. The $^9$Be$^+$ ion is subsequently recooled to its axial ground state.

Assume the proton is initially in the lower energy state $|\uparrow\rangle$. We now apply an oscillating magnetic field at a frequency $\omega_m$ which is near the proton spin flip frequency $\omega_o = g_p \mu_B B_o/\hbar$ where $g_p$ is the proton gyromagnetic ratio expressed in terms of the Bohr magneton ($g_p \simeq 1.521 \times 10^{-3}$). The amplitude and duration of this field is adjusted to make a $\pi$ pulse if the resonance condition $\omega_m = \omega_o$ is satisfied; that is, the proton spin undergoes the transition $|\uparrow\rangle \rightarrow |\downarrow\rangle$. We now apply an inhomogeneous magnetic field at a frequency $\omega_a$ by driving a current $I\cos\omega t$ through the vertical wire, shown on the left in Fig. 12. If we neglect the shielding effects of the trap electrodes, this provides a field at the site of ion 1 equal to $\vec{B}(t) = \hat{x}\mu_o I\cos\omega_a t/(2\pi\rho)$ and a field gradient $\partial B_x/\partial z = -I\cos\omega_a t/(2\pi\rho^2)$ where $\rho$ is the distance from the wire to the proton. This oscillating field gives rise to a coupling which has the form of Eq. (28). If the resonance condition $\omega_a = \omega_o - \omega_z$ is satisfied, this coupling takes the form of Eq. (27) with $\Omega_1 = g_p \mu_B \mu_o I z_o/(4\pi\hbar\rho^2)$ where $z_o$ is the zero-point motion for the proton (as in Sec. 2). We assume that the duration $t_{coupling}$ of this inhomogeneous field is adjusted to give a complete transfer $|\downarrow\rangle|0\rangle \rightarrow |\uparrow\rangle|1\rangle$ ($\Omega_1 t_{coupling} = \pi/2$). We now couple the axial modes for a time $t_{exch}$ so that if the proton was in the $|n=1\rangle$ state after the last step, the $^9$Be$^+$ ion is now in the $|n=1\rangle$ state. Finally, the $|n=1\rangle$ motional state is detected on the $^9$Be$^+$ ion using the methods as discussed above and in Sec. 3. If $\omega_m$ is nonresonant, the modes remain in the $|n=0\rangle$ state.

If the proton is initially in the $|\downarrow\rangle$ state and $\omega_m$ is resonant, then the field at frequency $\omega_a = \omega_o - \omega_z$ has no effect, and the ion remains in the $|n=0\rangle$ motional state and gives a false "no signal." However, when $\omega_m$ is reapplied the $|n=1\rangle$ state is produced giving a signal. After reinitializing (that is, preparing both ions in their axial ground states), signals are always produced if $\omega_m$ remains resonant. If $\omega_m$ is nonresonant and the proton initial state is $|\uparrow\rangle$, no signals are ever produced. However if $\omega_m$ is nonresonant and the proton initial state is $|\downarrow\rangle$, we produce a false signal. After reinitializing and repeating the experiment, we produce no signals. This is true on subsequent tries as long as $\omega_m$ remains nonresonant. Therefore to reduce the effects of false signals, we should repeat each try several times for each value of $\omega_m$ and discard the first measurement. Interlaced with measurements of the proton spin flip frequency, we measure the $^9$Be$^+$ spin flip frequency by driving with resonant rf radiation and using the techniques outlined in Sec. 3 for detection. This allows us to monitor and correct for magnetic field drifts on a fairly short time scale.

The antiproton/$^9$Be$^+$ comparison is accomplished similarly except for the important difference that axial potentials must be trapping for one species and nontrapping for the other at a given location. Apparently, the biggest liability in the scheme presented here is the same as that of the proposal of Quint and Gabrielse [322], namely, the weakness of the parametric coupling between proton magnetic moment and axial motion. For $\omega_z = 1$ MHz, we find $z_o$(proton) = 71 nm. With I = 1 A, and $\rho = 1$ mm, we find $\Omega_1/2\pi \simeq 0.15$ Hz. For these same conditions and d = 0.5mm, we have $t_{exch} \simeq 27$ ms. At a field of 5 T, $\omega_o/2\pi \simeq 106$ MHz. At very high resolutions we want resonance linewidths of less than 1 Hz which, in turn, requires long resonance times for the proton spin flip. Therefore the long times required for $t_{coupling}$ need not be a serious liability.

We have not considered details of proton, antiproton, or $^9$Be$^+$ transfer into the trap, however this might be accomplished by adapting a scheme similar to that described in Sec. 4.1. The trapping arrangement we show in Fig. 12 is essentially the same as the coupled trap idea of Ref. [48]. By removing the common endcap in the coupled-trap scheme, we arrive at the situation depicted in Fig. 12. In either case, we can show that the coupling between ions is approximately given by $q_1 q_2 z_1 z_2 / 2\pi\epsilon_o d^3$ where d is the overall distance between ions 1 and 2. The potential advantage of the scheme described here is that d can probably be made smaller than in a coupled trap, thereby reducing $t_{exch}$. We have assumed the use of $^9$Be$^+$ ions, but many other ions would work. A potential advantage of $^9$Be$^+$ ions is that the axial potential wells required to make $\omega_1 = \omega_2$ are not as different as for other choices of ion 2. In a similar spirit, we could measure the proton (or antiproton) magnetic moment in terms of other atomic parameters. For example, the proton spin flip frequency vs. $^9$Be$^+$ spin flip frequency could be combined with a separate measurement of the $^9$Be$^+$ spin flip frequency compared to its cyclotron frequency to yield a measurement of the proton spin flip frequency to $^9$Be$^+$ cyclotron frequency in the same magnetic field. Coupled with an accurate value of the electron to $^9$Be$^+$ mass ratio, these measurements yield an accurate value of $g_p$. An important systematic effect to consider in these measurements is magnetic field inhomogeneity. Field homogeneity could be checked to high accuracy by moving the $^9$Be$^+$ ion to various locations in the trap, thereby mapping the magnetic field. Equally important in the proton/antiproton comparison is to insure that the proton and antiproton are in the same location. It appears this could be accomplished by insuring the $^9$Be$^+$ is always at the same location (using optical means) and adjusting the trap potentials to always yield the same values of $\omega_z$ and $\delta\omega$. By this method, it appears that accuracy significantly better that 1 part in $10^9$ could be achieved.

6.4. Quantum state manipulation of mesoscopic mechanical resonators

Much of what we have discussed concerns manipulation of the mechanical oscillation of atoms or atomic ions. It is perhaps interesting to speculate on the possibility of applying similar techniques to the manipulation of macroscopic or mesoscopic mechanical resonators in the quantum regime [119,120]. If studies can be performed at the quantum level, new sensors at the single phonon level could be built [324]. One approach is to make the mechanical resonators small enough and the temperature low enough that $\hbar\omega > k_B T$ [325]; this approach may obviate the idea discussed here. The idea we examine here is an extension of the idea of coupling the oscillatory motion of two ions together; here we consider coupling the motion of an ion (or the COM mode of a collection of ions) to a single mode of mesoscopic mechanical resonator. Although current technology appears to prohibit performing such experiments at the quantum level, the development of mesoscopic resonators is rapid and such experiments may be possible in the future. Similar considerations regarding coupling of ions to piezoelectric resonators were discussed in Ref. [48].

To be specific, we consider the situation sketched in Fig. 13. The mechanical resonator is assumed to be a silicon beam resonator, fixed at both ends; we will assume the conditions realized in the experiments of Cleland and Roukes [325]. The beam resonator has a length D, thickness in the z direction equal to $\Delta z$ and a thickness in the x direction (out of the plane of Fig. 13) equal to $\Delta x$. If we take the conditions of Fig. 3 of Cleland and Roukes [325] as a guide, we have D = 7.7 μm, $\Delta z$ = 0.8 μm, and $\Delta x$ = 0.33 μm. We assume that the beam is metalized near its center and can support a charge $q_1$. For simplicity, we will assume the metalization is confined to a spherical shell of radius $R_1$, but the exact geometry is not so important. Neglecting the dielectric effects of the Si beam (T = 4 K), the capacitance of the metallized sphere is approximately equal to $C_b = 4\pi\epsilon_o R_1$. The sphere can therefore support a charge $q_1 = C_b V_b$ where $V_b$ is the potential on the sphere. An atomic ion of charge $q_2$ and mass $m_2$ is trapped by a combination of electrodes at a distance d from the cantilever. One of the electrodes is the charged, metallized sphere; the other electrodes are indicated schematically as A, B, and C. To estimate the mass $m_2$ of the mechanical resonator we assume it has a mass equal to half of the beam's mass and that this mass is concentrated in the metallized sphere. For the conditions assumed here, and $\rho(Si) = 2.33$ g-cm$^{-3}$, we find $m_1 \simeq 2.4 \times 10^{-12}$ g. We arrange the trapping potentials so that the z-oscillation frequency of the ion is equal to the beam oscillation frequency; in this case, we realize two coupled oscillators as described in the previous section. If we take d = 5 μm, $V_b$ = 1000 V, $\bar{\omega}/2\pi$ = 70 MHz, $q_2$ = q(proton), $m_2$ = 9 u ($^9$Be$^+$), we find from Eq. (155) that $t_{exch} \simeq 6.4$ μs. To reliably work in the quantum regime of the mechanical oscillator, $t_{exch}$ must be smaller than $t^*$, the time for the mechanical oscillator to make a transition from its $|n=0\rangle$ ground state to the $|n=1\rangle$ state. If we assume the conditions of Cleland and Roukes [325] where T = 4 K, and Q = $2\times10^4$, Eq. (64) gives $t^* \simeq 0.0382$ μs. This is clearly too short to sympathetically cool the beam resonator's mode to the ground state, although some cooling could be achieved as outlined by in Ref. [48]. In the future, it might be possible to make the ratio $t^*/t_{exch}$ larger than 1, perhaps through higher Q's and lower ambient temperatures (or perhaps if more exotic ion species with $Z_2 \gg 1$ become available (Sec. 5.2), it may be possible to manipulate the quantum motion of mesoscopic mechanical resonators by these techniques.

7. Summary/conclusions

We have attempted to identify some of the important practical effects which must be taken into account in order to create arbitrary, entangled quantum states of trapped ions. We have taken a "passive stabilization" approach in that we try to anticipate and suggest ways to guard against the physical effects causing decoherence. Ultimately, complicated manipulations, such as lengthy quantum computations, are expected to benefit from and/or require some form of active error correction. Indeed, some of the near-term future experiments will probably demonstrate some of these schemes. In section IV, we have listed some of the potential sources of decoherence in trapped ion experiments; here we speculate on what appear to be the most important of these.

Motional decoherence is discussed in Sec. 4.1. In the NIST single $^9$Be$^+$ ion experiments [17,21,45,47,131,132,211], heating appears to be the most important source of decoherence, primarily because, at the present time, its source is still unidentified. Various possibilities were discussed and, although the heating may be caused by some fundamental effect, we speculate that it is probably caused by some, as yet undetected, source of added electronic noise. (Because the current experiments employ electronic filtering at the motional frequencies, direct observation of fluctuating potentials on the electrodes has been precluded.) Future experiments will be able to resolve this. Moreover, once this source of noise is understood, the ion becomes an extremely sensitive detector of noise potentials appearing across the electrodes. In any case, it will probably be desirable to eventually operate the ion trap at cryogenic temperatures in order to, for example, reduce the effects of ion loss due to background gas collisions. A cryogenic environment will have the added benefit of reducing possible sources of thermal electronic noise and associated heating.

Various sources of internal-state decoherence are discussed in Sec. 4.2. In the current experiments, decoherence is dominated by fluctuations in magnetic fields acting on qubit transitions which are strongly field dependent. It appears that this problem can be highly suppressed by the use of magnetic shielding and, eventually, use of qubit transitions which are insensitive to magnetic field to first order. We expect that internal state decoherence will be negligible compared to motional decoherence.

Decoherence induced during application of the logic pulses (Sec. 4.4) may be the most troublesome. Many of the sources of this type of decoherence are primarily technical, for example, caused by intensity fluctuations in the laser pulses which induce transitions, or stray light impinging on ions not directly addressed. The more fundamental causes of decoherence are (1) coupling to internal or motional states other than the intended ones (spectator level problem), (2) coupling to unintended motional modes (cross-mode coupling), and (3) fluctuations in the Rabi rates due to excitation of the 3L-1 unused motional modes (Debye-Waller factors). The first two of these effects appear to be a question of speed. Coupling to unwanted spectator modes and cross-mode coupling can always be avoided by making the operations slow enough that the extraneous couplings are removed by spectral selection. This has the negative effect of allowing more time for decoherence and increasing the required time for a given computation. Fluctuations caused by fluctuating Debye-Waller factors are, in principle, reduced as the number of ions increases because of the averaging effects of many modes (Eq. (128)). However, it is also likely that excitation of these modes is harder to avoid as the number of ions (and unwanted spectator modes) increases. It will therefore be desirable to laser-cool all modes to the zero-point state.

In a single-ion trap, as the number of ions increases, it will become increasingly difficult to avoid these three types of effects. Therefore, some sort of multiplexing scheme will be necessary when large numbers of qubits are involved. In Sec. 5.1, we have presented one possible solution where the ions are broken up into smaller numbers of independent groups or registers. The ions are then connected by moving ions around between registers. It may also be advantageous to multiplex quantum information within multiple internal states of ions; this is briefly described in Sec. 5.2.

Stimulated-Raman transitions between long-lived qubit states (such as ground-state hyperfine levels) appear to offer significant advantages over single-photon optical transitions. Single-photon transitions require high laser frequency stability, whereas stimulated-Raman transitions require only high relative frequency stability between the Raman beams, which is technically easier to accomplish. Stimulated-Raman transitions also provide the ability to select the magnitude and direction of the effective k vector ($\vec{k}_{eff}$) by choosing different directions for the beams. This has the advantage of, for example, allowing $\vec{k}_{eff}$ to be parallel to the axis of ions in a linear trap, thereby suppressing coupling to radial modes, while still allowing the Lamb-Dicke parameter to be controlled by adjusting the angle of the beams. At the same time, spatial selection of ions along the axis of the trap can be good since each Raman beam can be at a relatively steep angle relative to the trap axis. With single photon transitions, selection of ions and modes by appropriately choosing the direction of $\vec{k}$ can only be obtained for the radial modes in a linear trap. Stimulated-Raman transitions have the potential disadvantage of inducing significant AC-Stark shifts (Sec. 4.4.3). However, for anticipated operating conditions, the effects of Stark shifts can be suppressed relative to the effects of laser amplitude fluctuations, which affect both single-photon and stimulated-Raman transitions.

Decoherence in the ion trap system can probably be substantially reduced over what has been obtained in experiments so far. How far this reduction can be carried is an issue which must be resolved experimentally. Decohering effects may eventually be controlled to such a level that fault-tolerant error correction schemes might be employed to achieve computations of arbitrary length. This may only require a single operation fidelity of $10^{-5}$ [326]. If this condition is met, speed will become an important issue because of the potentially large amount of overhead (increase in number of required qubits and operations in fault tolerant schemes). As discussed in Sec. 4.4.6, the Rabi rate for any operation is limited to approximately the motional mode frequency. In principle, mode frequencies can be substantially increased beyond what is currently achieved ($\simeq$ 10 MHz). An obvious direction to pursue is to make smaller traps with higher trapping potentials; however, this aggravates the problem of addressability, and will increase the coupling of the ions to the electrodes thereby increasing decoherence. The optimum conditions must, again, be resolved experimentally.

8. Appendices

Appendix A: Entangled states and atomic clocks

We compare the use of entangled vs. nonentangled states in an atomic clock under conditions and constraints different than those considered in Sec. 3.4. We assume that the resources available are a given number of atoms L and a total observation time which is longer than any other time scale in the problem. We assume that decoherence of the internal states is negligible during the Ramsey free precession time $T_R$. To make an atomic clock, we want to steer, or "lock," a reference oscillator ("local oscillator") to the atoms' resonance. Typically, we can find a local oscillator whose rms frequency fluctuations $\Delta\omega_{LO}(\tau)$ over short averaging times $\tau$ are smaller than the intrinsic fluctuations given by projection noise. However, for long times, the frequency fluctuations of the local oscillator are worse than those given by projection noise. (If this were not the case, the local oscillator would be a better clock, obviating the need for the atoms.) By measuring the atomic populations after each application of the Ramsey radiation [3,9,100], we can generate an error signal which steers the local oscillator to the center of the atomic resonance with a servo time constant $\tau_s$. The servo can make a correction after a few independent measurements on the atoms; for simplicity, we assume $\tau_s \propto T_R$ (Eq. (A4) below). To make the error signal as large as possible, we want the atomic linewidth $\Delta\omega_a$ as small as possible. Since the atomic linewidth can be expressed as $\Delta\omega_a = \pi/(L^{(2\epsilon-1)}T_R)$ [3,9,100], we therefore want $T_R$ as large as possible. However, if we make $T_R$ too large, the local oscillator fluctuations $\Delta\omega_{LO}(T_R)$ will be larger than $\Delta\omega_a$, thereby giving no useful signal. This is the constraint which tells us whether entangled or nonentangled states are more useful.

To analyze this problem, we make the following assumptions: (1) We assume the (free-running or unlocked) local oscillator has fluctuations over averaging time $\tau$ equal to

$$\Delta\omega_{LO}(\tau) = C\tau^n. \tag{A1}$$

This is assumed to hold over the range of values of $\tau$ which include $T_R$ and $\tau_s$. We will consider only values of $n \geq -\frac{1}{2}$, since, for $n < -\frac{1}{2}$, the local oscillator will better than the atoms for $\tau$ sufficiently long. (2) We assume the intrinsic atomic clock stability, limited by projection noise, is given by (Sec. 3.4)

$$\Delta\omega_{meas.}(\tau) = \frac{L^{-\epsilon}}{\sqrt{T_R \tau}}, \tag{A2}$$

where $\epsilon = \frac{1}{2}$ for nonentangled states and $\epsilon = 1$ for entangled states. (3) For a given servo time constant $\tau_s$, we have

$$\Delta\omega_{LO}(\tau_s) = K_1 \Delta\omega_{meas.}(\tau_s). \tag{A3}$$

Normally, in the case where $\Delta\omega_{LO} < \Delta\omega_{meas.}$ for short times $\tau$, we would think of adjusting $\tau_s$ so that $K_1 \leq 1$; that is, the local oscillator is locked to the atomic resonance at or before the time $\tau$ when the local oscillator fluctuations become worse than the projection noise. However, when $K_1 > 1$, the locked local oscillator stability can eventually reach the stability given by projection noise (approaching it as $1/\tau$) given adequate servo gain [327-329]. Therefore we will allow $K_1$ to be larger than 1.

Along with these assumptions, we impose two constraints: (1) The servo time constant $\tau_s$ must be longer than the sampling time $T_R$. (We will assume $T_R$ is much larger than the dead time, that is, the time for optical pumping, detection, etc.) This is expressed by the condition

$$\tau_s = K_2 T_R, \qquad K_2 > 1. \tag{A4}$$

For entangled states, each measurement gives one of two possible values (Sec. 3.4). Therefore, $K_2$ must be larger for entangled states than nonentangled states. (2) For $\tau \simeq T_R$, the atomic linewidth must be greater than the local oscillator fluctuations or else the error signal used to correct the local oscillator frequency is ambiguous. This is expressed by the condition

$$\Delta\omega_a(T_R) = \pi/(L^{(2\epsilon-1)}T_R) = K_3 \Delta\omega_{LO}(T_R), \qquad K_3 > 1. \tag{A5}$$

Eq. (A5) gives the sampling time $T_R$ as a function of $K_3$

$$T_R = \left(\frac{\pi}{CK_3 L^{2\epsilon-1}}\right)^{\frac{1}{n+1}}. \tag{A6}$$

Plugging this into Eq. (A2), we find that the stability of the locked local oscillator at long times $\tau$ is

$$\Delta\omega_{meas.} = \left[\frac{CK_3}{\pi}\right]^{\frac{1}{2(n+1)}} L^{-\frac{(n\epsilon+\frac{1}{2})}{n+1}} \frac{1}{\sqrt{\tau}}. \tag{A7}$$

This is the main result of Appendix A. From this expression, we see that $\Delta\omega_{meas.}$ is smaller for entangled states when $n > 0$; however, the gain is not as significant as when we assumed $T_R$ to be fixed. We also see that $\Delta\omega_{meas.}$ is smaller for nonentangled states when $n < 0$. These results are due to the constraint expressed in Eq. (A5). From the above, we also find

$$K_1 = \frac{\pi K_2^{n+\frac{1}{2}}}{K_3} L^{1-\epsilon} \propto K_2^{n+\frac{1}{2}} L^{1-\epsilon}. \tag{A8}$$

For nonentangled states on large numbers of atoms, $K_1$ and $T_R$ are both much larger than for entangled states, and a very long time may be required to achieve the intrinsic atomic clock stability given by Eqs. (A2) and (A7). Therefore, as a practical issue, we may wish to constrain $K_1$ to be equal to 1. In that case we find, from Eqs. (A3) and (A5) two values of $T_R$. We must take the smaller of these, which results from the value derived from Eq. (A3) (with $K_1 = 1$). Plugging this value of $T_R$ into Eq. (A2), we find the stability of the locked oscillator to be

$$\Delta\omega_{meas.} = \left[CK_2^{n+\frac{1}{2}}\right]^{\frac{1}{2(n+1)}} L^{-\epsilon\left(\frac{2n+1}{2n+2}\right)} \frac{1}{\sqrt{\tau}} . \tag{A9}$$

For large L, entangled states will give a smaller value of $\Delta\omega_{meas.}$ for all values of n > -½; however this result is not fundamental and simply comes from the constraint that $K_1$ is equal to 1.

Appendix B : Master equation for the density matrix of a radiatively damped harmonic oscillator

Equation (62) is the master equation for the density matrix r of a single harmonic oscillator, including radiative damping terms. This equation is equivalent to the following system of coupled first-order differential equations for the diagonal and off-diagonal number-state matrix elements of r:

$$\begin{aligned}\langle n|\dot{\rho}|n\rangle \equiv \dot{\rho}_{nn} &= \gamma(\bar{n}+1)(n+1)\rho_{n+1\,n+1} - \gamma(2n\bar{n}+n+\bar{n})\rho_{nn} + \gamma\bar{n}n\rho_{n-1\,n-1} \\ \langle m|\dot{\rho}|n\rangle \equiv \dot{\rho}_{mn} &= \gamma(\bar{n}+1)\sqrt{(m+1)(n+1)}\rho_{m+1\,n+1} + \gamma\bar{n}\sqrt{mn}\rho_{m-1\,n-1} \\ &\quad -\frac{1}{2}\gamma[2\bar{n}(m+n+1)+(m+n)]\rho_{mn}\end{aligned} \tag{B1}$$

Some special cases of these equations were given in Eq. (63), where they were used to estimate the time for an ion to make a transition from the state $|0\rangle$. Cohen-Tannoudji [330] treated a similar system, in which the incoherent excitation was absent ($\bar{n} = 0$), but a monochromatic perturbation was present. The time derivative of the thermal average of the number of vibrational quanta is

$$\begin{aligned}\frac{d\langle\hat{n}\rangle}{dt} \equiv \sum_{n=0}^{\infty} n\dot{\rho}_{nn} &= \gamma(\bar{n}+1)\sum_{n=0}^{\infty} n(n+1)\rho_{n+1\,n+1} - \gamma\bar{n}\sum_{n=0}^{\infty} n\rho_{nn} \\ &\quad -\gamma(2\bar{n}+1)\sum_{n=0}^{\infty} n^2\rho_{nn} + \gamma\bar{n}\sum_{n=1}^{\infty} n^2\rho_{n-1\,n-1}\end{aligned} \tag{B2}$$

The sums can be simplified to

$$\sum_{n=0}^{\infty} n(n+1)\rho_{n+1\,n+1} = \sum_{n=0}^{\infty} (n+1-1)(n+1)\rho_{n+1\,n+1}$$
$$= \sum_{n=0}^{\infty} (n+1)^2 \rho_{n+1\,n+1} - \sum_{n=0}^{\infty} (n+1)\rho_{n+1\,n+1} \quad \text{(B3)}$$
$$= \sum_{n=0}^{\infty} n^2 \rho_{nn} - \sum_{n+0}^{\infty} n\rho_{nn}$$

and

$$\sum_{n=1}^{\infty} n^2 \rho_{n-1\,n-1} = \sum_{n=1}^{\infty} [(n-1)^2 + 2n - 1]\rho_{n-1\,n-1}$$
$$= \sum_{n=1}^{\infty} [(n-1)^2 + 2(n-1) + 1]\rho_{n-1\,n-1}$$
$$= \sum_{n=0}^{\infty} n^2 \rho_{nn} + 2\sum_{n=0}^{\infty} n\rho_{nn} + \sum_{n=0}^{\infty} \rho_{nn} \quad \text{(B4)}$$
$$= \sum_{n=0}^{\infty} n^2 \rho_{nn} + 2\sum_{n=0}^{\infty} n\rho_{nn} + 1 \;.$$

Collecting terms, we have

$$\sum_{n=0}^{\infty} n\dot{\rho}_{nn} = [\gamma(\bar{n}+1) - \gamma(2\bar{n}+1) + \gamma\bar{n}]\sum_{n=0}^{\infty} n^2 \rho_{nn} + [-\gamma(\bar{n}+1) - \gamma\bar{n} + 2\gamma\bar{n}]\sum_{n=0}^{\infty} n\rho_{nn} + \gamma\bar{n}$$
$$= -\gamma \sum_{n=0}^{\infty} n\rho_{nn} + \gamma\bar{n}, \quad \text{(B5)}$$

or

$$\frac{d\langle \hat{n} \rangle}{dt} = -\gamma \langle \hat{n} \rangle + \gamma \bar{n} \;. \quad \text{(B6)}$$

In the steady state, $\langle \hat{n} \rangle = \bar{n}$, independent of $\gamma$ (see also, Ref. [121], p. 97).

Acknowledgements

We gratefully acknowledge the support of the National Security Agency, the Army Research Office, and the Office of Naval Research. We thank C. Myatt and C. Wood for comments and suggestions on the manuscript. We acknowledge useful discussions with P. Bardroff, R. Blatt, I. Cirac, T. Darling, L. Davidovich, A. Despain, D. DiVincenzo, A. Ekert, B. Esry, N. Gisin, S. Haroche, M. Holland, M. Holzscheiter, R. Hughes, D. James, J. Kimble, P. Knight, S. Lloyd, G. Milburn, J. Preskill, W. Schleich, A. Steane, W. Vogel, P. Zoller, and W. Zurek.

***About the authors:*** *David Wineland, Christopher Monroe, Wayne Itano, and Dawn Meekhof are staff physicists in the Time and Frequency Division of NIST, Boulder. Brian King is a graduate student in physics at the University of Colorado. Diedrich Leibfried was a guest researcher during the preparation of the manuscript; he is currently a physicist at the Institute for Experimental Physics in Innsbruck, Austria.*

Figure Captions

FIG. 1. The upper part of the figure shows a schematic diagram of the electrode configuration

for a linear Paul-rf trap (rod spacing ≈ 1 mm). The lower part of the figure shows an image of a string of $^{199}$Hg$^+$ ions, illuminated with 194 nm radiation, taken with a UV-sensitive, photon counting imaging tube [74]. The spacing between adjacent ions is approximately 10 μm. The "gaps" in the string are occupied by impurity ions, most likely other isotopes of Hg$^+$, which do not fluoresce because the frequencies of their resonant transitions do not coincide with those of the 194 nm $^2S_{1/2} \rightarrow ^2P_{1/2}$ transition of $^{199}$Hg$^+$.

FIG. 2. Experimental plot of the probability $P_\downarrow(t)$ of finding a single $^9$Be$^+$ ion in the $|\downarrow\rangle$ state after first preparing it in the $|\downarrow\rangle|0\rangle$ state and applying the first blue side band coupling (Eq. (16), for $\delta = +\omega_z$) for a time $\tau$. If there were no decoherence in the system, $P_\downarrow(t)$ should be a perfect sinusoid as indicated in Eq. (26). Decoherence causes the signal to decay as discussed in Sec. 3.2.1. The solid line is a fit to an exponentially decaying sinusoid as indicated in Eq. (43). Each point represents an average of 4000 observations [21].

FIG. 3. Schematic diagram relevant to stimulated-Raman transitions between internal states $|\downarrow\rangle$ and $|\uparrow\rangle$. Two plane wave radiation fields couple to a third state $|3\rangle$. The radiation fields are typically at laser frequencies; they are characterized by frequencies and wavevectors $\omega_{Li}$ and $\vec{k}_i$, i ∈ {1,2}. The couplings are typically described by electric dipole matrix elements. For simplicity, we assume field 1 only couples states $|\downarrow\rangle$ and $|3\rangle$; and field 2 only couples states $|\uparrow\rangle$ and $|3\rangle$. In this diagram, we do not show the additional energy level structure of the 3L modes of motion.

FIG. 4. Schematic diagram relevant to laser cooling using stimulated-Raman transitions. In (a), we show that when $\omega_{L1} - \omega_{L2} = \omega_o - \omega_z$, stimulated-Raman transitions can accomplish the transition $|\downarrow\rangle|n\rangle \rightarrow |\uparrow\rangle|n-1\rangle$. In the figure, the transition for n = 2 is shown. In (b), spontaneous-Raman transitions, accomplished with radiation tuned to the $|\uparrow\rangle \rightarrow |3\rangle$ transition, pumps the atom back to the $|\downarrow\rangle$ state, thereby realizing the transition $|\uparrow\rangle|n-1\rangle \rightarrow |\downarrow\rangle|n-1\rangle$. When atomic recoil can be neglected, one application of steps (a) and (b) reduces the atom's motional energy by $\hbar\omega_z$ unless n = 0, in which case the atom is in it's motional ground state.

FIG. 5. Hyperfine levels of the 2s $^2S_{1/2}$ ground state of $^9$Be$^+$ in a weak magnetic field (not to scale). The energy levels are designated by horizontal lines. Above the lines, the levels are represented by atomic physics labels (F,$M_F$) where F is the total angular momentum (electron plus nuclear angular momentum) and $M_F$ is the projection of the angular momentum along the magnetic field axis. The separation of Zeeman substates in the different F manifolds is approximately equal to $0.7 \times 10^{10} B_o$ Hz where $B_o$ is expressed in teslas. The separation of the F = 1 and F = 2 manifolds is approximately 1.25 GHz at $B_o = 0$. For simplicity of notation, in most of the paper we make the identifications $|F=2,M_F=2\rangle \equiv |\downarrow\rangle$, $|1,1\rangle \equiv |\uparrow\rangle$, $|2,0\rangle \equiv |aux\rangle$.

FIG. 6. Schematic diagrams of the lumped circuit equivalents for the trap electrodes and trapped ion(s). (a) The left part of the diagram shows, schematically, the electrodes for a Paul trap with hyperbolic electrodes and a collection of trapped ions. On the right is shown the corresponding lumped circuit equivalent; $C_T$ represents the inter-electrode capacitance (the combined effects of the capacitances shown in the left part of the figure); r represents the resistive losses in the electrodes and connecting wires, $\ell_L$ and $c_L$ represent the equivalent inductance and capacitance for the COM mode of oscillation in the z (vertical) direction [103]. (b) A schematic diagram of the endcaps electrode for the trap of Jefferts, *et al.* [211] which was used in the NIST experiments (the ring is not shown). Induced currents in the z (vertical) direction are assumed to follow a path indicated by shading; the resistance in this path represents r in part (a) of the figure. (c) The rf potential between ring and endcaps electrode (or between pairs of rods as indicated in Fig. 1) is typically generated with a resonant rf step-up transformer. The resistance in this transformer can, in principle, couple to the ion motion as discussed in the text.

FIG. 7. Ramsey signal of the 2s $^2S_{1/2}$ ($M_I = -\frac{1}{2}$, $M_J = +\frac{1}{2}$) → 2s $^2S_{1/2}$ (-3/2, +½) hyperfine transition in $^9Be^+$ at a magnetic field of 0.8194 T). This resonance was obtained using a free precession time of 550 s. The data are the result of one sweep (one measurement per frequency point). The fluctuations in the data were due to the instability of the reference oscillator used to take the spectrum. These kinds of measurements indicated that the coherence time for superpositions between the two hyperfine states (which could be used as qubit levels) was longer than 10 minutes [76].

FIG. 8. Scheme for differential AC Stark shifting of neighboring ions. Equal intensity counter propagating beams 1 and 2 are centered on ions j and k, respectively. A fraction $\epsilon$ of the peak intensity $I_o$ of each beam is applied to the other ion. This results in a differential AC Stark shift of ions j and k, allowing the possibility of individually accessing the ions by tuning the frequency of the laser beams.

FIG. 9. Simplified energy level diagram for characterization of coupling to, and spontaneous emission from, off-resonant "spectator" levels. We assume coherent radiation is tuned near the transition frequency for the $|\downarrow\rangle|n\rangle \to |\uparrow\rangle|n'\rangle$ transition (for simplicity we have not shown the motional substructure for the $|\downarrow\rangle$ and $|\uparrow\rangle$ states). The state $|s\rangle$ is assumed to be the nearest spectator state from which off-resonant coupling can occur.

FIG. 10. Schematic diagram of an ion trap "super-register" containing few-ion accumulators in which logic operations are performed. We assume the trap is generically equivalent to the linear trap of Fig. 1. The rods of the trap in Fig. 1 are replaced by segmented electrodes in which adjacent segments are at the same rf potential but where the segments support different static potentials. This allows ions to be selectively moved around and in and out of ion accumulators. In the accumulator shown, a logic operation between ions k and m is indicated. The configuration of the laser beams is chosen to null the intensity on ion m while performing an operation on ion k. The beams must be reconfigured to perform an operation on ion m while leaving ion k unaffected. The auxiliary ion may be required to perform laser cooling each time new logic ions are moved into the accumulator.

FIG. 11. Hyperfine levels of the 2s $^2S_{1/2}$ ground state of $^9Be^+$ in a weak magnetic field (not to scale). The energy levels are designated by horizontal lines. Above the lines, the levels are represented by atomic physics labels (F,$M_F$) where F is the total angular momentum and $M_F$ is the projection of the angular momentum along the magnetic field axis. The separation of Zeeman substates in the different F manifolds is approximately equal to $0.7 \times 10^{10} B_o$ Hz where $B_o$ is expressed in teslas. The separation of the F = 1 and F = 2 manifolds is approximately 1.25 GHz at $B_o = 0$. For simplicity of notation, in Sec. 5.2, we make the identifications $|F, M_F\rangle = |2,2\rangle \equiv |0\rangle$, $|1,1\rangle \equiv |1\rangle$, $|2,1\rangle \equiv |0'\rangle$, $|1,0\rangle \equiv |1'\rangle$.

FIG. 12. Schematic diagram of a trap for simultaneous storage of two ions with different charge to mass ratios, $q_1/m_1$ and $q_2/m_2$. We assume the trap is generically equivalent to the linear trap of Fig. 1. The rods of the trap in Fig. 1 are replaced by segmented electrodes in which adjacent segments are at the same rf potential but where the segments support different static potentials in order to make a double well potential along z. Therefore in this figure, the view is from above, where we see two (of four) segmented electrodes which have replaced two of the rods in the trap of Fig. 1. On the left is a cross section of a wire through which an rf oscillating current is sent; this rod generates an rf magnetic field at the site of the ions as explained in the text.

FIG. 13. Schematic diagram of an ion coupled to mechanical resonator. The bridge-type cantilever is essentially the same as that reported by Cleland and Roukes [325], but is assumed to support a metallized sphere supporting charge $q_1$. An ion $q_2$ is confined by the potentials on the metallized sphere and additional electrodes shown schematically as A, B, and C. The Coulomb coupling between the charges provides the coupling between the two harmonic oscillators.

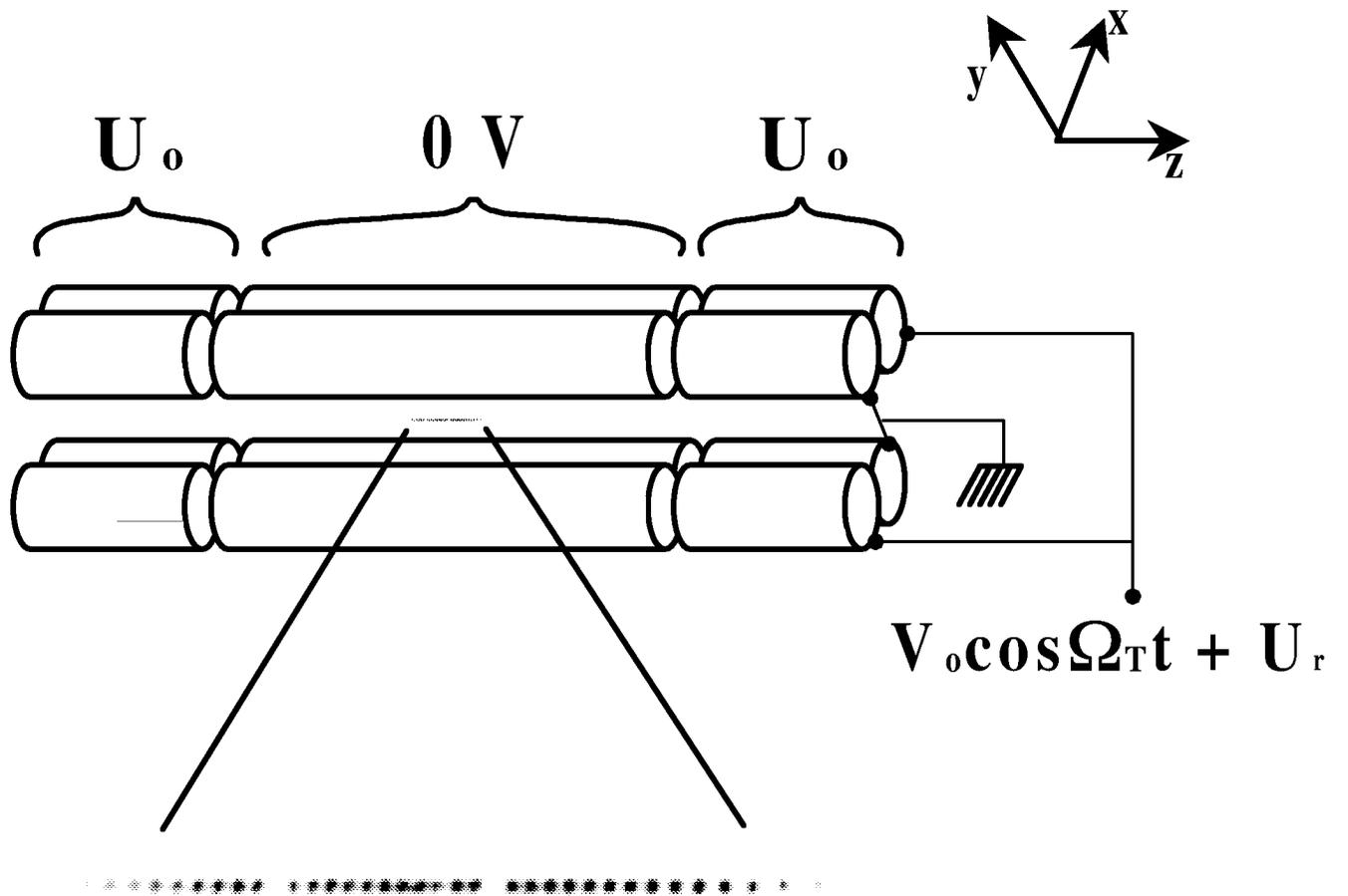

Fig. 1

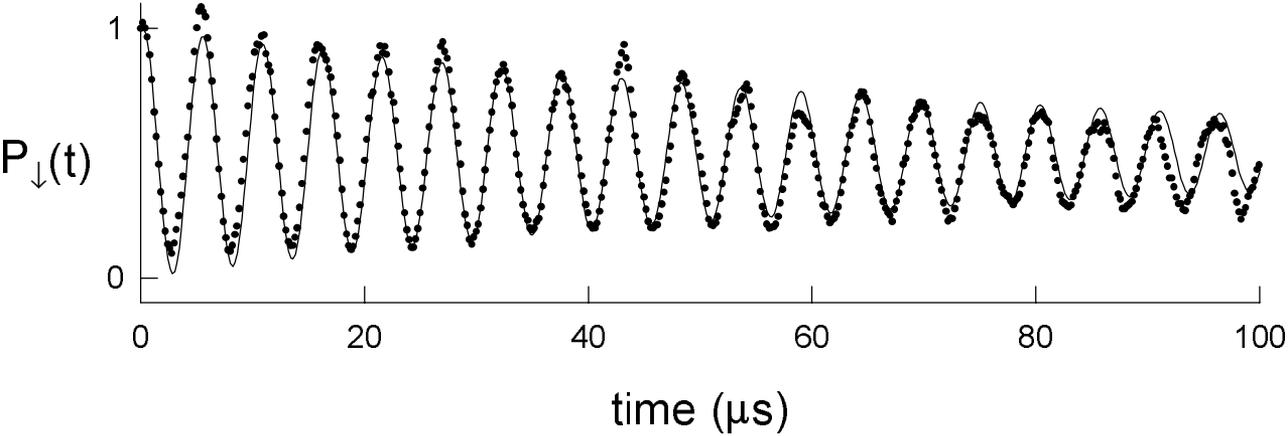

**Fig. 2**

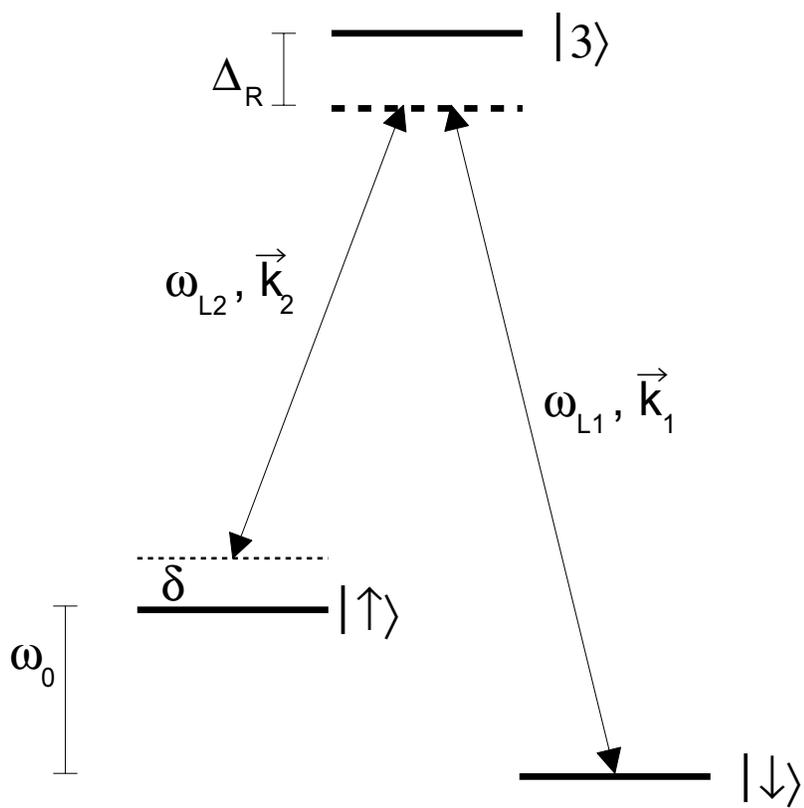

**Fig. 3**

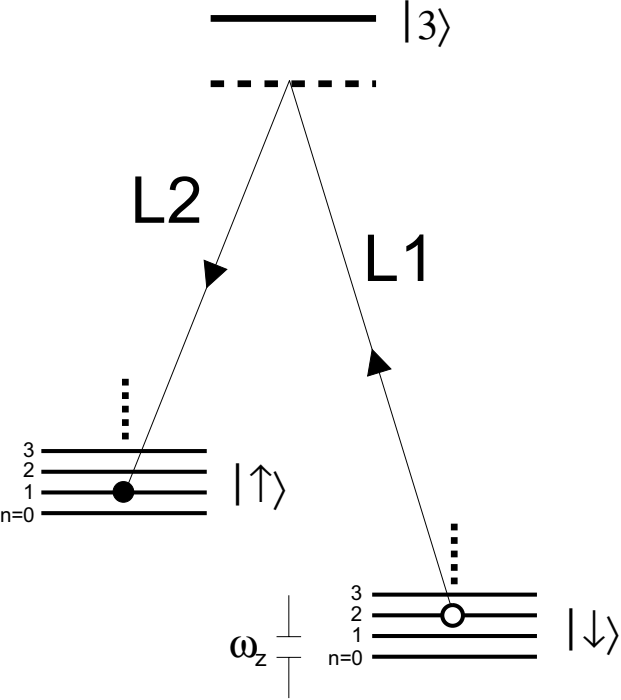 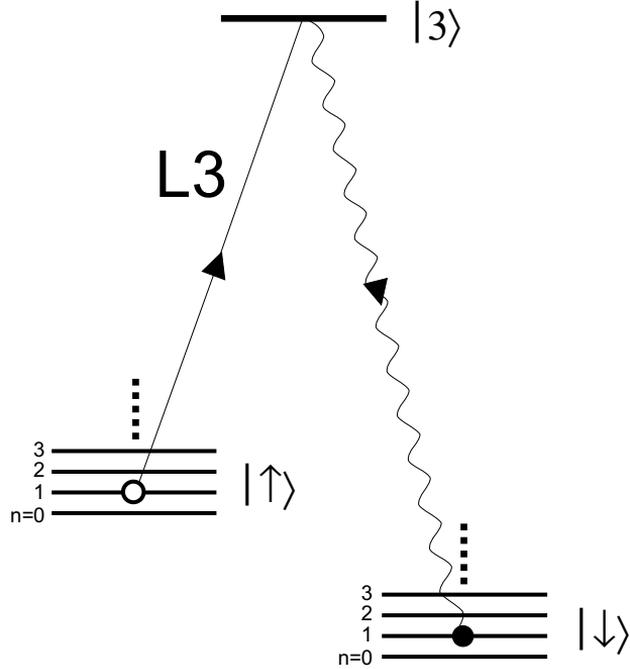

(a) stimulated Raman (Δn = -1)  (b) spontaneous Raman (Δn ≈ 0)

**Fig. 4**

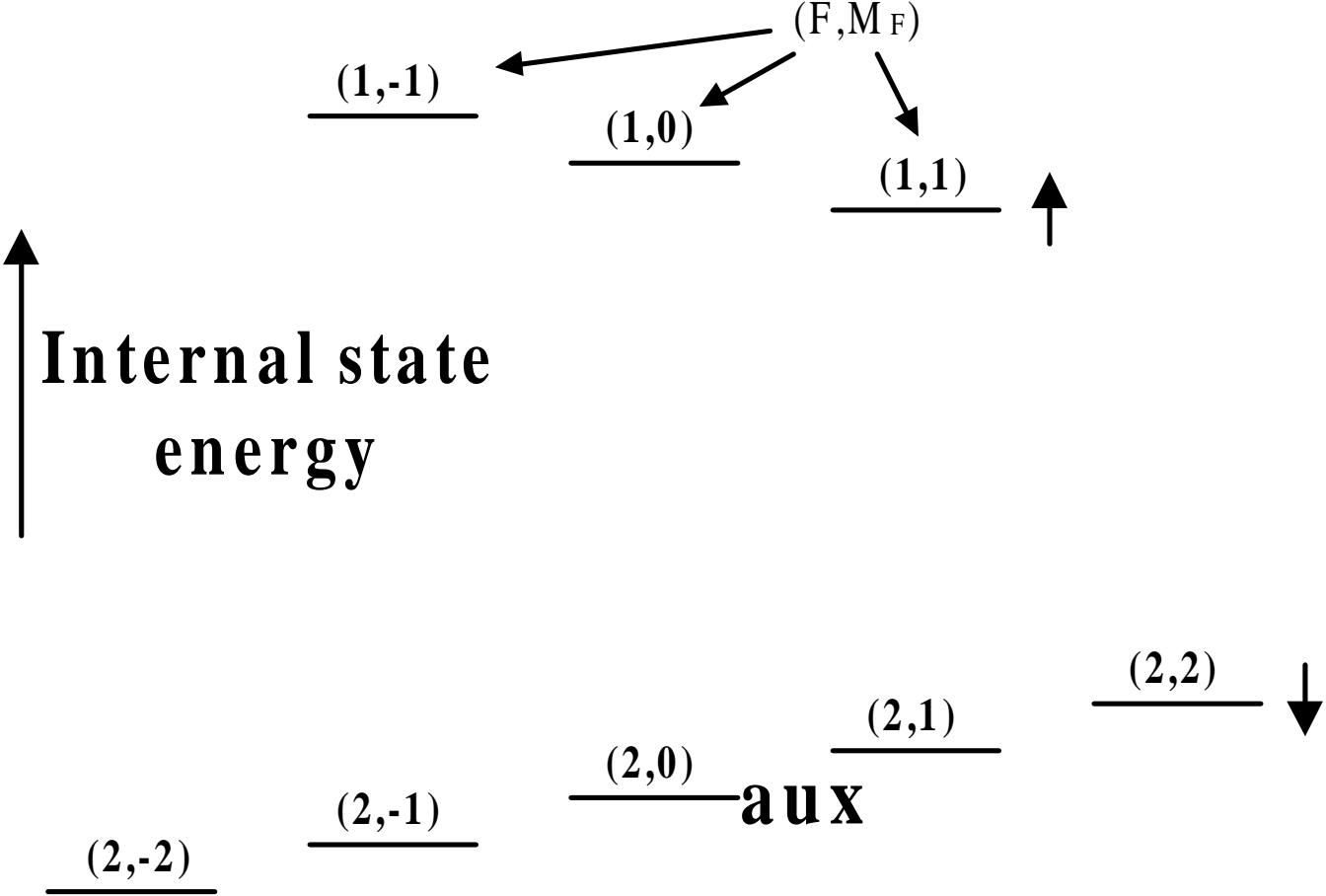

Fig. 5

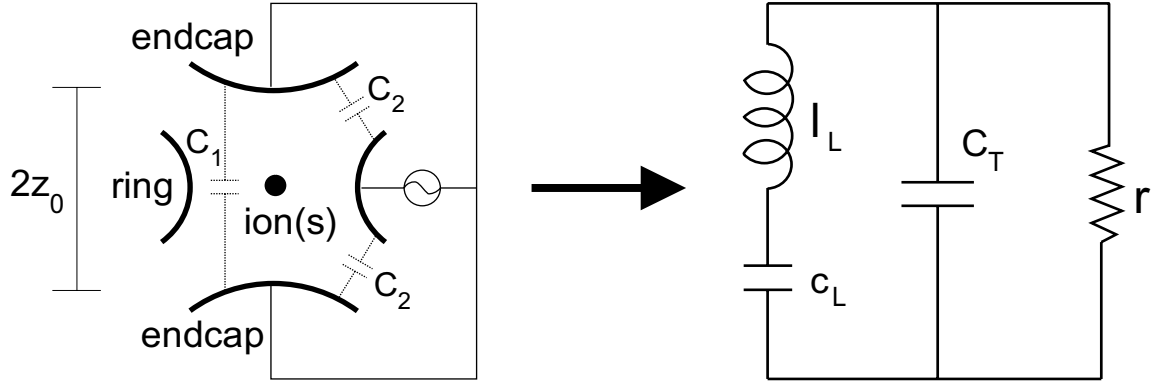

(a)

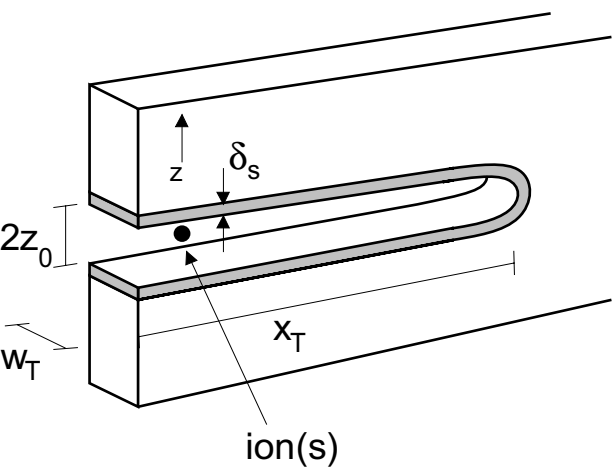

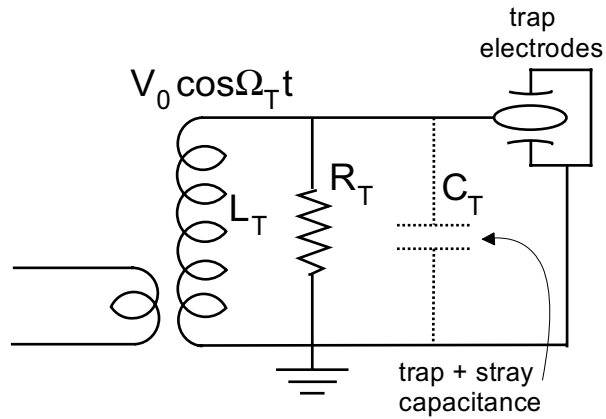

(b) (c)

**Fig. 6**

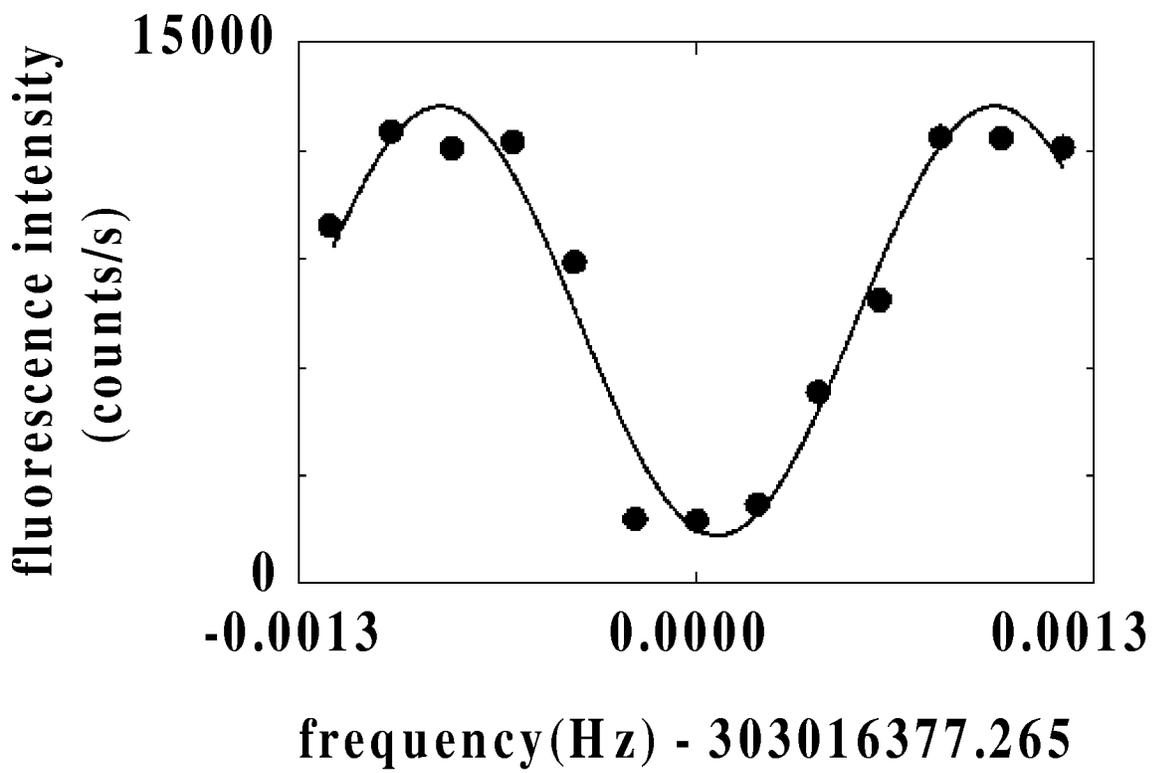

Fig. 7

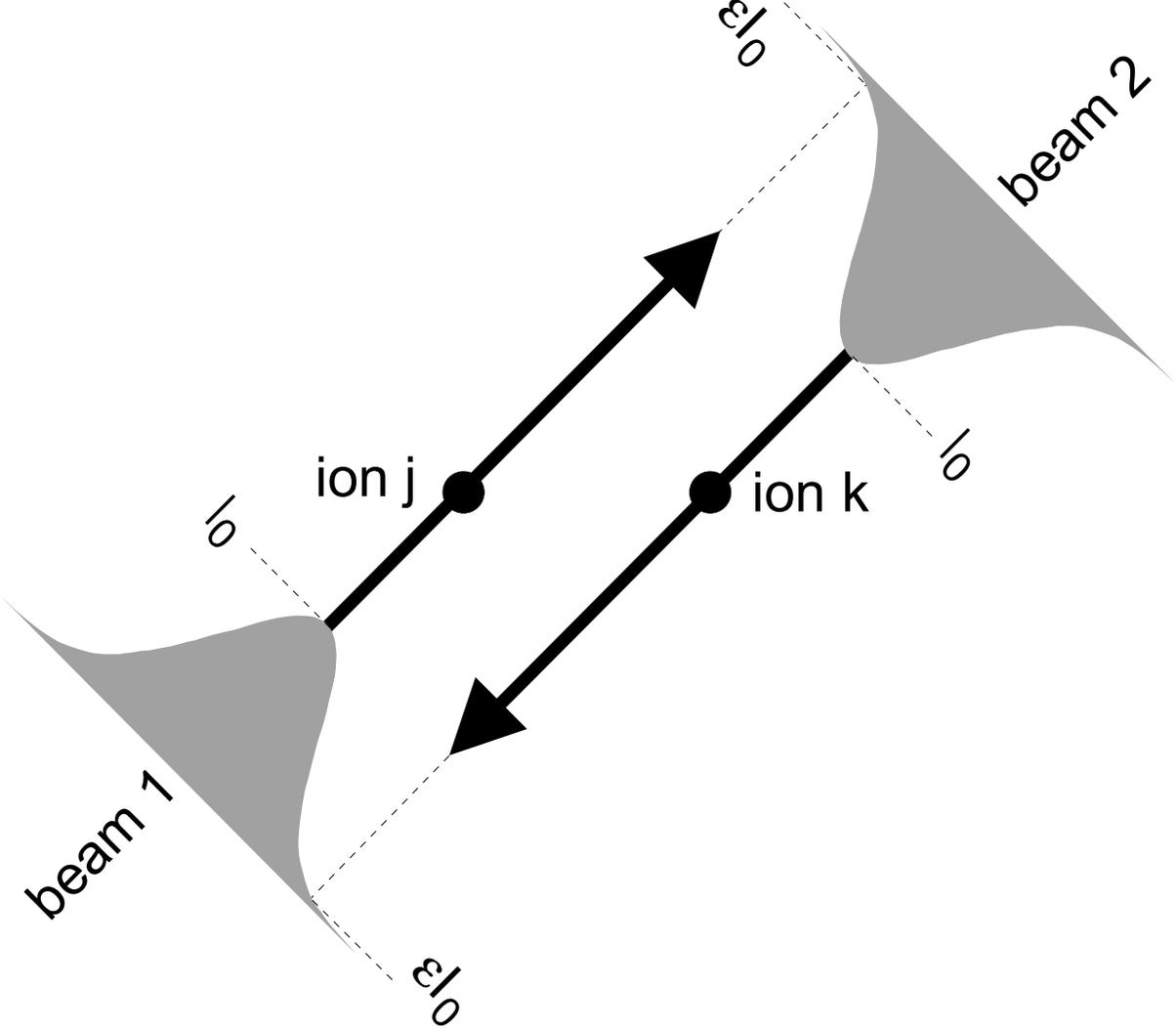

**Fig. 8**

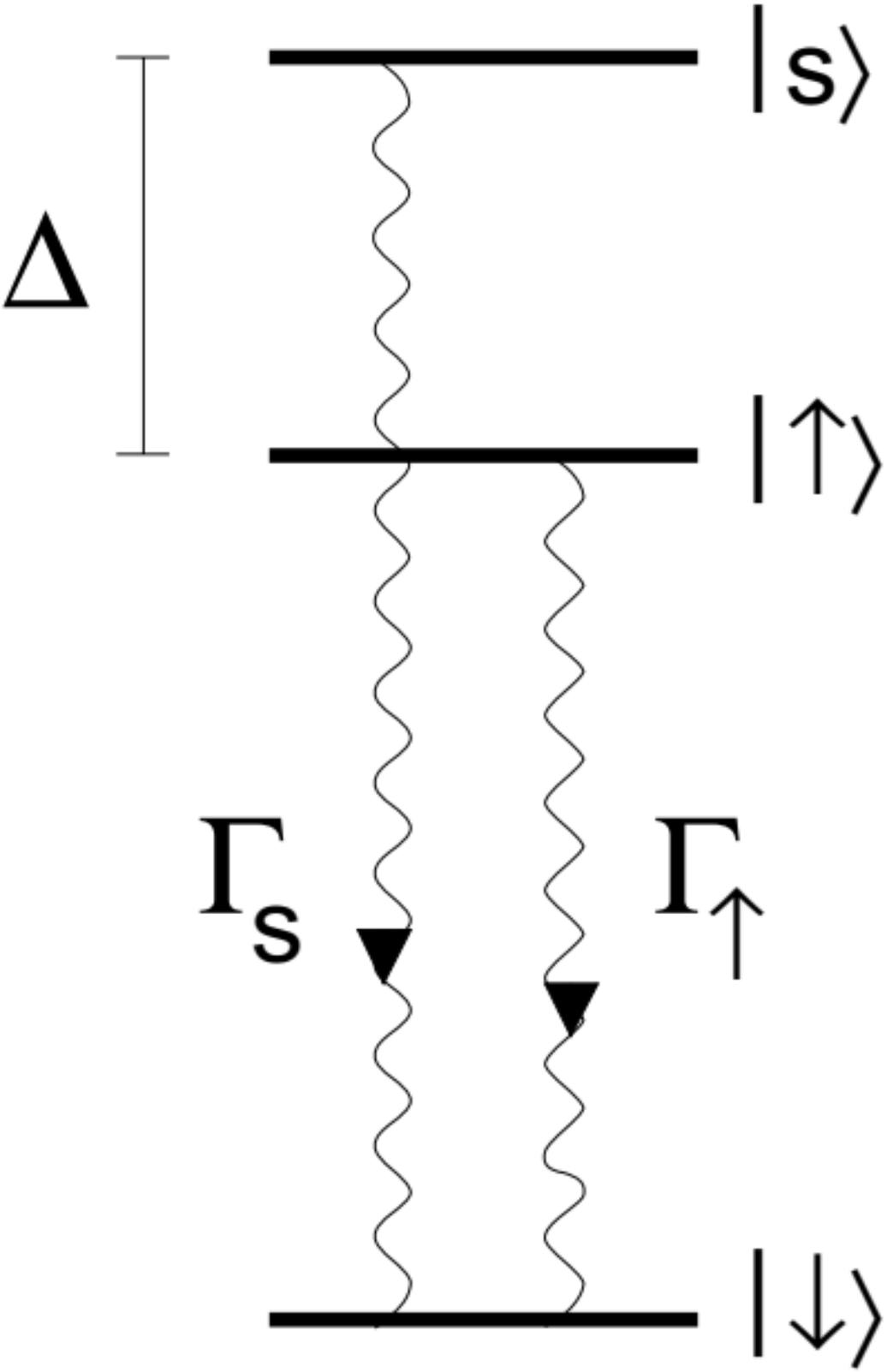

**Fig. 9**

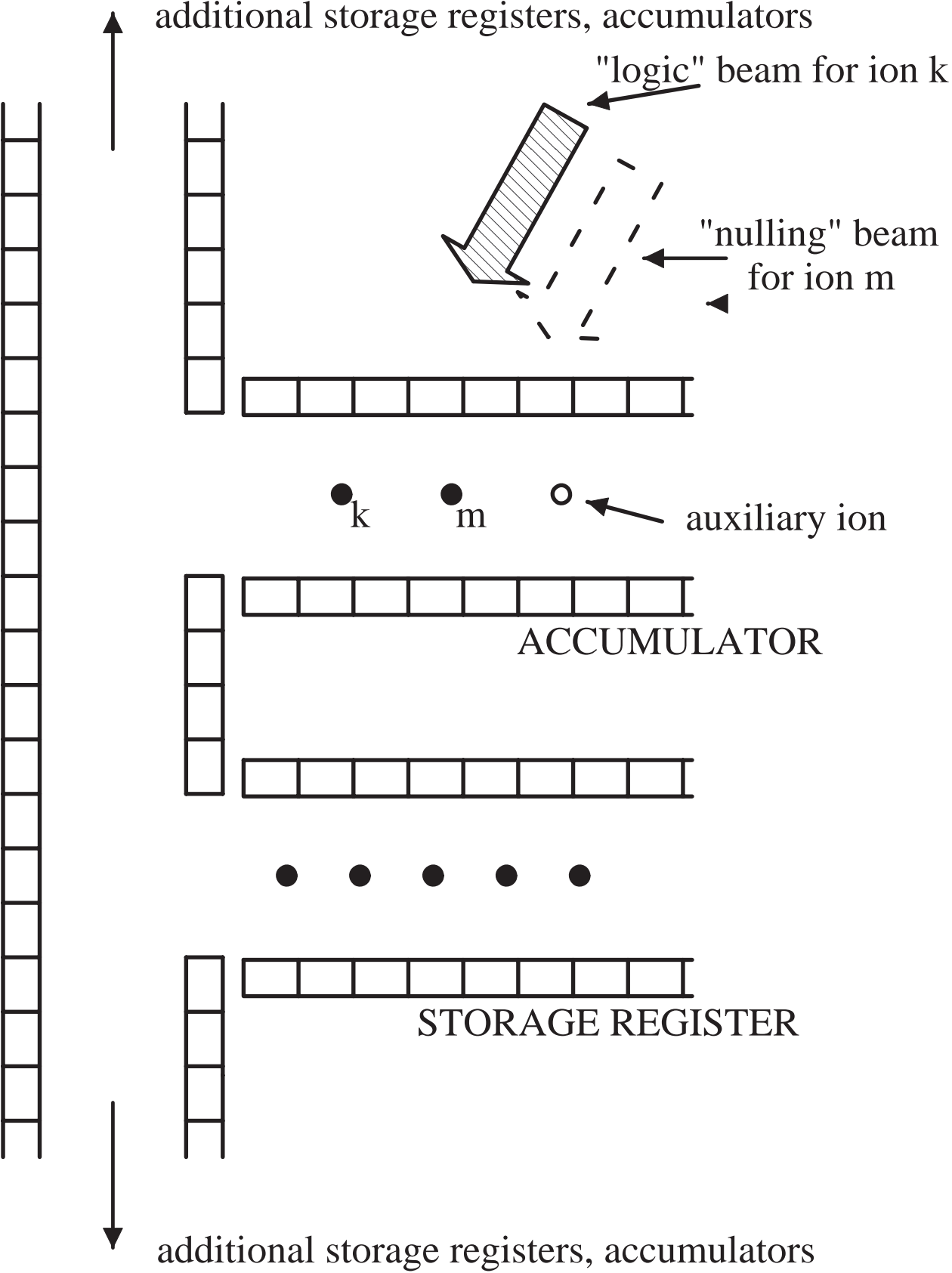

Fig. 10

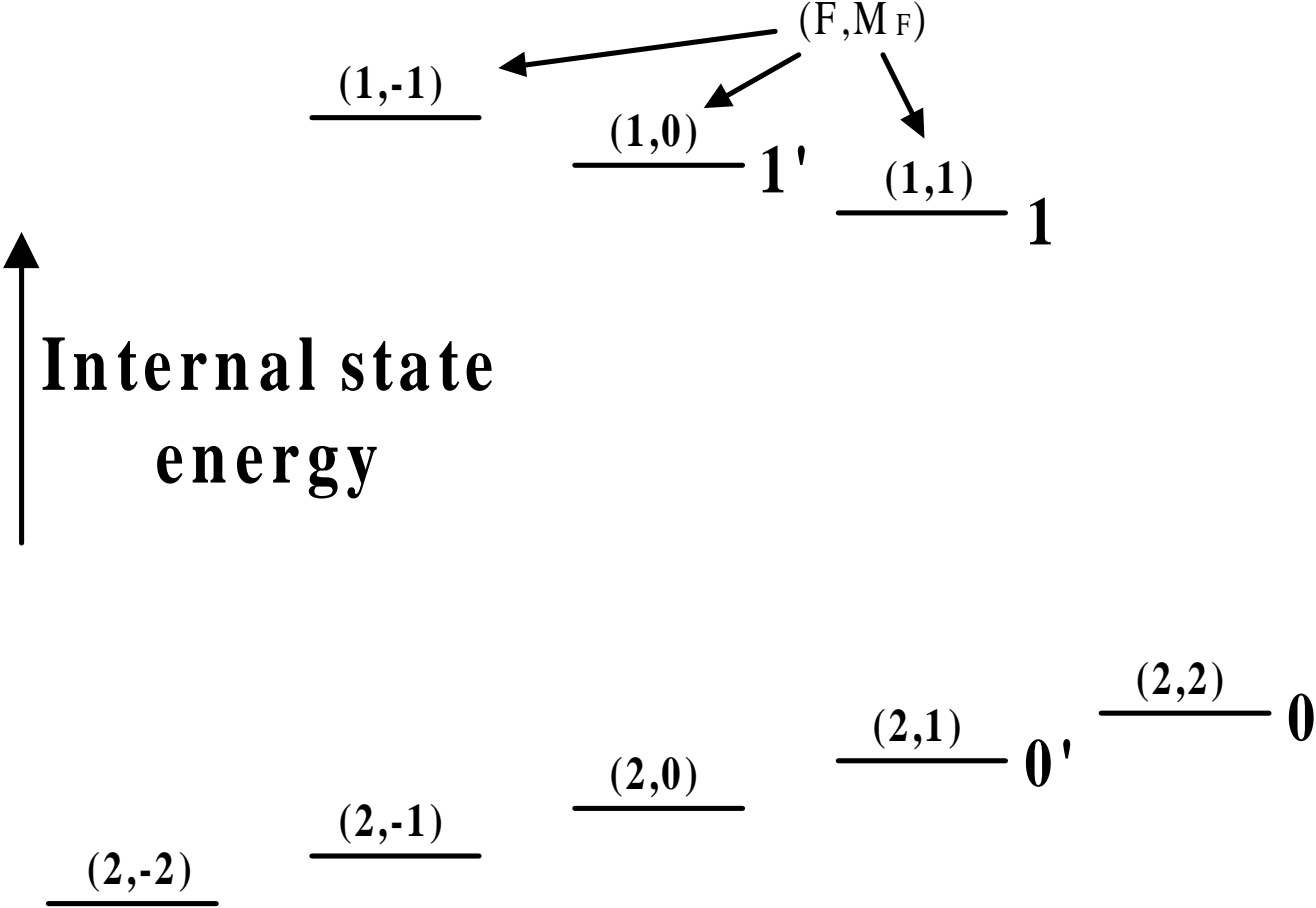

**Fig. 11**

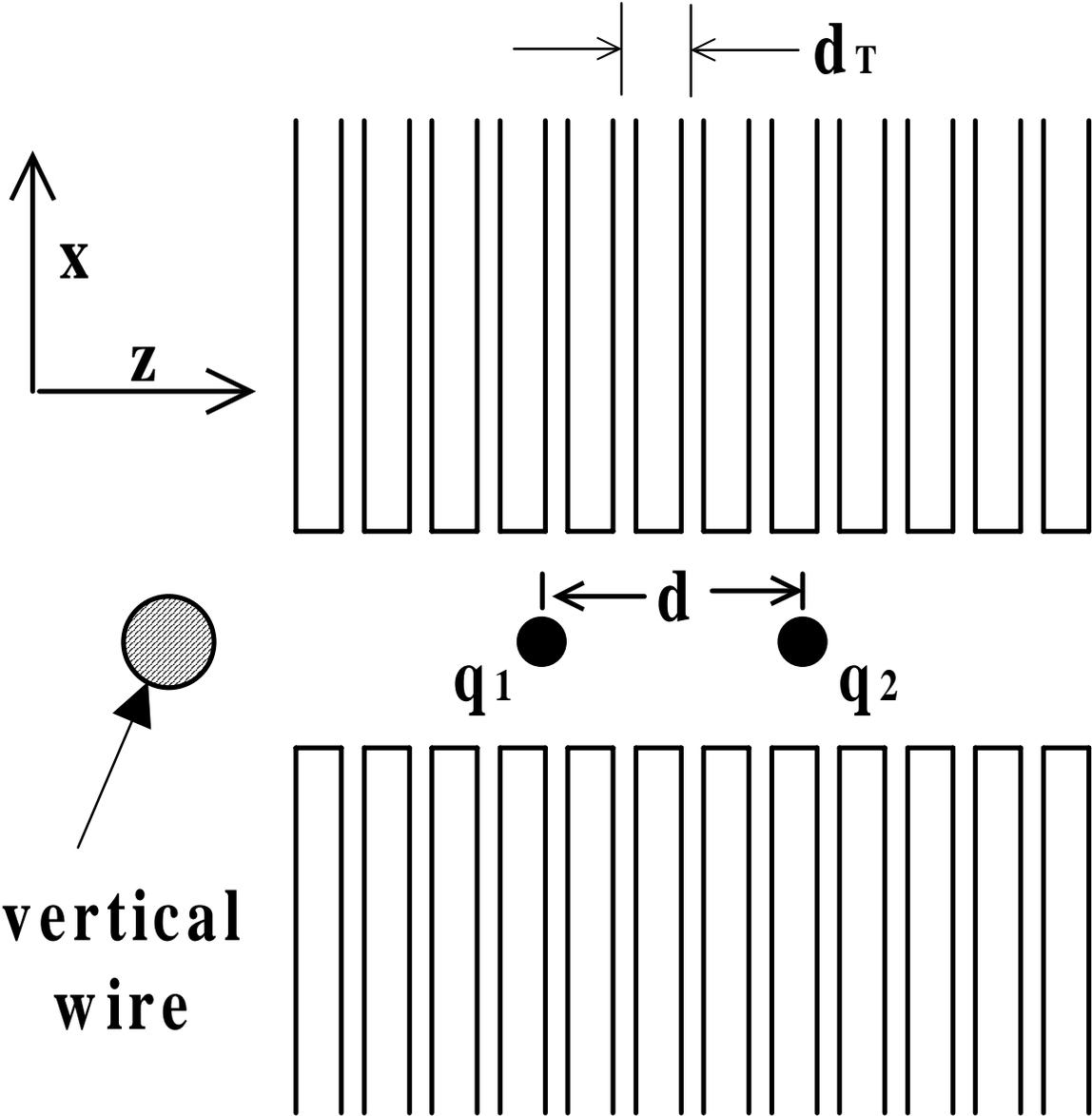

Fig. 12

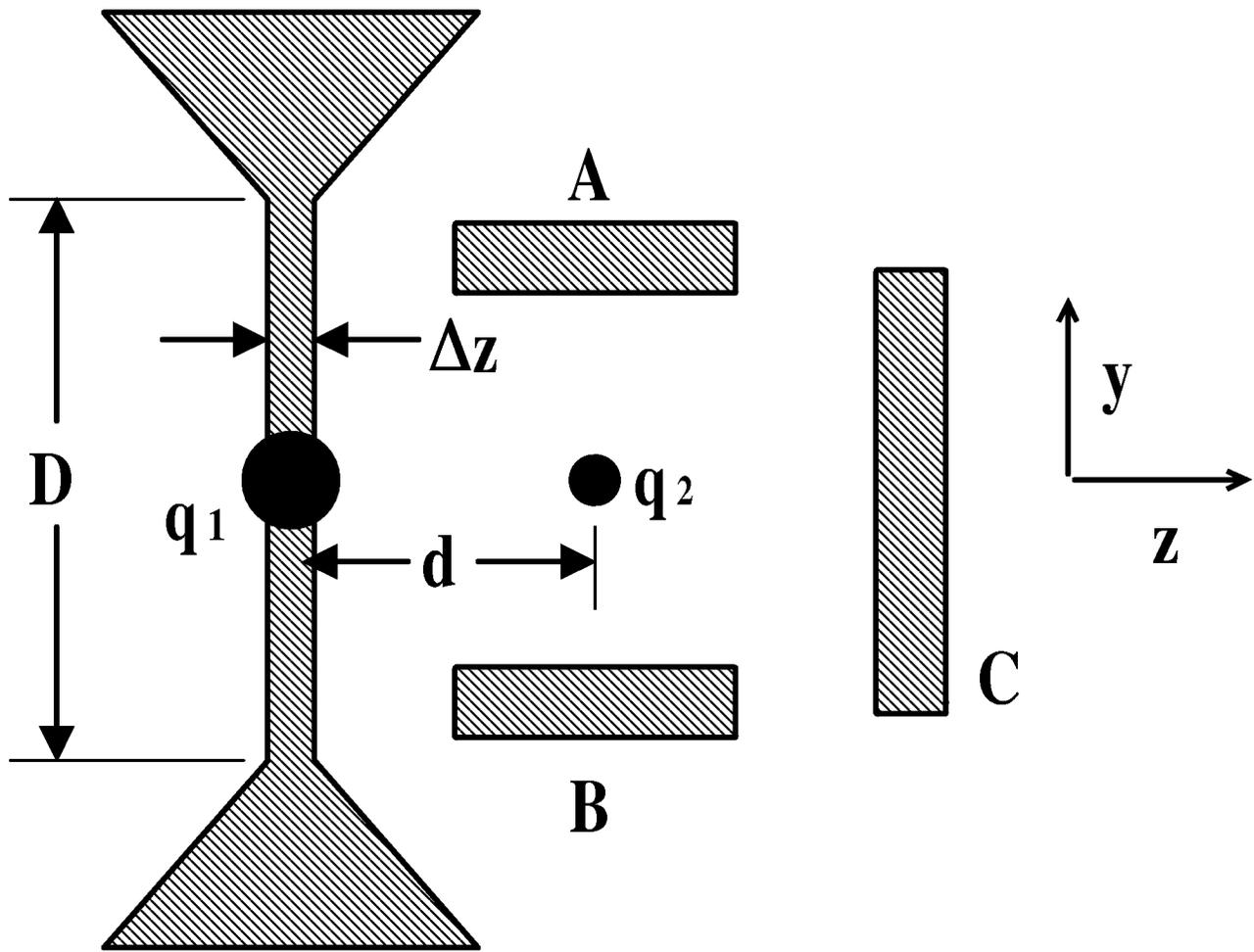

Fig. 13